\documentclass[runningheads]{llncs}

\usepackage{makeidx} 
\usepackage{multirow}
\usepackage{graphicx}
\usepackage{amsmath}
\usepackage{amssymb}
\usepackage{amsfonts}
\usepackage{url}
\usepackage{rotating}
\usepackage[vlined,boxed,commentsnumbered,ruled,linesnumbered]{algorithm2e}
\usepackage{caption}
\usepackage{subcaption}
\usepackage{lscape}
\usepackage{bm}
\usepackage{braket} % The Dirac bra-ket notations

\usepackage{amsthm}
\usepackage{float}

\usepackage{booktabs} 

\usepackage[colorlinks, citecolor=blue]{hyperref} % citation hyperref

\usepackage{listings}
\usepackage{color}

\usepackage{pgfplots, pgfplotstable}
\usepgfplotslibrary{fillbetween}
\usetikzlibrary{patterns}

\allowdisplaybreaks[4] 
\usepackage{soul}
\usepackage{autobreak}
\usepackage{pifont}
\usepackage{rotfloat}
\usepackage[bottom]{footmisc}
\usepackage{ifsym}
\usepackage{subcaption}
\usepackage{amsmath} %%
\usepackage{mathtools}
\usepackage{graphicx} % 旋转核心包
\usepackage{amssymb}
\usepackage{booktabs}   %用于更美观的表格线
\usepackage{makecell}   %用于单元格内换行
\usepackage{threeparttable}

\usepackage{comment}
\usepackage{ulem}
\usepackage{xcolor}
\usepackage{colortbl}
\usepackage{marvosym}
\usepackage{algorithmicx}
\usepackage{algpseudocode}

\usepackage{amssymb} 
\usepackage{lineno}

\usepackage{tikz}
\usetikzlibrary{positioning, calc, backgrounds, shapes.geometric}

\begin{document}

% \zirui{Add Rebuttal section here, remember to delete}
% \section*{Rebuttal}
% \input{sections/Rebuttal}

%\linenumbers
\title{Algebraic Attack on Convolutional Neural Networks with Max Pooling}

% \author{Anonymous Author(s)}
% \institute{Anonymous Institute}
\author{Zirui Chen\inst{1} \and
Shi Tang\inst{2} \and
Zhengchao Gao\inst{2} \and
Yongjia Su\inst{2} \and \\
Lingyue Qin\inst{1,3,4}\textsuperscript{(\Letter)} \and
Xiaoyang Dong\inst{1,3,4}\textsuperscript{(\Letter)}}
\authorrunning{Z. Chen et al.}

\institute{Tsinghua University, Beijing, P. R. China\\
% \email{chenzr25@mails.tsinghua.edu.cn}\\
% \email{qinly@tsinghua.edu.cn}\\
% \email{xiaoyangdong@tsinghua.edu.cn}\\
\email{chenzr25@mails.tsinghua.edu.cn}, \email{\{qinly,xiaoyangdong\}@tsinghua.edu.cn}\\
\and
Shandong University, Qingdao, P. R. China
\email{\{shi.tang,chao\_qwq,yongjia.su\}@mail.sdu.edu.cn}\\
% \email{shi.tang@mail.sdu.edu.cn}\\
% \email{chao\_qwq@mail.sdu.edu.cn}\\
% \email{yongjia.su@mail.sdu.edu.cn}
\and
Zhongguancun Laboratory, Beijing, P. R. China\\
\and 
State Key Laboratory of Cryptography and Digital Economy Security, Tsinghua
University, Beijing, P. R. China
}

\titlerunning{}

\maketitle         

\begin{abstract}

Recovering the weights and biases of deep neural networks (DNNs) via black-box input-output queries --- known as parameter extraction attacks --- has been extensively studied for ReLU-based fully connected neural networks (FCNNs), but remains unexplored for convolutional neural networks (CNNs) with the max pooling function, a core architecture for computer vision and multimedia processing. The key challenge lies in the CNN’s max pooling layer, which introduces an additional non-linearity and hides ReLU critical points, rendering existing FCNN extraction methods inapplicable. To address this gap, we propose the first cryptanalytic extraction attack tailored for CNNs with the max pooling function.

First, we establish an algebraic representation of CNNs, formally proving that CNNs are piecewise linear functions --- enabling the extension of linearity-based extraction principles. % while accounting for pooling operations. 
We then identify two novel types of critical points in CNNs: (1) ReLU-Pooling Critical Points (RPCPs), where a ReLU neuron is at its zero-input critical point and its output is selected by max pooling; and (2) Pooling Switching Points (PSPs), where two neurons within a local receptive field yield identical maximum outputs, triggering a switch in the pooling selection.

Leveraging these critical points, we design complementary extraction techniques: a pattern matching method for RPCPs to recover partial signatures and signs (exploiting the property that unselected pooling neurons have negative outputs), and an internal differential extraction attack for PSPs --- inspired by cryptographic internal differential analysis --- to recover high-accuracy signatures. 
%Given that PSPs are far more abundant and efficient than RPCPs (verified by experiments), and RPCPs are capable of bias recovery,
Given that PSPs are far more abundant than RPCPs and yield a highly efficient extraction method (verified by experiments), and that RPCPs are indispensable for bias recovery,
we integrate both methods: the PSP method enables efficient signature extraction, while a single RPCP recovers the sign and bias.

We evaluate our attack on multiple CNN architectures, including modern adaptations of LeNet-5, trained on random data, MNIST, and CIFAR-10. Experimental results demonstrate that our approach achieves high extraction accuracy with polynomial query complexity and runtime, even for deep CNN layers. This work fills a research gap in CNN security. %and provides a foundation for analyzing the robustness of CNNs against black-box extraction attacks.

\end{abstract}

 \section{Introduction}
\label{Introduction}

Deep Neural Networks (DNNs) compute a function from inputs
to outputs, and the architecture, weights and biases of the network determine the function that is expressed. Treating DNN as a black box and knowing its inputs and outputs, how to extract weights and biases has been a long-standing  problem, dating back as far as 30 years ago \cite{blum1988training,fefferman1994reconstructing}. In 2005, Lowd and Meck proposed an adversarial reverse engineering attack in \cite{lowd2005adversarial}.  Recently, new ideas from research teams in industry and academia have continued to emerge regarding this problem \cite{batina2019csi,jagielski2020high,oliynyk2023know,rolnick2020reverse,tramer2016stealing}. Targeting DNNs with alternating fully-connected linear layers and Rectified Linear Units (ReLU) activation (denoted as {\em ReLU-based fully connected neural network (FCNN)}), at CRYPTO 2020, Carlini, Jagielski, and Mironov achieved a breakthrough with a {\em cryptographic differential attack} \cite{DBLP:conf/crypto/CarliniJM20}, which requires querying a polynomial number of the raw outputs of FCNN to extract the unsigned weights. However, the sign recovery is a  brute-force guessing method, which makes  the overall time complexity of Carlini {\em et al.}'s algorithm exponential \cite{DBLP:conf/crypto/CarliniJM20}. 
At EUROCRYPT 2024, 
Canales-Mart{\'{\i}}nez {\em et al.} \cite{DBLP:conf/eurocrypt/CanalesMartinezCHRSS24} introduced the {\em  neuron wiggle technique} to recover the neuron signs in polynomial time. At NIPS 2024, Foerster {\em et al.} \cite{foerster2024beyond} built an end-to-end attack on practical models by combining the methods of Carlini {\em et al.} \cite{DBLP:conf/crypto/CarliniJM20} and  Canales-Mart{\'{\i}}nez et al. \cite{DBLP:conf/eurocrypt/CanalesMartinezCHRSS24}. Liu {\em et al.} \cite{DBLP:journals/iacr/LiuSELBP26} introduced more techniques to extract parameters in deep layers. %At ASIACRYPT 2025, Chen {\em et al.} \cite{chen2025delving} studied the parameter extraction on  FCNN with  Parametric Rectified Linear Unit (PReLU)  function \cite{he2015delving}. 
Recently, more model extraction attacks \cite{he2015delving,asselineau2026nonlinearactive,qi2026various,wei2026rnn} were proposed against different settings of neural networks. 

In another direction, still targeting {\em ReLU-based FCNNs}, where the attackers can only access the final classification labels ({\em e.g.}, ``dog'' or ``car''), {\em i.e.}, the {\em hard-label scenario}, Chen {\em et al.} \cite{DBLP:conf/asiacrypt/ChenDGSWW24} first proposed an extraction method requiring a polynomial number of queries but an exponential execution time. At EUROCRYPT 2025, Carlini {\em et al.} \cite{DBLP:conf/eurocrypt/CarliniCHRS25} introduced an extraction attack in the hard-label setting with a polynomial number of queries and polynomial time by analyzing the geometric shape of its decision boundaries. Later, Canales{-}Mart{\'{\i}}nez \cite{DBLP:conf/latincrypt/CanalesMartinezS25} proposed to recover the output layer, where no ReLU functions exist. In 2025, Ito, Miura, and Todo found that as the depth of the attack-target grows, the attack hypothesis of Carlini {\em et al.} \cite{DBLP:conf/eurocrypt/CarliniCHRS25} may become impractical, and they proposed a {\em cross-layer extraction} method to solve this problem. Until now, all the cryptanalytic extraction attacks focus on FCNNs, where the ReLU-like activation functions are the only non-linear factor in the network. In this paper, we consider the cryptanalytic extraction attacks on the {\em Convolutional Neural Networks (CNNs)} with the max pooling function for the first time.  

\subsection{CNNs and the Challenges in the Model Extraction}
 The concept of convolutional neural networks (CNNs) can be traced back to the concept of receptive fields in the 1960s. In the 1980s, Fukushima {\em et al.} \cite{fukushima1983neocognitron} proposed the concept of the neurocognitive machine. In 1998, LeCun officially proposed convolutional neural networks and designed the well-known  LeNet-5 \cite{lecun2002gradient}  network for handwritten digit recognition. 
CNNs have been among the most popular and modern neural networks, playing critical roles in the processing of images \cite{krizhevsky2012imagenet},  video \cite{Tran2015C3D}, time-series signals \cite{vanDenOord2016WaveNet}, text data \cite{Zhang2015CharCNN}, and other domains. Well-known architectures of CNNs include LeNet-5 \cite{lecun2002gradient}, AlexNet \cite{krizhevsky2012imagenet}, VGG \cite{Simonyan2015very}, GoogLeNet \cite{szegedy2015going}, ResNet \cite{he2016deepresidual}, DenseNet \cite{huang2017densenet}, etc. 

While deeper and larger CNN models are continually developed for higher feature extraction performance, shallow architectures remain highly relevant and are actively deployed in scenarios with strict constraints on computational resources, power, or cryptographic overhead, such as edge computing \cite{lai2018cmsis,banbury2021mlperf}, low-power wearable devices \cite{kim2022lightweight}, and privacy-preserving inference \cite{liu2017oblivious,pmlr-v48-gilad-bachrach16}.
% For instance, in edge computing, CMSIS-NN \cite{lai2018cmsis} relies on a compact architecture of 3 CNN and 1 FCNN layers for efficient local inference on microcontrollers; 
% in privacy-preserving inference, frameworks like MiniONN \cite{liu2017oblivious} often evaluate on shallow networks (e.g., 2 CNN and 2 FCNN layers) to mitigate the massive computational overhead introduced by cryptographic protocols.

% For instance, in edge computing, CMSIS-NN \cite{lai2018cmsis} uses 3 CNN + 1 FCNN layers for efficient local inference on microcontrollers; 
% in low-power wearable devices,  \cite{kim2022lightweight} utilizes 5 CNN + 1 FCNN for real-time ECG classification;
% and in privacy-preserving inference, MiniONN \cite{liu2017oblivious} employs 2 CNN + 2 FCNN to mitigate the massive computational overhead of cryptographic protocols.

\begin{figure}
	\centering
        \includegraphics[width=0.8\linewidth]{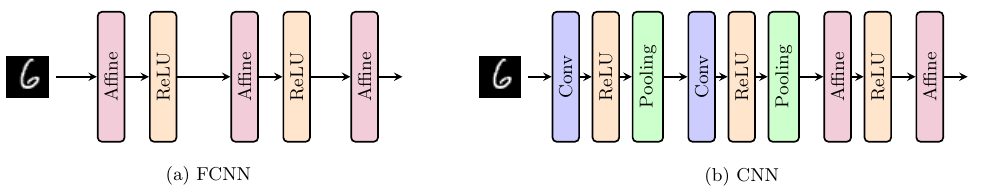}
	\caption{FCNN and CNN}
	\label{fig:fnn_cnn}
\end{figure}

In Figure \ref{fig:fnn_cnn},  the key difference between the CNN and the ReLU-based FCNN is that the CNN includes a pooling layer (note that the convolutional layer can be transformed into a linear affine layer in Sect.  \ref{sect:def_notations}). In the pooling layer,  a pooling function (usually using the max or average function) is used to compress the outputs of all the neurons in a  local receptive field, and output a single value. Typically, the max pooling function is widely used in practice due to its unique advantages in feature extraction, anti-interference ability, and prevention of overfitting, which have been verified and applied in many classic CNN papers \cite{lecun2002gradient,boureau2010theoretical,krizhevsky2012imagenet,Simonyan2015very,matoba2023benefits}. However, when max pooling function\footnote{In fact, different from max pooling,  the average pooling does not introduce additional non-linearity to the network, and Carlini {\em et al.}'s technique \cite{DBLP:conf/crypto/CarliniJM20} can be trivially applied to this case.}  is applied, there will be another non-linear function in the network besides the ReLU function, which brings two problems when applying Carlini {\em et al.}'s parameter extraction attacks. 
\begin{itemize}
    \item The first problem is that the critical point of the ReLU function will be hidden by max pooling in CNNs. 
    By varying the inputs in the tiny vicinity of the critical point, it is expected that the network's output will change non-linearly as a result of this ReLU transition. However, for CNNs, the outputs of the neurons will be selected by a max pooling function, and only the maximum output of the neurons in a local receptive field will be selected as the output. There is a significant probability that the output (whose value is around 0) of the neuron at the critical point will not be selected, and then the network's output will change linearly even though a neuron is at its critical point. This case happens easily when one of the outputs of the other neurons within the same local receptive field is greater than 0. 
      \item The second problem is that the max pooling layer introduces an additional non-linearity, and a similar critical point can be spotted when two outputs of the neurons within a local receptive field are the two maximum values ({\em e.g.}, $a=b$). By varying the inputs in the tiny vicinity of the critical point, $a$ will exceed $b$, or $b$ will exceed $a$. Then, the max pooling layer's selection will switch between $a$ and $b$. Obviously, this kind of critical point is quite different from Carlini {\em et al.}'s critical point. How to leverage this new critical point to extract parameters will be a new problem.
\end{itemize}

\subsection{Our Contributions}
\label{subsec:our contributions}
%We introduce the algebraic attack that is effective at performing functionally equivalent neural network model extraction attacks on Convolutional Neural Network. 
We introduce an algebraic representation of the CNN, transform all the states and operations into vectors and matrices, and thereby formally make it clear that the CNN is also a piecewise linear function, {\em i.e.}, within a tiny linear neighborhood of input $X$, the CNN acts as a linear function, and all operations in the CNN are matrix-vector multiplications. 
We formally introduce two kinds of critical points, {\em i.e.}, 
    \begin{itemize}
        \item[$\circ$] 
        {\bf ReLU-Pooling Critical Point (RPCP):} It is the point where a neuron is at its critical point ($\text{input}=0$), and subsequent max pooling layers select its output, having a non-linear effect on the CNN's output. 
        Based on RPCPs, Carlini {\em et al.}'s differential attack is applied to recover the partial signatures (some entries are impossible to recover with a single RPCP, since the ReLU functions suppress negative outputs \cite[Section 4.3.2]{DBLP:conf/crypto/CarliniJM20}). We propose a new {\em pattern matching method} for CNNs to merge multiple partial signatures to get the full one. Furthermore, additional RPCPs are needed to improve the accuracy by applying Carlini {\em et al.}'s least squares method. 
        Interestingly, the sign recovery is quite easy based on a property of RPCP, {\em i.e.}, the unselected neurons' outputs by the max pooling function should be negative.  

        \item[$\circ$]   
        {\bf Pooling Switching Point (PSP):} It is the point where the outputs of two neurons within a Local Receptive Field of Pooling Layer achieve the same maximum value. To leverage the PSP, we invent a new parameter extraction technique inspired by the cryptographic internal differential attack \cite{DBLP:conf/fse/DinurDS13,DBLP:conf/crypto/Peyrin10}, named {\em internal differential extraction} attack, where the propagation of the difference between different parts of a single PSP is studied through the CNN. 

\end{itemize}
 {\bf Advantages and limitations of the RPCP and PSP methods:} 
\begin{itemize}
    \item[$-$] According to our experiments, RPCPs are relatively rare in CNNs, since max pooling hides the output of the neuron at the critical point with significant probability. Therefore, merging partial signatures or improving the accuracy of signatures that require many RPCPs is difficult.  
    \item[$+$] The RPCP method can easily recover the sign and bias based on its property. 
    \item[$+$] According to our experiments, PSPs are more common in CNNs than RPCPs. %, as we have verified in our practical experiments on several CNNs. 
    Based on the {\em internal differential extraction} attack, the signature (without bias) can be extracted with higher accuracy, shorter runtime, and fewer model queries than using RPCP methods. 
    \item[$-$] The PSP method cannot recover the bias.  
\end{itemize}
\paragraph{\bf Unified Model Combining RPCP and PSP.} Given the above advantages and limitations, we integrate the RPCP and PSP methods into a unified parameter extraction model, where the PSP method is used to recover the signature (without bias) and sign, while only a single (or a few) RPCPs are needed to recover the bias. Besides, we introduce a method named the {\em targeted heuristic search strategy for RPCPs}, to handle the case that even one RPCP is hard to find. This method leverages the signature recovered by PSPs to directly detect the points close to an RPCP. 

\paragraph{\bf End-to-end Experiments.} We perform practical parameter extractions on several CNNs, including the popular LeNet-5 with modern architectures. Taking the full ($m+n$)-deep CNN ($m$ Convolutional Rounds and $n$ FCNN Rounds) as a black box, with the inputs and raw outputs, we present two kinds of experiments: 
\begin{itemize}
    \item[$\circ$] We recover all parameters of the $m$ Convolutional Rounds of the first 5 CNNs (including the modern LeNet-5) in Table \ref{tab:Attack on CNNs}, where (\textcolor{red}{$m$}+$n$) means only the parameters of the \textcolor{red}{$m$} Convolutional Rounds are recovered. 
    \item[$\circ$] We give an experiment recovering the full parameters of an (\textcolor{red}{$m+n$})-deep CNN, which is a reduced version of LeNet-5 with 2 standard Convolutional Rounds, 1 reduced FCNN Round (the modern LeNet-5 has 2 FCNN Rounds), and an output layer ({\em i.e.}, 3 hidden layers). To our knowledge, this matches the depth of the deepest architecture - 3 hidden layers \cite{DBLP:conf/crypto/CarliniJM20}, among existing end-to-end black-box full-model parameter extraction works.\\ 
    We also attack the standard modern LeNet-5 (specified in Sect. \ref{subsect: modern LeNet-5}), extracting the full Convolutional Block and 93.33\% weights of FCNN Round 1 (incomplete due to dead neurons and persistent neurons according to \cite{ito2025hard}). \\
    We present all results and the comparisons with Carlini {\em et al.}'s work \cite{DBLP:conf/crypto/CarliniJM20} and Foerster {\em et al.}'s work \cite{foerster2024beyond} in the last four rows of Table \ref{tab:Attack on CNNs}.
\end{itemize}
All the source codes and experiments can be found via:
\begin{center}
    % \url{https://anonymous.4open.science/r/Algebraic-Attack-on-CNN-7F2610}
    \url{https://github.com/czr-eric/Algebraic-Attack-on-CNN.git}
\end{center}

\subsubsection{Related and Concurrent Works.} At ACISP 2026, Sun {\em et al.}  \cite{cnn_average_pooling} proposed a cryptanalytic extraction of convolutional neural networks with {\bf average pooling}. However, on the commonly used max pooling function, they claimed that \cite[Section 5]{cnn_average_pooling}: 
\begin{quote}
    ``{\em However, the non-linear nature of max pooling layers, which only retains the maximum value within each region, inherently leads to significant information loss and makes the operation irreversible. As a result, reconstructing the original input or parameters from these layers presents considerable difficulties, making the handling of such layers a persistent challenge for future research.}''
\end{quote}
This paper addresses the challenges posed by the max-pooling function and, for the first time, recovers the parameters of a CNN employing max pooling layers.

Recently, Liu {\em et al.} \cite{liu2026cnn} also proposed a model extraction attack on max-pooling CNNs. They introduced a distinct localization strategy for critical points based on receptive-field analysis, interpreted the max-pooling CNN as a structural generalization of the ReLU setting, and extended the geometric extraction framework accordingly. However, we approach the attack from an algebraic perspective, introducing novel techniques such as the pattern matching method, internal differential \cite{DBLP:conf/crypto/Peyrin10} extraction, {\em etc}.

% Attack on Convolutional Block
\begin{table} [!b]
\centering
\vspace*{-18pt}
\captionsetup{labelfont=bf}
\caption{{\bf Experiments on different CNNs.} Convolution kernel dimensions are denoted as $(\mathsf{c}_{in}^{(k)}, \mathsf{c}_{out}^{(k)}, h_{c}^{(k)}, w_{c}^{(k)})$, with stride $s^{(k)}_c=1$. Pooling kernels are $2 \times 2$ with stride $s^{(k)}_{\rho}=2$. The accuracy term $(\varepsilon,0)$ is defined in Sect. \ref{sect:assumptions_1}, and $\max \lvert \theta - \hat{\theta} \rvert$ directly measures the maximum extraction error of model parameters.}
\label{tab:Attack on CNNs}
% 1. 设置表格整体行高 (Models 之间的距离)
\renewcommand{\arraystretch}{0.8} 
\setlength{\tabcolsep}{2pt}

% 2. 定义一个统一的换行间距命令，方便统一调整内部行高
% 修改这个值 (比如 6pt, 1em, 12pt) 可以同时改变两列的内部行间距
\newcommand{\ggap}{\\[0.8pt]} 

\begin{threeparttable}
    \resizebox{\textwidth}{!}{
    \begin{tabular}{l c l c l l l} % 将 Architecture 和 Kernel 改为左对齐(l)会更整齐
    \toprule
    \makecell[l]{\textbf{Models} \\
    ($m+n$) 
    }
    & \textbf{Trainset}
    & \makecell[c]{\textbf{Architecture}$^\dag$\\ 
    $d^{(k)}$--$d_f^{(k)}$--$d^{(k+1)}$}
    & \textbf{Kernel}
    & \textbf{Queries}
    & $(\varepsilon, 0)$
    & $\max \lvert \theta - \hat{\theta} \rvert$ 
    \\
    \midrule
    
    (\textcolor{red}{\bf 1}+1)
    & Random 
    & $C_1$: 64--36--9
    & (1,1,3,3)
    & $2^{9.98}$ & $2^{-43.19}$ & $2^{-47.52}$
    \\
    \midrule % 添加分割线让不同模型区分更明显
    
    (\textcolor{red}{\bf 2}+1)
    & MNIST 
    & \makecell[l]{ % 左对齐
        $C_1$: 1024--784--196 \ggap % 使用自定义间距
        $C_2$: 196--100--25
      }
    & \makecell[c]{ % 对应使用左对齐和相同的间距
        (1,1,5,5) \ggap
        (1,1,5,5)
      }
    & $2^{15.06}$ & $2^{-30.20}$ & $2^{-34.65}$
    \\
    \midrule
    
    (\textcolor{red}{\bf 2}+2)$^{\ast}$ & MNIST 
    & \makecell[l]{
        $C_1$: 1024--4704--1176 \ggap
        $C_2$: 1176--1600--400
      }
    & \makecell[c]{
        (1,6,5,5) \ggap
        (6,16,5,5)
      }
    & $2^{27.29}$ & $2^{-23.38}$ & $2^{-27.31}$
    \\
    \midrule
    
    (\textcolor{red}{\bf 3}+1)${}^{p_2}$
    & MNIST 
    & \makecell[l]{
        $C_1$: 1024--1024--256 \ggap
        $C_2$: 256--256--64 \ggap
        $C_3$: 64--64--16
      } 
    & \makecell[c]{
        (1,1,5,5) \ggap
        (1,1,5,5) \ggap
        (1,1,5,5)
      }
    & $2^{16.90}$ & $2^{-27.54}$ & $2^{-31.15}$
    \\
    \midrule

    (\textcolor{red}{\bf 2}+2)${}^{p_1}$  
    & CIFAR10 
    & \makecell[l]{
        $C_1$: 3072--3072--768 \ggap
        $C_2$: 768--1024--256
      }
    & \makecell[c]{
        (3,3,3,3) \ggap
        (3,4,3,3)
      }
    & $2^{18.37}$ & $2^{-29.13}$ & $2^{-36.19}$
    \\\midrule[0.9pt]
    
    (\textcolor{red}{\bf 2+1})$^\ddag$ & MNIST &
    \makecell[l]{
        $C_1$: 1024--4704--1176 \ggap
        $C_2$: 1176--1600--400 \ggap
        F: 400--20--10 
      } 
      
    & \makecell[c]{ % 对应使用左对齐和相同的间距
        (1,6,5,5) \ggap
        (6,16,5,5) \ggap
        --
      }
    & $2^{27.65}$ & $2^{-22.06}$ & $2^{-27.14}$
    \\ \midrule

    (\textcolor{red}{\bf 2+2})$^{\ast}$$^{\clubsuit}$ & MNIST &
    \makecell[l]{
        \textcolor{red}{$\bf C_1$}: 1024--4704--1176 \ggap
        \textcolor{red}{$\bf C_2$}: 1176--1600--400 \ggap
        F: \textcolor{red}{\bf 400--120}--84--10 
      } 
      
    & \makecell[c]{ % 对应使用左对齐和相同的间距
        (1,6,5,5) \ggap
        (6,16,5,5) \ggap
        --
      }
    & $2^{27.31}$ & -- & $2^{-26.03}$
    \\ \midrule

    (0+\textcolor{red}{\bf 3}) \cite{DBLP:conf/crypto/CarliniJM20} & MNIST 
    & \makecell[l]{
        F: 40--20--10--10--1 
      } 
    & --
    & $2^{17.8}$ & $2^{-23.4}$ & $2^{-27.1}$
    \\ 
    \midrule

    (0+\textcolor{red}{\bf 3})
    \cite{foerster2024beyond} $^{\spadesuit}$& MNIST &
    \makecell[l]{
        F: 784--16--\textcolor{red}{\bf 16--16}--1 
      } 
      
    & --
    & $2^{25.4}$ & -- & --\\

    \bottomrule
    \end{tabular}
    }
    \begin{tablenotes}[flushleft] 
        \item[$\dag$:] $C_i$ indicates Convolutional Round $i$; $F$ denotes FCNN Block with $d^{(1)}-\cdots - d^{(n+1)}$.
        \item[$\ast$:] LeNet-5 modern model, specified in Sect. \ref{subsect: modern LeNet-5}.
        \item[$p_1, p_2$:] $p_i$ indicates $pad^{(k)}=i$ defined as ``same padding'' in \textsf{Supplementary Material} \ref{supp:other_parameters}; otherwise, $pad^{(k)}=0$.
        \item[$\ddag$:] Reduced version of LeNet-5 with only one FCNN Round.
        \item[$\clubsuit$:] The red bold layer in Architecture indicates end-to-end extracting of the full Convolutional Block and 93.33\% weights of FCNN Round 1 (not full due to existing dead neurons and persistent neurons according to \cite{ito2025hard}).  
        \item[$\spadesuit$:] Red bold layer in Architecture indicates extracting the single layer, with prior layers' extractions assumed to be correct.
    \end{tablenotes}
\end{threeparttable}
\vspace*{-6pt}
\end{table}

\section{Preliminaries}
\label{Attack Overview}

% \input{figures/figure1.tex}
%Observing Convolutional Neural Network (CNN) from different perspectives, each angle emphasizes distinct aspects of description. This facilitates a more comprehensive and in-depth understanding of their underlying logic and essence.

%\textbf{Symbolic View.} This perspective directly traces the flow of information through the neural network. It explicitly illustrates the topological structure, including the quantity of neurons and the specific connectivity patterns between layers.

%\textbf{Operational View.} Unlike fully-connected networks, convolution and pooling are no longer intuitive multiplication or addition operations. Instead, the operations performed on inputs are often conducted through descriptions, demonstrations with specific examples, or even animations.

%\textbf{Geometric View.} Since neural networks operate within real vector spaces, their behavior can be visualized by plotting two-dimensional slices of the high-dimensional landscape. This geometric interpretation will clearly show the decision boundaries and helps us visually understand how the CNN transforms and manifolds the input space.

%\textbf{Algebraic View.} Finally, from an algebraic standpoint, we will see how our algebraic attack describes or reflects the information from all other perspectives.

\subsection{Basic Definitions and Notations}\label{sect:def_notations}

%\begin{table}[]
  %  \centering
 %   \begin{tabular}{rl}
 \begin{itemize}
     \item  {\bf FCNN}: Fully Connected Neural Network (FCNN) in Figure \ref{fig:fnn_cnn}.
     \item   {\bf CNN}: Convolutional Neural Network (CNN) in Figure \ref{fig:fnn_cnn}. 
       \item  {\bf ReLU}: Rectified Linear Unit. 
          \item {\bf FCNN Round}: Including a linear function $f^{(k)}$ called fully-connected layer, and a non-linear function $\sigma$ (component-wise ReLU function). 
       \item  {\bf Convolutional Round}: Including a convolutional layer $f_c^{(k)}$, a non-linear activation layer $\sigma_c^{(k)}$ (ReLU), and a pooling layer $\rho^{(k)}$ (max pooling). 
    \item {\bf $\mathbf{(m+n)}$-Deep Convolutional Neural Network (CNN)}: A CNN with $m$ Convolutional Rounds and $n$ FCNN Rounds, as well as an output layer with a  fully-connected linear layer. 
    \item  $\bf(\mathsf{c}_{in}^{(k)}, \mathsf{c}_{out}^{(k)}, h_{c}^{(k)}, w_{c}^{(k)})$: Convolutional layer with  $\mathsf{c}_{in}^{(k)}$ input and  $\mathsf{c}_{out}^{(k)}$ output channels, and the size of kernel matrices is $h_{c}^{(k)}\times w_{c}^{(k)}$. 
    \item  $\mathbf{X^{(k)},Y^{(k)},Z^{(k)}}$:  Input vectors of the operations $f_c^{(k)}$, $\sigma_c^{(k)}$, $\rho^{(k)}$ in the $k$-th Convolutional Round, $k\geq 1$.  $X^{(k)}\in \mathbb{R}^{d^{(k)}},~Y^{(k)},~Z^{(k)} \in \mathbb{R}^{d_f^{(k)}}$, since $\sigma_c^{(k)}$ does not change the dimension. $X^{(k)}[i]$ is the $i$-th ($0\leq i < d^{(k)}$) entry of  $X^{(k)}$. 
    \item   $\mathbf{A^{(k)}}$: Matrix whose entry in  $i$-th row and $j$-th is $A^{(k)}_{i,j}$, $i,j$ start from 0. 
%\item $A_c^{(i)}$
    \item {\bf RPCP, PSP, FCP}: ReLU-Pooling Critical Point, Pooling Switching Point, Critical points in the Fully-connected Block, respectively.
% \item {\bf PSP}: Pooling Switching Point 
% \item {\bf FCP}: Critical points in the Fully-connected Block 
    \item {\bf LRF-C}, Local Receptive Field of Convolutional Layer: A sub-matrix of $IN^{(k)}$,  which is multiplied by {\em Convolution Kernel Matrix} $C^{(k)}$ to produce each entry in $Y^{(k)}$. The LRF-C will be different for different elements of $Y^{(k)}$. Let $Y^{(k)}[t]$ correspond to the $t$-th ($0\leq t\leq d^{(k)}_f-1$) LRF-C. 
    \item {\bf LRF-P}, Local Receptive Field of Pooling Layer: A sub-matrix of $Z^{(k)}$, which is processed by a pooling function to produce each entry of $X^{(k+1)}$. The LRF-P will be different for different elements of $X^{(k+1)}$. Let $X^{(k+1)}[t]$ correspond to the $t$-th ($0\leq t\leq d^{(k+1)}-1$) LRF-P. 
\end{itemize}

  %  \end{tabular}
  %  \caption{Symbols and Notations}
  %  \label{tab:placeholder}
%\end{table}

%\subsection{Brief Comparison Between CNN and Fully Connected Network}

\subsection{Algebraic View of CNN}\label{sect:def_notations}

To facilitate the understanding of the algebraic formulations in this section, we provide a concrete toy example of a CNN in \textsf{Supplementary Material} \ref{sec:appendix_toy_example}.

Previous works
\cite{DBLP:conf/crypto/CarliniJM20,DBLP:conf/eurocrypt/CanalesMartinezCHRSS24,DBLP:conf/asiacrypt/ChenDGSWW24,DBLP:conf/eurocrypt/CarliniCHRS25,chen2025delving} concentrate on the Fully Connected Neural Network (FCNN) in Figure \ref{fig:fnn_cnn}, which is  composed of a sequence of functions alternating between linear functions $f^{(k)}: \mathbb{R}^{d^{(k)}}\mapsto \mathbb{R}^{d^{(k+1)}}$ ($k\geq 1$, called fully-connected layers), and a non-linear function $\sigma^{(k)}$ (component-wise ReLU function): 
\begin{equation}
    \mathcal{F}_{\text{fc}}=f^{(n+1)}\circ \sigma^{(n)}\circ f^{(n)}\circ \sigma^{(n-1)}\circ\cdots f^{(2)}\circ \sigma^{(1)}\circ f^{(1)}. 
\end{equation}

 %a typical CNN architecture consists of an initial sequence of convolutional and pooling layers, followed by a series of fully-connected layers.

In Figure \ref{fig:fnn_cnn}, a typical CNN architecture consists of an initial sequence of convolutional, non-linear ReLU, and pooling layers (this paper focuses on the max pooling), followed by a series of fully-connected layers, which are referred to as the {\em Convolutional Block} and the {\em Fully-connected Block}, respectively. 

%It is important to note that both convolutional and fully-connected layers are typically followed by non-linear activation functions. Without these non-linearities, adjacent linear layers could mathematically collapse into a single linear transformation.
    
\begin{definition} 
    {\bfseries\upshape ($(m+n)$-Deep Convolutional Neural Network)}
    A $(m+n)$-deep Convolutional Neural Network (CNN) is a function $\mathcal{F}_{\theta}$ parameterized by $\theta$ that takes inputs from an input space $\mathbb{R}^{d^{(1)}}$ and returns values in an output space $\mathbb{R}^{d^{(m+n+2)}}$. 
    The function $\mathcal{F}_{\theta}: \mathbb{R}^{d^{(1)}} \rightarrow \mathbb{R}^{d^{(m+n+2)}}$ is composed of two sequential blocks: the first block (Convolutional Block, denoted as $\mathcal{F}_{\text{conv}}$) consists of $m$ rounds of alternating convolutional layers $f_c^{(k)}$, non-linear activation layers $\sigma_c^{(k)}$, and pooling layers $\rho^{(k)}$ ($1\leq k\leq m$); the second block (Fully-connected Block, denoted as $\mathcal{F}_{\text{fc}}$) consists of $n$ rounds of alternating linear layers $f^{(k)}$ and activation functions $\sigma^{(k)}$. Then,
    \begin{align}
        \label{eqn:cnn_f}
        \resizebox{1.0\hsize}{!}{$%
        \mathcal{F}_{\theta} 
        =\mathcal{F}_{\text{fc}} \circ \mathcal{F}_{\text{conv}}  
        = \underbrace{f^{(n+1)} \circ \sigma^{(n)} \circ \cdots \circ \sigma^{(1)} \circ f^{(1)}}_{\mathclap{\text{Fully-connected Block},~\mathcal{F}_{\text{fc}}}} \,
        \circ 
        \underbrace{\rho^{(m)} \circ \sigma_c^{(m)} \circ f_c^{(m)} \circ \cdots \circ \rho^{(1)} \circ \sigma_c^{(1)} \circ f_c^{(1)}}_{\mathclap{\text{Convolutional Block},~\mathcal{F}_{\text{conv}}}}.
        $}%
    \end{align}
\end{definition}

%From a symbolic view, each unit in a layer receives inputs from a set of units located in a small neighborhood in the previous layer. CNNs adopt partial connectivity while FCNNs adopt full connectivity. Notably, when the input size matches the kernel size, this operation becomes equivalent to a fully-connected layer.

%\xiaoyang{Give definitions of Channels and only use the special case $c_{in}=c_{out}=1$ to simplify the symbols. @Lingyue}

%\xiaoyang{Figures}

\begin{figure}
    \vspace*{-12pt}
	\centering
        \includegraphics[width=0.95\linewidth]{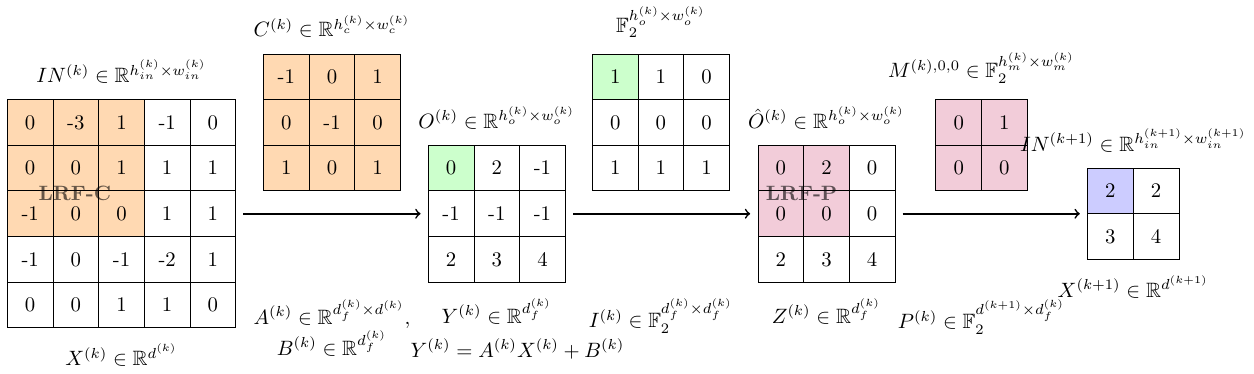}
	\caption{Symbolic View of CNN}
	\label{fig:symbol_cnn}
    \vspace*{-12pt}
\end{figure}

%\begin{definition} 
\subsubsection{Convolutional Layers.}  The CNN's inputs are formulated as matrices. The convolution operation of the convolutional layer $f_c^{(k)}$ employs a sliding window mechanism, where a small matrix of learnable parameters (the {\em Convolution Kernel Matrix}) traverses the input from left to right, top to bottom. At each position, the sum of element-wise products between the Convolution Kernel Matrix and the overlapping local input patch ({\em Local Receptive Field of Convolutional Layer}, abbreviated as {\em LRF-C}), is computed to generate a single output value. A shared bias is then added to the result at each position to constitute the output matrix ({\em Feature Map}) of the convolutional layer. 

In the algebraic view as shown in Figure \ref{fig:symbol_cnn},  the input matrix of $f_c^{(k)}$ is denoted as $IN^{(k)}$  with dimension $h_{in}^{(k)} \times w_{in}^{(k)}$, so $d^{(k)} = h_{in}^{(k)} \cdot w_{in}^{(k)}$. The convolution kernel matrix is defined by the shape $(\mathsf{c}_{in}^{(k)}, \mathsf{c}_{out}^{(k)}, h_{c}^{(k)}, w_{c}^{(k)})$, where $\mathsf{c}_{in}^{(k)}$ and $\mathsf{c}_{out}^{(k)}$ denote the number of input and output channels, and $h_{c}^{(k)}, w_{c}^{(k)}$ denote the height and width of the convolution kernel matrix $C^{(k)}$, respectively. 
%\end{definition}

For briefness, we will first consider the simple case of single input and output channel, {\em i.e.}, $\mathsf{c}_{in}^{(k)}=\mathsf{c}_{out}^{(k)}=1$, and the case of multiple channels will be analyzed in {\sf Supp.} \ref{sect:LeNet5}. Given a convolution stride $s_c^{(k)}$, the $h^{(k)}_{o} \times w^{(k)}_{o}$ output matrix or the {\em feature map} of $f_c^{(k)}$ is denoted as $O^{(k)}$, where 
    \begin{align}
        h^{(k)}_{o} = \left\lfloor \frac{h^{(k)}_{in} - h^{(k)}_{c}}{s^{(k)}_c} \right\rfloor + 1, \quad 
        w^{(k)}_{o} = \left\lfloor \frac{w^{(k)}_{in} - w^{(k)}_{c}}{s^{(k)}_c} \right\rfloor + 1.
    \end{align}
%Let $d^{(k)}_{f} = h^{(k)}_{c}\cdot w^{(k)}_{c}$ denote the dimension of the flattened output vector corresponding to the $k$-th convolutional layer.

For briefness, we let the convolution stride $s_c^{(k)}=1$. The input of $f_c^{(k)}$ can be a $h_{in}^{(k)} \times w_{in}^{(k)}$ $IN^{(k)}$ or a $d^{(k)}$-dimension vector $X^{(k)}$. Denote 
$IN^{(k)}={[a^{(k)}_{i,j}]}_{h_{in}^{(k)} \times w_{in}^{(k)}}$ and the $i$-th ($0\leq i\leq h_{in}^{(k)}-1$) row $IN^{(k)}_i=(a^{(k)}_{i,0},a^{(k)}_{i,1}, \cdots, a^{(k)}_{i,w_{in}^{(k)}-1})$. Then the vector $X^{(k)} = (IN^{(k)}_0, IN^{(k)}_1,\cdots,IN^{(k)}_{h_{in}^{(k)}-1})^{\mathsf{T}}$.

Denote the kernel matrix $C^{(k)} = [c^{(k)}_{i,j}]_{{h_{c}^{(k)}} \times w_{c}^{(k)}}$, where $C^{(k)}_i=(c^{(k)}_{i,0},c^{(k)}_{i,1}, \cdots, $ $c^{(k)}_{i,w_{c}^{(k)}-1})$. %In vector form, $Ker^{(k)}=(C^{(k)}_0, C^{(k)}_1,\cdots,C^{(k)}_{h_{in}^{(k)}-1})^{\mathsf{T}}$. 
Then convolution operation computes the output matrix $O^{(k)}$ as 
\begin{equation}
    O^{(k)}_{i,j}=
    \sum\limits _{p=0} ^{h^{(k)}_{c}-1} 
    \sum\limits _{q=0} ^{w^{(k)}_{c}-1} 
    IN^{(k)}_{i+p,j+q} C^{(k)}_{p,q} +b^{(k)}, 
\end{equation}
where the bias $b^{(k)}$ is the same for all the outputs.
In vector form, the output matrix  
$O^{(k)}$ will be a ($d^{(k)}_{f} = h^{(k)}_{o}\cdot w^{(k)}_{o}$)-dimension vector $Y^{(k)}=(O^{(k)}_{0},\cdots,O^{(k)}_{h^{(k)}_{o}-1})^{\mathsf{T}}$. 
%Then 
%\begin{equation}
%    Y^{(k)}_{i}=
%    \sum\limits _{j=0} ^{h^{(k)}_{ker} w^{(k)}_{ker} -1} 
%    X^{(k)}_{w_{in}^{(k)} \cdot \left\lfloor \frac{i}{w^{(k)}_c} \right\rfloor + i \bmod w^{(k)}_c 
%    + w_{in}^{(k)} \cdot \left\lfloor \frac{j}{w^{(k)}_{ker}} \right\rfloor + j \bmod w^{(k)}_{ker}} 
%    {Ker}^{(k)}_{j}
%\end{equation}

%\begin{definition} 
%    {\bfseries\upshape (Convolutional Kernel Matrix)}
%    If the input  matrix $h_{in}^{(k)} \times w_{in}^{(k)}$ is transformed into a $d^{(k-1)} = h_{in}^{(k)} \cdot w_{in}^{(k)}$ vector $X^{(k-1)}\in \mathbb{R}^{d^{(k-1)}}$, the convolution layer $f_c^{(k)}$ can be modeled as new $d^{(k)}_f \times d^{(k-1)}$  matrix $A^{(k)}$ by transforming the $h_{ker}^{(k)}\times w_{ker}^{(k)}$ kernel matrix. 

    %This pattern inserts zero-padding into the flattened kernel to ensure proper row boundary handling. The length of each zero-padding block is given by $w_{in} - w_{ker}$. Let $\mathcal{P}$ denote this kernel pattern vector:
    %$$
    %\mathcal{P} = (\,    \underbrace{c_1, \dots, c_w}_{\text{Row 1}}, \,    \underbrace{0, \dots, 0}_{\text{Padding}}, \,    \underbrace{c_{w+1}, \dots, c_{2w}}_{\text{Row 2}}, \,    \dots, \,    \underbrace{0, \dots, 0}_{\text{Padding}}, \,    \underbrace{c_{(h-1)w+1}, \dots, c_{hw}}_{\text{Row } h}    \,)    $$

 %   Consequently, the length of the kernel pattern $l_{pattern} = (h_{ker}-1) \cdot w_{in} + w_{ker}$.
%\end{definition}
%\zirui{The formula here is based on the case where step=1.}

\begin{definition} \label{def:conv_matrix}
    {\bfseries\upshape (Convolutional Matrix)}
    Let $\mathsf{c}_{in}^{(k)}=\mathsf{c}_{out}^{(k)}=s_c^{(k)}=1$.    
    In the $k$-th convolutional layer, given a convolution kernel matrix $C^{(k)}$, the function $f^{(k)}_c:\mathbb{R}^{d^{(k)}} \rightarrow \mathbb{R}^{d^{(k)}_{f}}$, is defined as 
    $f^{(k)}_c(X^{(k)})=A^{(k)} X^{(k)} + B^{(k)}$, where $X^{(k)} \in \mathbb{R}^{d^{(k)}}$ is the input vector.    $A^{(k)} = (A^{(k){\mathsf{T}}}_0, A^{(k){\mathsf{T}}}_1,\cdots,A^{(k){\mathsf{T}}}_{d_{f}^{(k)}-1})^{\mathsf{T}} \in \mathbb{R}^{{d_{f}^{(k)}} \times d^{(k)}}$, whose $i$-th row ($0\leq i\leq d^{(k)}_{f}-1$) is:
        \begin{equation}\label{eqn:A_row}\small
            A^{(k)}_i = (\underbrace{0, \cdots, 0}_{w^{(k)}_{in}\cdot \left\lfloor \frac{i}{w^{(k)}_{o}} \right\rfloor}, \underbrace{
        \underbrace{0, \cdots, 0}_{i\bmod w^{(k)}_o}, C^{(k)}_{0},     \underbrace{0, \dots, 0}_{padding}}_{w_{in}^{(k)}}, 
     \cdots, 
        \underbrace{
        \underbrace{0, \cdots, 0}_{i \bmod w^{(k)}_o}, C^{(k)}_{h_{c}^{(k)}-1} ,     \underbrace{0, \dots, 0}_{padding}}_{w_{in}^{(k)}}, 
        \underbrace{0, \cdots, 0}_{padding}),
        \end{equation}
     and $B^{(k)}=(b^{(k)},b^{(k)},\cdots,b^{(k)})^T \in \mathbb{R}^{d^{(k)}_{f}}$ (the bias for each neuron is the same). 
\end{definition}

%
  %  $$ 
 %       z^{(k)}_i = \left\lfloor \frac{i-1}{w_{conv}} \right\rfloor \cdot (w_{in} ~ s_c) + ((i-1) \bmod w_{conv}) \cdot s_c
 %   $$

\begin{figure}
    \vspace*{-6pt}
	\centering
        \includegraphics[width=0.4\linewidth]{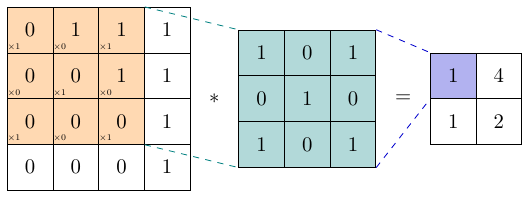}
	\caption{Example of Convolution Operation (bias $b^{(k)}$=0)}
	\label{fig:conv_f}
\end{figure}

Consider a concrete scenario with $\mathsf{c}_{in}^{(k)}=\mathsf{c}_{out}^{(k)}=s_c^{(k)}=1$ as shown in Figure \ref{fig:conv_f}, where the input matrix $IN^{(k)}$ has dimensions $h_{in}^{(k)} \times w_{in}^{(k)} = 4 \times 4$, i.e., $d^{(k)}=16$. Given a $h^{(k)}_{c} \times w^{(k)}_{c} = 3 \times 3$ kernel matrix $C^{(k)}= [c^{(k)}_{i,j}]_{3\times3}$, the output matrix $O^{(k)}$  is of dimension $h_o^{(k)}\times w_o^{(k)}=2 \times 2$, {\em i.e.}, $d^{(k)}_{f}=4$. According to Def. \ref{def:conv_matrix}, 
\begin{align*}
    A^{(k)} = \left(
    \begin{array}{*{16}{c}}
        c^{(k)}_{0,0} &c^{(k)}_{0,1} &c^{(k)}_{0,2} &0  & c^{(k)}_{1,0} &c^{(k)}_{1,1} &c^{(k)}_{1,2} &0 &c^{(k)}_{2,0} &c^{(k)}_{2,1} &c^{(k)}_{2,2} &0   &0   &0   &0   &0   \\
        0&c^{(k)}_{0,0} &c^{(k)}_{0,1} &c^{(k)}_{0,2} &0  & c^{(k)}_{1,0} &c^{(k)}_{1,1} &c^{(k)}_{1,2} &0 &c^{(k)}_{2,0} &c^{(k)}_{2,1} &c^{(k)}_{2,2}    &0   &0   &0   &0   \\
        0&0&0&0&c^{(k)}_{0,0} &c^{(k)}_{0,1} &c^{(k)}_{0,2} &0  & c^{(k)}_{1,0} &c^{(k)}_{1,1} &c^{(k)}_{1,2} &0 &c^{(k)}_{2,0} &c^{(k)}_{2,1} &c^{(k)}_{2,2}    &0      \\
        0&0&0&0&0&c^{(k)}_{0,0} &c^{(k)}_{0,1} &c^{(k)}_{0,2} &0  & c^{(k)}_{1,0} &c^{(k)}_{1,1} &c^{(k)}_{1,2} &0 &c^{(k)}_{2,0} &c^{(k)}_{2,1} &c^{(k)}_{2,2}         \\
    \end{array}
    \right).
\end{align*}

%We employ the convolution matrix $A^{(k)}_c$ to describe the convolution operation. Each row vector contains the Convolutional Kernel Pattern $\mathcal{P}$. After filtering out all zero entries, the row vector reduces to the kernel weights themselves. The index of the first non-zero element in a row serves as a spatial anchor. By identifying this index, we can extract precise positional information about the kernel's action, which locates the specific neurons in the previous layer involved in the calculation. We will see this also helps us determine the position of the neuron at the critical state. Furthermore, due to the weight-sharing mechanism, every row vector contains the same kernel weights. This implies that parameter information is consistently available across all neurons within the same layer, which also reflects the translation-invariant nature of the operation.    

\subsubsection{Non-linear Activation Layer.} 
   % {\bfseries\upshape (Activation Layer)}
    The $k$-th activation layer $\sigma_c^{(k)}$ of a CNN consists of a set of parallel $d^{(k)}_f$ non-linear activation functions $\mbox{ReLU}(x_i) = \max\{x_i, 0\}$ for $i=0,\cdots, d^{(k)}_f-1$. %In this paper, we focus on the ReLU activation function.
    %Denote $\sigma^{(k)}_i = \text{ReLU}(x_i) = \max\{x_i, 0\}$
%\end{definition} 
%Given an input to CNN, the input of $\sigma_c^{(k)}$ layer is determined, 
The input of 
$\sigma_c^{(k)}$ layer is the $h_o^{(k)}\times w_o^{(k)}$ matrix $O^{(k)}$, and we denote its output as $\hat{O}^{(k)}$ with the same dimension. In vector form, the input is denoted as $Y^{(k)}=(y_0,y_1,\cdots, y_{d_f^{(k)}-1})^{\mathsf{T}}$. Then $\sigma_c^{(k)}$ layer can be expressed as a matrix $I^{(k)}$ multiplied by $Y^{(k)}$, {\em i.e.}, $Z^{(k)} =(\hat{O}^{(k)}_0,\cdots,\hat{O}^{(k)}_{h_o^{(k)}-1})^{\mathsf{T}} = \sigma_{c}^{(k)}(Y)=I^{(k)}Y^{(k)}$, where
    \begin{align}\label{eqn:sigma_I}
        I^{(k)} = 
        \begin{pmatrix}
            \tau^{(k)}_0 &0&\cdots &0\\
            0&\tau^{(k)}_1&\cdots &0\\
            \vdots &\vdots & &\vdots \\
            0&0&\cdots &\tau^{(k)}_{d^{(k)}_f-1}
        \end{pmatrix} ,~
        0\leq i \leq d^{(k)}_f-1, ~
        \tau^{(k)}_{i} = \left\{ \begin{array}{l}
        1, ~~\mbox{if} ~~~y_i>0,\\
        0, ~~\mbox{if}~~~ y_i\leq 0.
        \end{array} \right.
    \end{align}

%The ReLU layer $\sigma_c^{(k)}$ does not reduce the dimension, {\em i.e.,} the dimensions of $Z^{(k)}$ and  $ Y^{(k)}$ are both $d^{(k)}_f$.     
%For briefness, we distinguish the activation functions of the $k$-th convolutional layer and the $k$-th fully connected layer by $I^{(k)}_c$ and $I^{(k)}$, respectively (although they are identical).

\subsubsection{Pooling Layers.}
In the pooling layer, a small fixed-size window (analogous to a convolution kernel, and referred to as a pooling kernel) slides across a feature map (input matrix $\hat{O}^{(k)}$) with a predefined stride.  The region of the input matrix covered by the pooling window is also referred to as the {\em Local Receptive Field of Pooling Layer} ({\em LRF-P} for short). 
For each window position, it applies a pooling function (usually use the max or average function) to compress the contents in the LRF-P into a single value, producing a dimension reduced output matrix (denoted as $IN^{(k+1)}$). Unlike convolution kernels, pooling kernels are parameter-free (no learnable weights and biases) — their behavior is determined solely by size, stride, and compressing rule. 
%Pooling works by sliding a window across the input and feeding the content of the window to a pooling function (usually use max or average function). The region of the input matrix covered by the pooling window is also referred to as the {\em local receptive field}. %Each unit in the pooling layer is connected to a local receptive field in the preceding convolutional layer. 
Usually, the max pooling function is more widely used than the average pooling in practice due to its unique advantages in feature extraction, anti-interference ability, and prevention of overfitting, which have been verified and applied in many classic CNN papers \cite{lecun2002gradient,boureau2010theoretical,krizhevsky2012imagenet,Simonyan2015very,matoba2023benefits}. Therefore, this paper focuses on the max pooling function. Algebraically, we reformulate the operation of the pooling layer as follows.

In the pooling layer $\rho^{(k)}$, we assume that the max pooling kernel is applied to an LRF-P of size  $h^{(k)}_{m} \times w^{(k)}_{m}$ with stride $s^{(k)}_{\rho}$. The input is $h_o^{(k)}\times w_o^{(k)}$ matrix $\hat{O}^{(k)}$ in Figure \ref{fig:symbol_cnn}.  The output is denoted as a $h^{(k)}_{\rho} \times w^{(k)}_{\rho}$ matrix $IN^{(k+1)}$, where:
    \begin{align*}
        h^{(k)}_{\rho} = \left\lfloor \frac{h^{(k)}_{o} - h^{(k)}_{m}}{s^{(k)}_{\rho}} \right\rfloor + 1, \quad 
        w^{(k)}_{\rho} = \left\lfloor \frac{w^{(k)}_{o} - w^{(k)}_{m}}{s^{(k)}_{\rho}} \right\rfloor + 1. 
    \end{align*} 
In particular, when the pooling kernel
 size is $h^{(k)}_{m} \times w^{(k)}_{m} = 2 \times 2$ and the stride is $s_{\rho}^{(k)}=2$, the windows partition the input into disjoint LRF-Ps, and the calculation of the output size reduces to: 
    \begin{align*}
        h^{(k)}_{\rho} = \left\lfloor \frac{h^{(k)}_{o}}{2} \right\rfloor, \quad
        w^{(k)}_{\rho} = \left\lfloor \frac{w^{(k)}_{o}}{2} \right\rfloor.
    \end{align*}
The entry of the $i$-th row and $j$-th column ($0\leq i\leq h^{(k)}_{\rho}-1$, $0\leq j\leq w^{(k)}_{\rho}-1$) of the output matrix $IN^{(k+1)}$ takes the maximum value of the elements in the $(i\cdot w^{(k)}_{\rho}+j)$-th LRF-P of $\hat{O}^{(k)}$, whose indexes are denoted as  $\Omega_{i,j}={\{(i\cdot s^{(k)}_{\rho}+ h,~ j\cdot s^{(k)}_{\rho}+ w), ~0\leq h\leq h_m^{(k)}-1, 0\leq w\leq w_m^{(k)}-1\}}$. Then, 
%For $0\leq i\leq h^{(k)}_{\rho}-1$, $0\leq j\leq w^{(k)}_{\rho}-1$,  
\begin{equation}\label{eqn:max_pool_1}
    IN^{(k+1)}_{i,j} = \max\{\hat{O}^{(k)}_{\Omega_{i,j}}\}. %, 0\leq h\leq h_m^{(k)}-1, 0\leq w\leq w_m^{(k)}-1. 
\end{equation}
%$IN^{(k+1)}_{i,j} = \max\{\hat{O}^{(k)}_{(i-1)s^{(k)}_{\rho}+h,(j-1)s^{(k)}_{\rho}+w}, 1\leq h\leq h_m^{(k)}, 1\leq w\leq w_m^{(k)}\}$.
When the input matrix $\hat{O}^{(k)}$ is given, the max pooling operation can be expressed as multiplying a series of Boolean matrices  $M^{(k),i,j}\in \mathbb{F}_2^{h_m^{(k)}\times w_m^{(k)}}$ ($0\leq i\leq h^{(k)}_{\rho}-1$, $0\leq j\leq w^{(k)}_{\rho}-1$) by different $h^{(k)}_{m} \times w^{(k)}_{m}$  LRF-P of $\hat{O}^{(k)}$,  
\begin{equation}\label{eqn:max_pool_2}
    IN^{(k+1)}_{i,j} = 
\sum\limits_{0\leq h\leq h_m^{(k)}-1,
0\leq w\leq w_m^{(k)}-1}M^{(k),i,j}_{\{h,w\}}\cdot \hat{O}^{(k)}_{\{i\cdot s^{(k)}_{\rho}+h,j\cdot s^{(k)}_{\rho}+w\}},
\end{equation}

% In the $(i\cdot w^{(k)}_{\rho}+j)$-th LRF-P of $\hat{O}^{(k)}$, %let $({h'}, {w'})$ be the lexicographically first index pair that achieves the maximum value:
% define a set of indexes $({h}, {w})$ that achieves the maximum value as: 
% \begin{align}
% U=\left\{ (h,w) 
% \mid \hat{O}^{(k)}_{\{i\cdot s^{(k)}_{\rho}+h, j\cdot s^{(k)}_{\rho}+w\}} = \max\{\hat{O}^{(k)}_{\Omega_{i,j}}\} \right\}
% \end{align}
% Arbitrarily select an index $({h'}, {w'})\in U$ to define $M^{(k),i,j}$ as:
% \begin{align}
% M^{(k),i,j}_{\{h,w\}}=\begin{cases} 1, & \text{if } (h,w) = ({h'}, {w'}), \\ 0, & \text{otherwise}. \end{cases}
% \end{align}
% \color{black}

In the $(i\cdot w^{(k)}_{\rho}+j)$-th LRF-P of $\hat{O}^{(k)}$, let $\Lambda_{i,j}$ be the set of all index pairs that achieve the maximum value:
\begin{align}
\label{eq:maximum_index_Lamda}
\Lambda_{i,j} = \left\{ (h,w) \mid \hat{O}^{(k)}_{\{i\cdot s^{(k)}_{\rho}+h, j\cdot s^{(k)}_{\rho}+w\}} = \max\{\hat{O}^{(k)}_{\Omega_{i,j}}\} \right\}. 
\end{align}
To ensure exactly one neuron is selected from the LRF-P, let $(h^*, w^*) \in \Lambda_{i,j}$ be an arbitrarily chosen index pair. The elements of $M^{(k),i,j}$  are defined as:
\begin{align}
\label{eq:pooling_selection_M}
M^{(k),i,j}_{\{h,w\}}=\begin{cases} 
1, & \text{if } (h,w) = (h^*, w^*), \\ 
0, & \text{otherwise}, 
\end{cases} 
\end{align}
with $0\leq h\leq h_m^{(k)}-1$, $0\leq w\leq w_m^{(k)}-1$. 
Note that any arbitrary choice of $(h^*, w^*) \in \Lambda_{i,j}$ yields a functionally equivalent matrix $M^{(k),i,j}$ that produces the exact same pooling output $IN^{(k+1)}_{i,j}$.

%with $\Omega_{i,j}={\{i\cdot s^{(k)}_{\rho}+\bar{h}, j\cdot s^{(k)}_{\rho}+\bar{w}, 0\leq \bar{h}\leq h_m^{(k)}-1, 0\leq \bar{w}\leq w_m^{(k)}-1\}}$ and $\Omega_{i,j}$ is the $(i,j)$-th local receptive field.  

In vector form, the input and output vector of  pooling layer $\rho^{(k)}$ are $Z^{(k)} =(\hat{O}^{(k)}_0,\cdots,\hat{O}^{(k)}_{h_o^{(k)}-1})^{\mathsf{T}}\in \mathbb{R}^{d^{(k)}_f=h_o^{(k)}\cdot w_o^{(k)}}$ and $X^{(k+1)} = (IN^{(k+1)}_0,IN^{(k+1)}_1,\cdots, $ $ IN^{(k+1)}_{h_{\rho}^{(k)}-1})^{\mathsf{T}}\in \mathbb{R}^{d^{(k+1)}},~d^{(k+1)}=h^{(k)}_{\rho} \cdot w^{(k)}_{\rho}$. 

\begin{definition}\label{def:pooling_matrix}{\bfseries\upshape (Boolean Pooling Matrix)}    
There exists a $d^{(k+1)}\times d^{(k)}_{f}$ Boolean pooling matrix $P^{(k)}$  so that $\rho^{(k)}:\mathbb{R}^{d^{(k)}_{f}} \rightarrow \mathbb{R}^{d^{(k+1)}}$ satisfying  $X^{(k+1)} = \rho^{(k)}(Z^{(k)}) = P^{(k)}\cdot Z^{(k)}$. Assuming the LRF-Ps are disjoint and square ({\em i.e.}, $s^{(k)}_{\rho}=w_m^{(k)}=h_m^{(k)}$), then the max pooling operation on the $r$-th LRF-P can be defined as the $r$-th row ($0\leq r\leq d^{(k+1)}-1$) of $P^{(k)}$: 
\begin{equation}\label{eqn:P_matrix}
    \resizebox{1.0\hsize}{!}{$%
    P^{(k)}_r = (\underbrace{0, \cdots, 0}_{w^{(k)}_{o}\cdot i \cdot s^{(k)}_{\rho}}, \underbrace{
        \underbrace{0, \cdots, 0}_{j\cdot s^{(k)}_{\rho}} , M^{(k),i,j}_{0},     \underbrace{0, \dots, 0}_{padding}}_{w_{o}^{(k)}}, 
     \cdots, 
        \underbrace{
        \underbrace{0, \cdots, 0}_{j\cdot s^{(k)}_{\rho}}, M^{(k),i,j}_{h_m^{(k)}-1},     \underbrace{0, \dots, 0}_{padding}}_{w_{o}^{(k)}}, 
        \underbrace{0, \cdots, 0}_{padding}),
    $}%
\end{equation}
where $i=\left\lfloor \frac{r}{w^{(k)}_{\rho}} \right\rfloor$, $j=r \bmod w_{\rho}^{(k)}$. 
\end{definition}

Obviously, similar to the matrix of $I^{(k)}$ for $\sigma_c^{(k)}$ layer, the pooling matrix $P^{(k)}$ varies with the input $X^{(1)}$. 
For example, the input matrix $\hat{O}^{(k)}$ has dimensions $h_o^{(k)}\times w_o^{(k)} = 4 \times 4$, i.e., $d_{f}^{(k)}=16$, the max pooling kernel is of $h_m^{(k)}\times w_m^{(k)} = 2 \times 2$ with stride $s_{\rho}^{(k)}=2$. The output matrix $IN^{(k+1)}$ will have dimension $h_{\rho}^{(k)}\times w_{\rho}^{(k)} = 2 \times 2$, i.e., $d^{(k+1)}=4$. The $d^{(k+1)}\times d^{(k)}_{f}=4\times 16$  pooling matrix $
    P^{(k)} = 
    \begin{pmatrix}
        &P^{(k)}_{0,0}   &{\bf 0}_{2 \times 8}  \\
        &{\bf 0}_{2 \times 8}  &P^{(k)}_{1,1} 
    \end{pmatrix}
$, where ${\bf 0}_{2\times 8}$ is $2\times8$ zero matrix, and
$$
    P^{(k)}_{0,0} =
    \left(
    \begin{array}{*{8}{c}} % 使用16列居中对齐
        M^{(k),0,0}_{0,0} &M^{(k),0,0}_{0,1} &0   &0   &M^{(k),0,0}_{1,0} &M^{(k),0,0}_{1,1} &0   &0  \\
        0   &0   &M^{(k),0,1}_{0,0} &M^{(k),0,1}_{0,1}
        &0   &0   &M^{(k),0,1}_{1,0} &M^{(k),0,1}_{1,1} 
    \end{array}
    \right),
$$
$$
    P^{(k)}_{1,1} =
    \left(
    \begin{array}{*{8}{c}} % 使用16列居中对齐
        M^{(k),1,0}_{0,0} &M^{(k),1,0}_{0,1} &0   &0   &M^{(k),1,0}_{1,0} &M^{(k),1,0}_{1,1} &0   &0   \\
        0  &0  &M^{(k),1,1}_{0,0} &M^{(k),1,1}_{0,1} 
        &0  &0  &M^{(k),1,1}_{1,0} &M^{(k),1,1}_{1,1} 
    \end{array}
    \right).
$$
    
% $$\rotatebox{270}{$\tiny
%     P^{(k)} = \left(
%     \begin{array}{*{16}{c}} % 使用16列居中对齐
%         M^{(k),0,0}_{0,0} &M^{(k),0,0}_{0,1} &0   &0   &M^{(k),0,0}_{1,0} &M^{(k),0,0}_{1,1} &0   &0   &0   &0   &0   &0   &0   &0   &0   &0   \\
%         0   &0   &M^{(k),0,1}_{0,0} &M^{(k),0,1}_{0,1}&0   &0   &M^{(k),0,1}_{1,0} &M^{(k),0,1}_{1,1} &0   &0   &0   &0   &0   &0   &0   &0   \\
%         0   &0   &0   &0   &0   &0   &0   &0   &M^{(k),1,0}_{0,0} &M^{(k),1,0}_{0,1} &0   &0   &M^{(k),1,0}_{1,0} &M^{(k),1,0}_{1,1} &0   &0   \\
%         0   &0   &0   &0   &0   &0   &0   &0   &0   &0   &M^{(k),1,1}_{0,0} &M^{(k),1,1}_{0,1} &0   &0   &M^{(k),1,1}_{1,0} &M^{(k),1,1}_{1,1} 
%     \end{array}
%     \right).$}$$

%The pooling layer performs down-sampling and blocks the propagation of unselected neurons. A significant property is that the order of applying the activation function $I_c$ and the pooling matrix $P_c$within the same layer is interchangeable without affecting the result. For each layer, the convolutional layer with the activation function and pooling layer could be treated as a single unit.

\begin{definition} 
    {\bfseries\upshape (Fully-connected Layer)}
    The $k$-th fully-connected layer within the Fully-connected Block of CNN is a function $f^{(k)}:\mathbb{R}^{d^{(k)}} \rightarrow \mathbb{R}^{d^{(k+1)}}$ given by an affine transformation:
    \begin{equation}
        f^{(k)}(X)=A_{\text{fc}}^{(k)}X_{\text{fc}}^{(k)}+B_{\text{fc}}^{(k)},
    \end{equation}
    where $X_{\text{fc}}^{(k)}\in\mathbb{R}^{d^{(k)}}$ is the input vector. The weight matrix $A_{\text{fc}}^{(k)}\in \mathbb{R}^{d^{(k+1)}\times d^{(k)}}$ and the bias vector $B_{\text{fc}}^{(k)}\in \mathbb{R}^{d^{(k+1)}}$ consist of floating-point numbers.
\end{definition}
The matrix for the non-linear layer $\sigma$ (ReLU functions) of the Fully-connected Block can be similarly defined by Eq. \eqref{eqn:sigma_I}. 
Given $X^{(1)}=x\in \mathbb{R}^{d^{(1)}}$, the Convolutional Rounds and FCNN Rounds of CNN ``collapse'' into an affine transformation. Assume that the function of the $m$ Convolutional Rounds is $\mathcal{F}_{\text{conv}}$, and the function of the $n$ FCNN Rounds is $\mathcal{F}_{\text{fc}}$, then
% \begin{equation}\label{eqn:affine_transformation_convolutional}
% \begin{array}{ll}
% y=\mathcal{F}_{1}(x) &= P^{(m)}(I^{(m)}(A^{(m)}  \cdots P^{(2)}(I^{(2)}(A^{(2)}(P^{(1)}(I^{(1)}(A^{(1)}x+B^{(1)})))+B^{(2)}))\cdots+B^{(m)}))\\
%        & = P^{(m)}I^{(m)}A^{(m)}  \cdots P^{(2)}I^{(2)}A^{(2)}P^{(1)}I^{(1)}A^{(1)}x+ \beta_1\\
%        & = \Gamma_1 \cdot x+\beta_1
%   \end{array}
% \end{equation}
% and
\begin{equation}
  \begin{array}{ll}
y=\mathcal{F}_{\text{conv}}(x) &= P^{(m)}(I^{(m)}(A^{(m)}  \cdots (P^{(1)}(I^{(1)}(A^{(1)}x+B^{(1)})))\cdots+B^{(m)}))\\
       & = P^{(m)}I^{(m)}A^{(m)}  \cdots P^{(2)}I^{(2)}A^{(2)}P^{(1)}I^{(1)}A^{(1)}x+ \beta_1\\
       & = \Gamma_1 \cdot x+\beta_1.
  \end{array}
\end{equation}
According to \cite[Section 3.1]{DBLP:conf/eurocrypt/CanalesMartinezCHRSS24}, given $y$, $ \mathcal{F}_{\text{fc}}$ collapses into 
\begin{equation}\label{eqn:affine_transformation_fcnn}
  \begin{array}{ll}
 \mathcal{F}_{\text{fc}}(y) =\Gamma_2 \cdot y+ \beta_2.
  \end{array}
\end{equation}
Consequently, the full CNN ``collapses'' into
\begin{equation}\label{eqn:affine_transformation}
  \begin{array}{ll}
\mathcal{F}_{\theta}(x) =\Gamma_2  (\Gamma_1 \cdot x+\beta_1)+ \beta_2 = \Gamma_2 \cdot \Gamma_1
\cdot  x + \Gamma_2 \beta_1 + \beta_2.
\end{array}
\end{equation}
\begin{definition}{\bfseries\upshape(Linear Neighborhood)}
    Given an input  $x\in  \mathbb{R}^{d^{(1)}}$ with the corresponding  output of the CNN computed by Eq. \eqref{eqn:affine_transformation}, the linear neighborhood of $x$ is defined as the set 
\begin{equation*}
       \{u\in  \mathbb{R}^{d^{(1)}}| \mathcal{F}_{\theta}(u)=  \Gamma_2 \cdot \Gamma_1
\cdot  u + \Gamma_2 \beta_1 + \beta_2\}, 
\end{equation*}
i.e., the same affine transformation is used for $u$ and $x$ to compute the output of the CNN. 
\end{definition}
If we make a change of $\Delta$ to the input $x$,  and $x+\Delta$ remains within the linear neighborhood of $x$,  we can observe the corresponding change of the output 
\begin{equation*}
   \mathcal{F}_{\theta}(x+\Delta)-\mathcal{F}_{\theta}(x) =  \Gamma_2 \cdot \Gamma_1
\cdot  (x+\Delta) + \Gamma_2 \beta_1 + \beta_2 - (\Gamma_2 \cdot \Gamma_1
\cdot  x + \Gamma_2 \beta_1 + \beta_2) =\Gamma_2 \cdot \Gamma_1
\cdot  \Delta.
\end{equation*}

\begin{definition} \label{def:layer_merging}
    {\bfseries\upshape (Layer Merging)}
    We focus on the Convolutional Block of the $(m+n)$ network.  
   We assume that we have complete knowledge of the first $k-1$ layers of the Convolutional Block, and we are currently recovering layer $k$. Let $F_x^{k-1}$ and $G_x^{k+1}$ represent, respectively, the fully recovered and non-recovered parts of the CNN.
        $$\mathcal{F}_{\theta} = \underbrace{f^{(n+1)} \circ \cdots \circ \sigma^{(1)}\circ f^{(1)} \circ 
        \rho^{(m)} \circ \cdots \circ f_c^{(k+1)}}_{G_x^{k+1}}
        \circ \rho^{(k)} \circ \sigma_c^{(k)} \circ f_c^{(k)} \circ 
        \underbrace{\rho^{(k-1)} \circ \cdots \circ f_c^{(1)}}_{F_x^{k-1}}.$$
   Given $X^{(1)} \in  \mathbb{R}^{d^{(1)}}$,  $G_x^{k+1}$ and $F_x^{k-1}$ become $G_x^{k+1}(X^{(k+1)})=
    G^{k+1}X^{(k+1)}+B^{k+1}$ and $F_x^{k-1}(X^{(1)})=F^{k-1}X^{(1)}+B^{k-1}$, where the matrices $F^{k-1}\in \mathbb{R}^{d^{(k)}\times d^{(1)}}$, $B^{k-1}\in \mathbb{R}^{d^{(k)}}$, $G^{k+1}\in \mathbb{R}^{d^{(m+n+2)}\times d^{(k+1)}}$, $B^{k+1}\in \mathbb{R}^{d^{(m+n+2)}}$,   respectively.
\end{definition}

\begin{definition} \label{def:model_parameters}
    {\bfseries\upshape (Model Parameters)}
    The parameters $\theta$ of a $(m+n)$-deep neural network $\mathcal{F}_{\theta}$ are the concrete assignments to the weights $C^{(k)}$, biases $B^{(k)}$ for $k \in \{1,\cdots,m\}$ in Convolutional Block and the concrete assignments to weights $A_{\text{fc}}^{(k)}$ and biases $B_{\text{fc}}^{(k)}$ for $k \in \{1,\cdots,n+1\}$ in Fully-connected Block. 
\end{definition}

\begin{definition} \label{def:signatures}
    {\bfseries\upshape (Signatures \cite{DBLP:conf/eurocrypt/CarliniCHRS25})}
    The signature of a convolutional kernel is equal to $\alpha \cdot C^{(k)}$, where $\alpha$ is an arbitrary rescaling of the corresponding parameter.
    %, while the signature of a neuron $i$ in Fully-connected Block is equal to $\alpha \cdot A_i^{(k)}$, where $\alpha$ is an arbitrary rescaling of the corresponding parameter.
\end{definition}

% \begin{definition}
%     \label{def:conv signatures}
%     {\bfseries\upshape (Signatures \cite{DBLP:conf/crypto/CarliniJM20})}
%     Let the convolutional kernel weight $C^{(k)}$ be described as $(c_1,c_2,\dots,c_{h^{(k)}\omega^{(k)}})$. The signature in layer $k$ is the tuple
%     \begin{align}
%         (\frac{c_1}{c_1}=1, \frac{c_2}{c_1}, \dots, \frac{c_{h^{(k)}\omega^{(k)}}}{c_1}).
%     \end{align}
% \end{definition}

\begin{definition} 
    {\bfseries\upshape (Neuron State \cite{DBLP:conf/crypto/CarliniJM20})}
    Let $\mathcal{V}(\eta;X)$ denote the value that neuron $\eta$ takes with $X \in \mathbb{R}^{d^{(1)}}$ before applying its corresponding activation function $\sigma$. If $\mathcal{V}(\eta;X)>0$ (respectively, $\mathcal{V}(\eta;X)<0$), the neuron state of $\eta$ is activated (respectively, inactivated). If $\mathcal{V}(\eta;X)=0$, the neuron state is critical.
\end{definition}

\subsection{Adversarial Goals and Assumptions}\label{sect:assumptions_1}
%The adversarial goal is to extract all secret parameters (as formalized in Def. \ref{def:model_parameters}) of a target CNN, using only black-box input-output queries. Our objective is to achieve an $(\varepsilon,\xi)$-functionally equivalent extraction, rather than replicating the exact original parameters.
The adversarial goal is to achieve an $(\varepsilon,\xi)$-functionally equivalent parameter extraction, rather than replicating the exact original parameters \cite{DBLP:conf/crypto/CarliniJM20,DBLP:journals/iacr/LiuSELBP26}. 

\begin{definition}
    {\bfseries\upshape ($(\varepsilon,\xi)$-Functional Equivalence \cite{DBLP:conf/crypto/CarliniJM20})}
    Two neural networks $\mathcal{F}_{\theta}$ and $\mathcal{F}_{\hat{\theta}}$ are $(\varepsilon,\xi)$-functional equivalent on the input dataset $S$ if 
    $$Pr_{x\in S}(\lvert \mathcal{F}_{\theta}(x)-\mathcal{F}_{\hat{\theta}}(x) \rvert \leq \varepsilon) \geq 1-\xi.$$    
\end{definition}

\subsubsection{$(\varepsilon,0)$-Functional Equivalence \cite{DBLP:conf/crypto/CarliniJM20}.}
To evaluate the precision of our attack, we adopt the concept of $(\varepsilon,\xi)$-Functional Equivalence proposed by Carlini et al. \cite{DBLP:conf/crypto/CarliniJM20} and subsequently used by Chen et al. \cite{DBLP:conf/asiacrypt/ChenDGSWW24,chen2025delving}. 

In our analysis, we specifically focus on the $(\varepsilon,0)$-functional equivalence given by  Carlini {\em et al.} \cite{DBLP:conf/crypto/CarliniJM20} which represents an upper bound on the maximum error of the output. We adopt the error bounds propagation method introduced by Carlini {\em et al.} \cite[Section 6.2]{DBLP:conf/crypto/CarliniJM20}  to compute the $(\varepsilon,0)$-functional equivalence. %Detailed calculation procedures are presented in \textsf{Supplementary Material}  \ref{supp:carlini_error_eval}.
%The difference is when the convolutional layer has only one output channel, permutation alignment is unnecessary.

\paragraph{Assumptions.} We make the following assumptions about the Oracle and the capabilities of the attacker: 
\textbf{Architecture knowledge:} We require knowledge of the model architecture of the neural network, including dimensions of kernel matrix, max pooling kernel, and all the strides $s_c^{(k)}$, $s_{\rho}^{(k)}$. 
\textbf{Full-domain inputs:} The attacker can query arbitrary inputs from $\mathbb{R}^{d^{(1)}}$. 
\textbf{Precise Computation:} The CNN is specified and evaluated using sufficiently high precision floating-point arithmetic. 
\textbf{ReLU Activations:} All activation functions $\sigma^{(k)}$ are the ReLU function. 
\textbf{Raw Output Accessible\footnote{When the CNN applies a Softmax function to the raw outputs (logits) to generate probability scores, the neuron fusion technique proposed by Chen {\em et al.} \cite{chen2025delving} at ASIACRYPT 2025 allows us to use the scores to compute the raw output of a CNN$^*$, which adds an additional and known linear transformation after the last layer of the original CNN. Then our raw-output based attack still works.}:} Given input $X$, the raw outputs of the CNN is accessible to the adversary. The same assumption is also used by Carlini {\em et al.}'s attack \cite{DBLP:conf/crypto/CarliniJM20}.

% We make the following assumptions about the Oracle and the capabilities of the attacker:
% %Architecture knowledge. We require knowledge of the model architecture of the neural network. Full-domain inputs. The attacker can query arbitrary inputs from $R_{d_0}$. The DNN is specified and evaluated using a sufficiently high precision floating-point arithmetic. 
% \begin{itemize}
% \item \textbf{Architecture knowledge.} We require knowledge of the model architecture of the neural network, including dimensions of kernel matrix, max pooling kernel, and all the strides $s_c^{(k)}$, $s_{\rho}^{(k)}$.  
% \item \textbf{Full-domain inputs.} The attacker can query arbitrary inputs from $\mathbb{R}^{d^{(1)}}$.
% \item \textbf{Precise Computation.} The CNN is specified and evaluated using sufficiently high precision floating-point arithmetic. 
% \item \textbf{ReLU Activations.} All activation functions $\sigma^{(k)}$ are the ReLU function.
% \item \textbf{Raw Output Accessible.} Given input $X$, the raw output of the CNN is accessible to the adversary. The same assumption is also used by Carlini {\em et al.}'s attack \cite{DBLP:conf/crypto/CarliniJM20}. %Note that the last softmax function is fixed and invertible, therefore, the attacker knows the full output of $\mathcal{F}_{\theta}(X)$ of Eq. \eqref{eqn:cnn_f}. 
% \end{itemize}

\section{Input Space Partition and Critical Points Recognition}\label{sect:critical_points}
The input space $\mathbb{R}^{d^{(1)}}$ is partitioned into many linear regions. Within each distinct linear region, the activation states of all neurons and the selection indices of the pooling layers remain constant, and the neural network computes a fixed linear function over the region. However, when linearly varying inputs cross the boundaries between these regions, the output exhibits non-linear changes. According to the definitions in Section \ref{sect:def_notations}, it is evident that both convolutional and fully-connected layers represent linear operations. The non-linearity is jointly introduced by the activation layers and the pooling layers:
\begin{itemize}
    \item 
\textbf{ReLU Activation.} This determines whether a neuron's input is propagated to the subsequent layer, where a non-linear transition between a continuous zero state (inactivated) and a linear non-zero state (activated) occurs. For a fixed input, the activation matrix is deterministic. 

Carlini {\em et al.}'s seminal attack \cite{DBLP:conf/crypto/CarliniJM20} and all the follow-up results \cite{DBLP:conf/eurocrypt/CanalesMartinezCHRSS24,DBLP:conf/eurocrypt/CarliniCHRS25,DBLP:conf/asiacrypt/ChenDGSWW24,chen2025delving} are based on the non-linearity introduced by ReLU at specific inputs, which are called {\em critical points}. The adversary studies the behavior of the FCNN in the vicinity of critical points, which are defined as inputs in whose tiny vicinity exactly one ReLU input in one of the layers changes sign. It is expected that the FCNN's output will change non-linearly as a result of this ReLU transition, which makes this event noticeable when we observe the network's outputs by varying the inputs in those tiny vicinities.

It is important to note that for the critical point to be effective, the neuron must be selected by the subsequent pooling layers; otherwise, its activation state has no impact on the final output.

\item 
\textbf{Max Pooling.} This operation selects the maximum value within each local receptive field. Since different neurons have different gradients in their linear outputs, a shift in the pooling selection results in a non-linear change in the final output. Similar to the activation layer, the pooling matrix is fixed when the input is determined. 
\end{itemize}

\subsection{Does Carlini {\em et al.}'s Critical Point Work for CNN?}
At EUROCRYPT 2024,  Canales-Martinez {\em et al.} \cite{DBLP:conf/eurocrypt/CanalesMartinezCHRSS24} stated that the attack based on Carlini {\em et al.}'s critical points also works for CNNs without modification: 
\begin{quote}
    ``{\em Our attack can
be applied without modifications to convolutional networks since they can be
described as a special form of a fully connected network.}''
\end{quote}
In the FCNN, a critical point can be identified by changing the inputs in its tiny vicinity and testing whether the network's output changes non-linearly. However, unlike  FCNN, the non-linearity of CNN is introduced through ReLU or max pooling, which may lead to two problems when applying Carlini {\em et al.}'s critical point methods.

\subsubsection{\bf Problem I:} 
\begin{figure}[htbp] % htbp为浮动位置参数，保证图片排版合理
	\centering
	\begin{minipage}{0.48\linewidth}
		\centering
		\includegraphics[width=\linewidth]{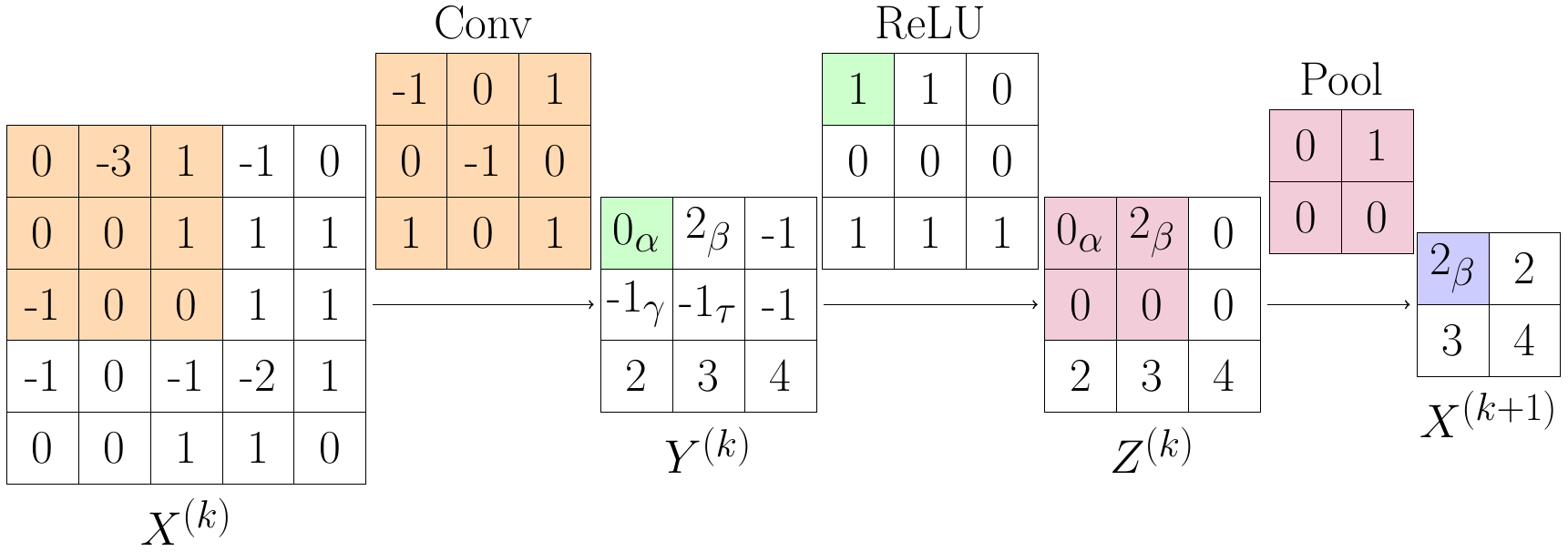}
		\caption{Undetectable Critical Point}
		\label{fig:critical_conv_no}
	\end{minipage}
	\vrule width 1pt 
	\hspace{1pt}     
	\begin{minipage}{0.48\linewidth}
		\centering
		\includegraphics[width=\linewidth]{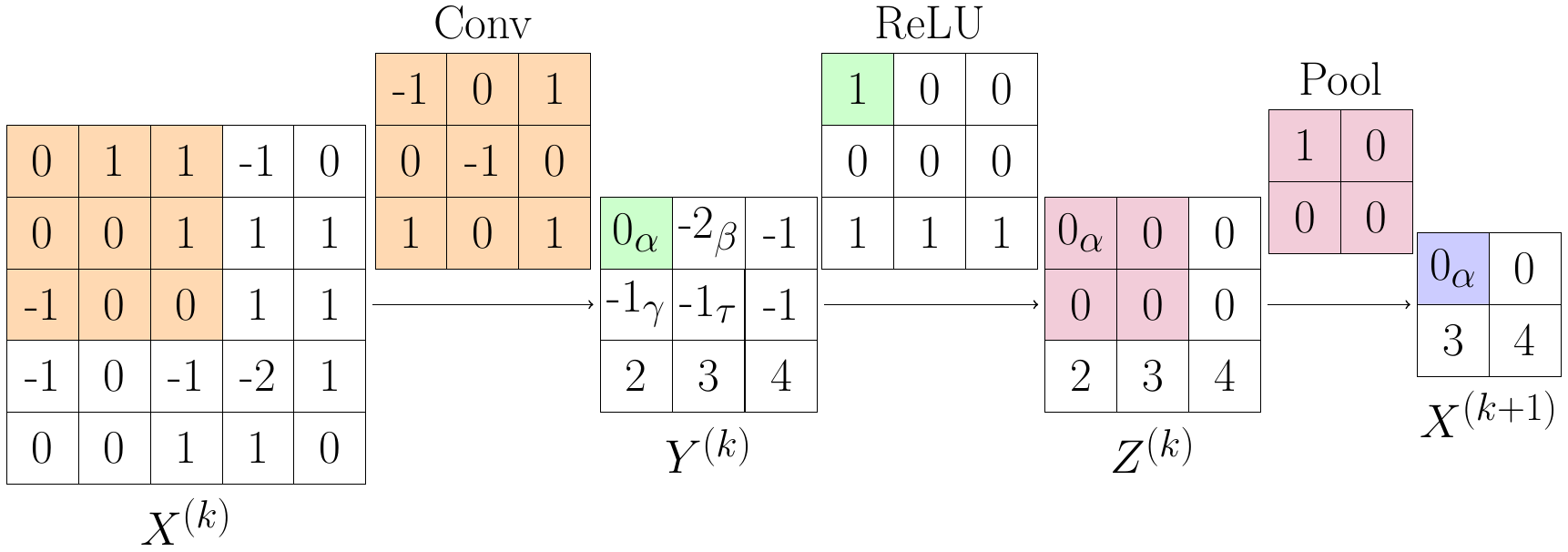}
		\caption{Detectable Critical Point}
		\label{fig:critical_conv_yes}
	\end{minipage}
\end{figure}

Take Figure \ref{fig:critical_conv_no} as an example. The critical point $\bar{X}^{(1)}$ causes the $0$-th entry of the input vector $Y^{(k)}$ of the $\sigma_c^{(k)}$ layer to be zero. However, the $2\times 2$ max pooling layer does not select the output of  $Y^{(k)}[0]$. It selects the maximum value among the neurons $Y^{(k)}[0,1,3,4]$, which outputs the value $2$ in Figure \ref{fig:critical_conv_no}. 

Therefore, changing the inputs within a tiny vicinity of the critical point $\bar{X}^{(1)}$ will cause the network's output to change linearly, making the critical point $\bar{X}^{(1)}$ undetectable. Only when the max pooling layer selects the output of neuron $\alpha$, the network's output will be affected non-linearly by changing the inputs around $\bar{X}^{(1)}$. As shown in Figure \ref{fig:critical_conv_yes}, the values in neuron %$\alpha$, $\beta$, $\gamma$, $\tau$ of 
$Y^{(k)}[0,1,3,4]$ are 0, -2, -1, -1, and neuron $\alpha$ is at the critical point. The max pooling layer is expected to select the output of  neuron $\alpha$ when changing the inputs near $\bar{X}^{(1)}$, and to cause the network's output to change non-linearly, which  makes the critical point $\bar{X}^{(1)}$ detectable.
%\begin{figure}
%	\centering
%        \includegraphics[width=0.6\linewidth]{figures/CritConvNo.pdf}
%	\caption{Undetectable Critical Point}
%	\label{fig:critical_conv_no}
%\end{figure}
%\begin{figure}
%	\centering
%        \includegraphics[width=0.6\linewidth]{figures/CritConvYes.pdf}
%	\caption{Detectable Critical Point}
%	\label{fig:critical_conv_yes}
%\end{figure}
Formally, the critical point $\bar{X}^{(1)}$ makes $Y^{(k)}_t=0$ ($0\leq t\leq d_f^{(k)}-1$) of $\sigma_c^{(k)}$. According to Eq. \eqref{eqn:sigma_I}, $\tau_t=0$. 
Let $\bar{X}^{(1)}_+ = \bar{X}^{(1)}+\epsilon$ and $\bar{X}^{(1)}_- = \bar{X}^{(1)}-\epsilon$ with small enough $\epsilon$, and assuming the corresponding $Y^{(k)}_{t+}>0$ and $Y^{(k)}_{t-}<0$. Hence, the corresponding  $\tau_{t+}=1$, $\tau_{t-}=0$. Considering the matrix $P^{(k)}\cdot I^{(k)}$, in order to apply Carlini {\em et al.}'s attack, the entries $\tau_{t+}=1$, $\tau_{t-}=0$ should result in different matrices  $P^{(k)}\cdot I^{(k)}$, else the two inputs $\bar{X}^{(1)}_+$ and $\bar{X}^{(1)}_-$ will have a linear impact on the output of the CNN. Note that the entries of $P^{(k)}\cdot I^{(k)}$ related to $\tau_{t}$ are the $t$-th column of $P^{(k)}\cdot I^{(k)}$. Only when $(P_{0,t}^{(k)},P_{1,t}^{(k)},\cdots, P_{d^{(k+1)}-1,t}^{(k)})^{\mathsf{T}}\neq {\bf 0}$, there exists $i$ such that  $P_{i,t}^{(k)}=1$. Then,  
$(P^{(k)}\cdot I^{(k)})_{i,t}=\tau_{t}$, which makes the $P^{(k)}\cdot I^{(k)}$ different for $\tau_{t+}=1$ and $\tau_{t-}=0$. Moreover, in the following layers' operations, $\tau_{t}$ should also appear in the operations matrix to affect the CNN's output, {\em i.e.}, the matrix $G^{k+1}\cdot P^{(k)}\cdot I^{(k)}$ should also be affected by $\tau_{t}$, otherwise, the critical point is undetectable as well. We give a definition on this kind of detectable critical point. 
\begin{definition} 
    {\bfseries\upshape (ReLU-Pooling Critical Point, RPCP)}
    A ReLU-Pooling Critical Point (RPCP) is an input $X^{(1)} \in \mathbb{R}^{d^{(1)}}$ that makes the input to the ReLU function in $\sigma^{(k)}_c$ layer 0, and by varying the inputs in the tiny vicinity of $X^{(1)}$, the output of CNN changes non-linearly. 
\end{definition} 

\begin{property}\label{pro:rpcp}
   If $X^{(1)}$ is a RPCP corresponding to the $t$-th neuron in layer $k$, then $A^{(k)}_t X^{(k)} + B_t^{(k)}=0$, and the inputs of other neurons in the same Local Receptive Field of Pooling Layer are smaller than 0. 
 %  and $A^{(k)}_{j} (F^{k-1} X^{(k)}) + B_j^{(k)}<0, \forall j \in \Omega_i,j\neq i$
\end{property}

%For instance, x is a CPCP corresponding to the $i$-th neuron in layer $k$ if: $A^{(k)}_{i} X^{(k)} + b^{(k)}=0 \text{ and } A^{(k)}_{j} X + b^{(k)}<0, \forall j \in \Omega_i,j\neq i$

%\xiaoyang{I am writting here}

\subsubsection{\bf Problem II:} 
\begin{figure}
	\centering
        \includegraphics[width=0.75\linewidth]{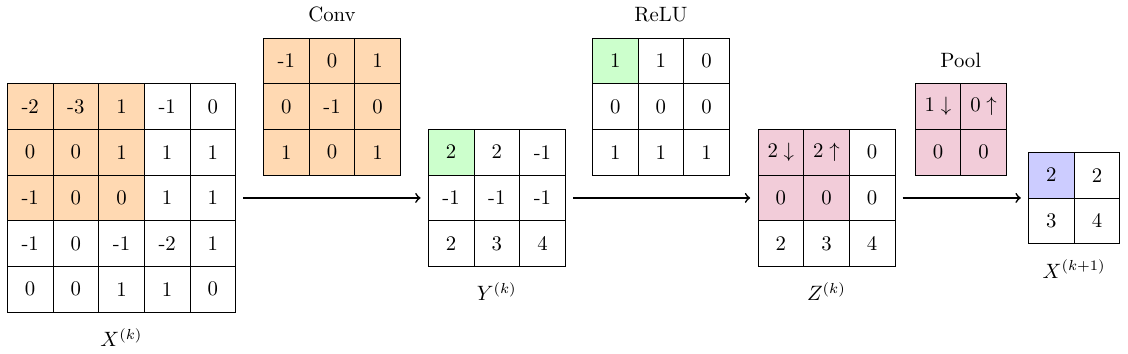}
	\caption{Pooling Switch Point (PSP)}
	\label{fig:crit_pool}
    \vspace*{-12pt}
\end{figure}

Besides RPCPs, the network's output may also change non-linearly because the pooling layer selection changes. As shown in Figure \ref{fig:crit_pool}, in layer $k$, there is no 0 in $Y^{(k)}$, {\em i.e.}, no RPCP. However, in $Z^{(k)}$, the input $X^{(1)}$ makes $Z_{(k)}[0]=Z_{(k)}[1] \geq Z_{(k)}[3]=Z_{(k)}[4]$, hence the max pooling kernel matrix $M^{(k),0,0}=\left[\begin{array}{cc}
   1  &0  \\
    0 & 0
\end{array}\right]$. 
Let $\bar{X}^{(1)}_+ = \bar{X}^{(1)}+\epsilon$ and $\bar{X}^{(1)}_- = \bar{X}^{(1)}-\epsilon$ with small enough $\epsilon$, and assuming the corresponding ($Z^{(k)}_+[0]<2$, $Z^{(k)}_+[1]>2$) and ($Z^{(k)}_-[0]>2$, $Z^{(k)}_-[1]<2$). Then, the selection matrix $M^{(k),0,0}$ for $\bar{X}^{(1)}_+ $ and $\bar{X}^{(1)}_-$ will be different. Thus, the output of CNN will be non-linearly impacted.

\begin{definition} 
    {\bfseries\upshape (Pooling Switching Point, PSP)}
    A Pooling Switching Point (PSP) is an input $X^{(1)} \in \mathbb{R}^{d^{(1)}}$ such that two inputs within a local receptive field achieve the same maximum value.  By varying the inputs in the tiny vicinity of $X^{(1)}$, the output of CNN changes non-linearly. 
\end{definition}

\begin{property}\label{pro:psp}
   If $X^{(1)}$ is a PSP corresponding to the $i$-th and $j$-th neurons in layer $k$, then $A^{(k)}_{i} X^{(k)} + b^{(k)} = A^{(k)}_{j} X^{(k)} + b^{(k)}$, and the inputs of  other neurons in the same LRF-P are smaller than it. 
 %  and $A^{(k)}_{j} (F^{k-1} X^{(k)}) + B_j^{(k)}<0, \forall j \in \Omega_i,j\neq i$
\end{property}

Thus, non-linearity in CNNs is jointly induced by the changes in neuron activation states via ReLU and the shifts in neuron selection via pooling operations. Meanwhile, note that in the Fully-connected Block, critical points of each neuron lead to the existence of a corresponding ReLU state change.

\begin{definition} 
    {\bfseries\upshape (Fully-connected Critical Point, FCP \cite{DBLP:conf/crypto/CarliniJM20})}
    A Fully-connected Critical Point (FCP) is an input $X^{(1)} \in \mathbb{R}^{d^{(1)}}$ that makes the input of ReLU function $\sigma^{(k)}$ of a neuron in the Fully-connected Block equal to 0. 
\end{definition}
% For instance, if $X^{(1)}$ is a FCP corresponding to the $i$-th neuron in layer $k$ in the Fully-connected Block, then $A^{(k)}_i X^{(k)} + b_i^{(k)}=0$.

We collect RPCPs, PSPs and FCPs from all layers by continuously varying inputs along any straight lines in $\mathbb{R}^{d^{(1)}}$, and searching for the breakpoints where non-linear changes happen.

\paragraph{Remarks.} In the Fully-connected Block, the FCPs will definitely affect the output. However, the Convolutional Block is in fact a series of partially connected layers. Therefore, the RPCPs and PSPs should satisfy the properties \ref{pro:rpcp} and \ref{pro:psp} respectively, and they should also be selected by subsequent Convolutional Rounds (which must be activated and selected through subsequent activation and pooling operations until the model's output) to affect the final output of CNN.

\section{Model Extraction with ReLU-Pooling Critical Point}
\label{sect:RPCP_Method}
%\subsection{Signature Extraction}

%\subsubsection{Calculate the row vector of Convolutional Kernel Matrix.}

\subsection{Signature Extraction: Algebraic View of Carlini {\em et al.}'s Differential Attack}
\label{subsect: RPCP_signature}
The signature extraction is mainly based on Carlini {\em et al.}'s differential attack \cite{DBLP:conf/crypto/CarliniJM20}. We formalize it in an algebraic view. 
Consider a specific input $X\in \mathbb{R}^{d^{(1)}}$ identified as a ReLU-Pooling Critical Point (RPCP)\footnote{Step 3 in Sect. \ref{sect:attack_in_practice} discusses how to distinguish between RPCP, PSP, and FCP.} corresponding to the $i$-th neuron in the $k$-th convolutional layer. $X$ is on the boundary between its two linear neighborhoods where only the activation state of neuron $i$ changes among all neurons in the network by slightly altering $X$. A small perturbation $\vec{\delta}$ is introduced such that the points $X+\vec{\delta}$ and $X-\vec{\delta}$ fall into these two respective linear neighborhoods. Consequently, the corresponding activation matrices, denoted as $I_{+}^{(k)}$ (for $X+\vec{\delta}$) and $I_{-}^{(k)}$ (for $X-\vec{\delta}$), differ only at the $i$-th diagonal entry. Specifically, the $i$-th entry is $1$ (assuming activation under $X+\vec{\delta}$) and $0$ (assuming inactivation under $X-\vec{\delta}$).

Assume that the $r=(p\cdot w^{(k)}_{\rho}+q)$-th LRF-P of $\hat{O}^{(k)}$ covers the critical neuron $i$, where ($0\leq p\leq h^{(k)}_{\rho}-1$, $0\leq q\leq w^{(k)}_{\rho}-1$, $0\leq r< d^{(k+1)}$). 
According to Eqs.~\eqref{eq:maximum_index_Lamda} and \eqref{eq:pooling_selection_M}, on the active side of $X$, the selection Boolean matrix  $M^{(k),p,q}$ corresponding to the activated state $X+\vec{\delta}$ only selects the positive activation state of neuron $i$, {\em i.e.},  the corresponding selected  position is 1 and all others are 0 in $M^{(k),p,q}$. However, on the inactive side of $X$, the $M^{(k),p,q}$ of the inactive state $X-\vec{\delta}$ can be any Boolean matrix with only one element being 1 and the rest being 0, because $X-\vec{\delta}$ makes all neurons in $(p\cdot w^{(k)}_{\rho}+q)$-th LRF-P  be 0 by the ReLU functions due to Prop. \ref{pro:rpcp} of RPCP.  Hence, we choose the same selection matrix $M^{(k),p,q}$ for the inactive side as for  the active side. Consequently, the pooling matrix $P^{(k)}$ remains identical in the both sides of $X$. 

Then, the output differences can be expressed as:
\begin{align}
    \mathcal{F}_{\theta}(X+\vec{\delta})-\mathcal{F}_{\theta}(X) = G^{k+1} P^{(k)} I_{+}^{(k)} A^{(k)} F^{k-1} \vec{\delta},\label{eqn:high_differential_1}\\
    \mathcal{F}_{\theta}(X)-\mathcal{F}_{\theta}(X-\vec{\delta}) = G^{k+1} P^{(k)} I_{-}^{(k)} A^{(k)} F^{k-1} \vec{\delta}.\label{eqn:high_differential_2}
\end{align}
Define the second-order differential function as $\mathcal{H}(X;\vec{\delta})$ by the subtraction of Eqs.~\eqref{eqn:high_differential_1} and \eqref{eqn:high_differential_2}, {\em i.e.,} $\mathcal{H}(X;\vec{\delta}) = \mathcal{F}_{\theta}(X+\vec{\delta}) + \mathcal{F}_{\theta}(X-\vec{\delta}) - 2\mathcal{F}_{\theta}(X)$, therefore
\begin{align}\label{eqn:2nd_rpcp}
    \mathcal{H}(X;\vec{\delta}) = G^{k+1} P^{(k)} (I_{+}^{(k)}-I_{-}^{(k)}) A^{(k)} F^{k-1} \vec{\delta},
\end{align}
where the term $I_{+}^{(k)}-I_{-}^{(k)}$ yields a diagonal matrix with a single non-zero entry (value 1) at the entry $(i,i)$, effectively acting as a selection operator.  The product $P^{(k)}(I_{+}^{(k)} - I_{-}^{(k)})$ extracts the specific column of the pooling matrix $P^{(k)}$ associated with the critical neuron $i$. Since $X$ is an RPCP, the output of the neuron $i$ is the unique maximum within its  LRF-P, implying that the corresponding entry of  $P^{(k)}$ is $1$ ({\em i.e.}, it is selected), while the contributions from other neurons in the window are zero. 

Assume that the LRF-Ps are disjoint and the network has a single output (targeting one of the components of the output vector). % or just target a specific element $y_j$ of the final output vector $y$. 
The matrix $G^{k+1}$ reduces to a single row vector, denoted as $g = (g_0, g_1, \cdots, g_{d^{(k+1)}-1})$, then %Note that: the operation simplifies to extracting the $i$-th row of the convolution kernel matrix:
\begin{align}\label{eqn:rpcp_linear_eqn}
    G^{k+1} P^{(k)} (I_{+}^{(k)} - I_{-}^{(k)}) A^{(k)}
    =(g_r A_{i,0}^{(k)}, g_r A_{i,1}^{(k)}, \cdots, g_r A_{i,d^{(k)}-1}^{(k)}).
\end{align}
By sampling $n$ different directions $\vec{\delta}^{(j)}$ ($0\leq j\leq n-1$) and querying the model to get $\mathcal{H}(X;\vec{\delta}^{(j)})$, we construct a system of equations:
\begin{align}\label{eqn:rpcp_linear_system}
    \begin{pmatrix}
    (F^{k-1} \vec{\delta}^{(0)})^{\sf T} \\
    (F^{k-1} \vec{\delta}^{(1)})^{\sf T} \\
    \vdots  \\
    (F^{k-1} \vec{\delta}^{(n-1)})^{\sf T}
    \end{pmatrix}
     \begin{pmatrix}
    g_r A_{i,0}^{(k)} \\
    g_r A_{i,1}^{(k)} \\
    \vdots  \\
    g_r A_{i,d^{(k)}-1}^{(k)}
    \end{pmatrix}
    =
    \begin{pmatrix}
    \mathcal{H}(X;\vec{\delta}^{(0)}) \\
    \mathcal{H}(X;\vec{\delta}^{(1)}) \\
    \vdots  \\
    \mathcal{H}(X;\vec{\delta}^{(n-1)})
    \end{pmatrix} =\vec{\mathcal{H}}.
\end{align}
Denote $S=({\vec{\delta}^{(0)}}, {\vec{\delta}^{(1)}},\cdots, {\vec{\delta}^{(n-1)}})^{\sf T}$, the system above can be rewritten as:
\begin{align}\label{eqn:diff_sign}
    S {F^{k-1}}^{\sf T} g_r {A_{i}^{(k)}}^{\sf T} = \vec{\mathcal{H}}.
\end{align}
%\paragraph{Remark.}  Eqs. \eqref{eqn:high_differential_2}-\eqref{eqn:rpcp_linear_eqn}. XXX
%\color{black}

\subsection{Pattern Matching Method: Recover Full Signature with Multiple RPCPs}\label{sect:rpcp_nan}
To ensure the linear system above possesses a unique solution, the coefficient matrix of Eq.~\eqref{eqn:diff_sign} should maintain full column rank, i.e., $\mathrm{rank}(S (F^{k-1})^{\sf T}) = d^{(k)}$.
As discovered by Carlini {\em et al.} \cite[Section 4.3.2]{DBLP:conf/crypto/CarliniJM20} in the attack on FCNN, the rank of the matrix is usually not full since the ReLU functions suppress negative output, and therefore some weights  are impossible to recover (denoted as {\sf NaN} values) with only one critical point. They have to find at least two critical points for the same $i$-th neuron to derive several partial signatures and merge them to extract the full signature \cite{DBLP:conf/crypto/CarliniJM20,DBLP:journals/iacr/LiuSELBP26}.  

Similarly, given one RPCP, the $\mathrm{rank}(S (F^{k-1})^{\sf T})$  of Eq.~\eqref{eqn:diff_sign} in CNNs is also affected by ReLU functions (i.e., the rank of $I^{(k-1)}$ may not be full), leading to ${rank}(S (F^{(k-1)})^{\sf T})<d^{(k)}$. 
%Consider each component of the transition matrix ${F^{k-1}}=P^{(k-1)} I^{(k-1)} \cdots A^{(1)}$, 
Note that according to Definitions \ref{def:conv_matrix} and \ref{def:pooling_matrix}, the convolutional matrix $A^{(k)}$ and the pooling matrix $P^{(k)}$ are full row rank. Therefore, some entries of $g_r {A_{i}^{(k)}}^{\sf T}$ are impossible to recover with only one RPCP. Similar to  Carlini {\em et al.}'s method \cite{DBLP:conf/crypto/CarliniJM20}, we also need multiple RPCPs to recover several partial signatures and merge them, since ${rank}(S (F^{k-1})^{\sf T})<d^{(k)}$ in most cases. 

The difference is that our RPCPs do not necessarily correspond to the same neuron, since according to Eq.~\eqref{eqn:A_row}, each row of $A^{(k)}$ contains the full information about the kernel matrix $C^{(k)}$, whose entries are to be recovered. 
According to Eq.~\eqref{eqn:A_row}, different RPCPs for different neurons in layer $k$ recover different rows of $A^{(k)}$. {\em The problem is to determine which rows of $A^{(k)}$ the RPCPs recover, and then merge them according to Eq.~\eqref{eqn:A_row} to get the full signature.} In theory, given a row  $\hat{A}_i^{(k)}$ recovered by an RPCP, the row number $i$ could also  be determined by the non-zero and zero positions in Eq.~\eqref{eqn:A_row}. However, in practice, both the zero-values and kernel weights (nonzero) in $A^{(k)}_i$ have possibility to be recovered as {\sf NaN}. Consequently, it cannot be directly recognized which row the RPCP recovers, {\em i.e.}, one cannot directly determine which neuron is associated with the RPCP. 
% In practice, given an RPCP, we can not trivially distinguish which ReLU function it corresponds to.
Therefore, it is not trivial to align the partially recovered $C^{(k)}$ and merge them. We propose the {\em Pattern Matching Method} to solve this problem.

\subsubsection{Pattern Matching Method.}
According to Eq.~\eqref{eqn:A_row}, for the $i$-th row of $A^{(k)}$, define the corresponding $i$-th convolution kernel pattern vector $\vec{V}_i\in \mathbb{F}_2^{d^{(k)}}$ as 
\begin{equation}\label{eqn:rpcp_pattern_V}
    \vec{V}_i[j]=\left\{ \begin{array}{l}
1, ~~~\mbox{if}~~ A_i^{(k)}[j]\neq 0,\\
0, ~~~\mbox{else}.
\end{array} \right.
\end{equation}

Define its bitwise inverse vector as $\vec{\check{V}}_i\in \mathbb{F}_2^{d^{(k)}}$ with $\vec{\check{V}}_i[j]=\vec{{V}}_i[j]\oplus 1$. 

Suppose we have an RPCP corresponding to the $i$-th ReLU function of layer $k$, denote the computed vector from Eq.~\eqref{eqn:diff_sign} as $\hat{A}_i^{(k)} \approx  {g_rA_{i}^{(k)}}$, where there exists {\sf NaN} values. According to Eq.~\eqref{eqn:A_row_nan}, the recovered $\hat{A}_i^{(k)}$ could be 
\begin{equation}\label{eqn:A_row_nan}\tiny
            \hat{A}^{(k)}_i = (\underbrace{\bf 0}_{w^{(k)}_{in}\cdot \bar{i}},\underbrace{
        \underbrace{\bf 0}_{\underline{i}}, \hat{C}^{(k)}_{0,0}, {\sf NaN}, \cdots, \hat{C}^{(k)}_{0,w_{c}^{(k)}-1},    \underbrace{\bf 0}_{pad}}_{w_{in}^{(k)}}, 
     \cdots, 
        \underbrace{
        \underbrace{\bf 0}_{\underline{i}}, %\hat{C}^{(k)}_{h_{ker}^{(k)}-1, 0},
        {\sf NaN}, {\sf NaN}, \cdots, \hat{C}^{(k)}_{h_{c}^{(k)}-1,w_{c}^{(k)}-1},   \underbrace{\bf 0}_{pad}}_{w_{in}^{(k)}},  
        \underbrace{\bf 0}_{pad}),
        \end{equation}
        where $\bar{i}=\left\lfloor \frac{i}{w^{(k)}_{o}} \right\rfloor$, $\underline{i}=i\bmod w^{(k)}_o$. Please note that any entry in $\hat{A}_i^{(k)}$ may contain a {\sf NaN} value, including those entries whose actual value should be 0. Then, we define two binary mask vectors for $\hat{A}_i^{(k)}$ as $\vec{\alpha}, \vec{\alpha}'\in \mathbb{F}_2^{d^{(k)}}$, where 
     \begin{equation*}
     \resizebox{1.0\hsize}{!}{$%
         \vec{\alpha}[j]=\left\{ \begin{array}{l}
1, ~~~\mbox{if}~~ A_i^{(k)}[j]\neq 0~~ \mbox{or}~~ A_i^{(k)}[j]= {\sf NaN} \\
0, ~~~\mbox{else},
\end{array} \right., ~~~\vec{\alpha}'[j]=\left\{ \begin{array}{l}
1, ~~~\mbox{if}~~ A_i^{(k)}[j]= 0~~ \mbox{or}~~ A_i^{(k)}[j]= {\sf NaN} \\
0, ~~~\mbox{else}.
\end{array} \right.
$}%
     \end{equation*}

%The calculated row vector is denoted as $\hat{A}_i^{(k)} = \widehat{g_rA_{i}^{(k)}}$. In an ideal scenario where every element of the calculated vector is precise, the entire kernel signature could be retrieved simply by extracting all non-zero values. However, as previously discussed, the majority of the row vectors are incomplete, containing NaN values. To match and recover We not only need the non-zero values, but their corresponding position information. To address this challenge, we propose a method to robustly extract the Convolutional Kernel Pattern from these vectors contaminated with NaNs.

Given an RPCP, $\vec{\alpha}, \vec{\alpha}'\in \mathbb{F}_2^{d^{(k)}}$ are computed. Then, we use the following \textit{matching principle} to detect which ReLU function the RPCP corresponds to. %we sequentially match the Convolutional Kernel Pattern $\mathcal{P}$ against the extracted row vector $\hat{A}_{i}^{(k)}$ via a sliding window approach from left to right. A valid alignment must satisfy the following \textit{Matching Principle}: 
Assuming the RPCP corresponds to the $i$-th ReLU function, we consider the $i$-th convolution kernel pattern vector $\vec{V}_i$ and the extracted vector $\hat{A}_i^{(k)}$. 
\begin{enumerate}
    \item Non-zero Consistency Principle: Obviously, when $\vec{V}_i[j]=1$, $\hat{A}_i^{(k)}[j]$ will be either non-zero or {\sf NaN}, {\em i.e.}, $\vec{V}_i[j]=\vec{\alpha}[j]$ and 
    \begin{equation}\label{eqn:rpcp_principle_1}
        \vec{V}_i\cdot \vec{\alpha}^{\sf T}=w_{c}^{(k)}\cdot h_{c}^{(k)}.
    \end{equation}

    %and the masking vector $\vec{A}$,  
    
    %At indices where the pattern $\mathcal{P}$ is non-zero(kernel weights), the corresponding element in the extracted vector $\hat{A}_i^{(k)}$ must be either non-zero or $\text{NaN}$;
    \item Zero Consistency Principle: Similarly, when $\vec{\check{V}}_i[j]=1$, $\vec{\alpha}'[j]=1$  we have  \begin{equation}\label{eqn:rpcp_principle_2}
    \vec{\check{V}}_i\cdot \vec{\alpha}'^{\sf T}=d^{(k)}-w_{c}^{(k)}\cdot h_{c}^{(k)}.
    \end{equation}

    %At indices where the pattern $\mathcal{P}$ is zero, the corresponding element in $\hat{A}_i^{(k)}$ must be either $0$ or $\text{NaN}$.
\end{enumerate}
We iteratively compute Eqs.~\eqref{eqn:rpcp_principle_1} and \eqref{eqn:rpcp_principle_2} for each $0\leq i\leq d_f^{(k)}-1$ and check if these two equations hold. If Eqs.~\eqref{eqn:rpcp_principle_1} and \eqref{eqn:rpcp_principle_2} hold, we deduce that the RPCP corresponds to the $i$-th ReLU function, and then extract partial values of the $\hat{C}^{(k)}$ from Eq.~\eqref{eqn:A_row_nan}. For multiple RPCPs, we recover the partial $\hat{C}^{(k)}$ and merge them to build a full $\hat{C}^{(k)}$. 
%\xiaoyang{why not just pick the nonzero or nan values from Eq.~\eqref{eqn:A_row_nan} to merge???}

 \paragraph{Remarks.} %\xiaoyang{Detail why here}
{By employing the pattern matching method, we identify the specific neuron and determine the constituent elements of the kernel matrix from a single RPCP. This makes the reconstruction of kernel weights simple. 
The identification mechanism is both intuitive and robust: results derived from noisy points (non-RPCPs) fail to yield a kernel pattern vector $\vec{V}_i$ that conforms to the matching principle, causing the pattern matching algorithm to inherently reject them.}

\subsubsection{Bias Recovery.}
%Once the starting position of the Convolutional Kernel Pattern $\mathcal{P}$ is determined, the specific position of the convolutional kernel activates on the input matrix is confirmed, and 
In layer $k$, after the neuron $i$ at the RPCP is identified, 
%Assuming the first $k-1$ layers are fully recovered (we can reconstruct $F^{(k-1)}$) and the convolution matrix $\hat{A}_c^{(k)}$ is reconstructed from the solved kernel signatures. Then 
the bias (note that the bias for each neuron is the same according to Definition \ref{def:conv_matrix}) is computed, since the input of the neuron $i$ is zero: 
\begin{equation}\label{eqn:rpcp_bias}
    \hat{b}^{(k)}=-\hat{A}_i^{(k)} ({F}^{k-1} x+B^{k-1}) \ .
%    \label{eq:bias recovery}
\end{equation}

\subsection{Sign Recovery}\label{sect:rpcp_sign}
%The Sign Recovery procedure leverages the property of the RPCP: 
%within the corresponding local receptive field, the critical neuron is at a critical state (value 0), while all other competing neurons must strictly exhibit negative values. This property serves as a verification oracle. 

After recovering $\hat{A}^{(k)}$ and $\hat{B}^{(k)}$ without sign, we compute the input vector of the neurons as
$\hat{A}^{(k)} (F^{k-1} x+B^{k-1})+\hat{B}^{(k)}$, where $x$ is the RPCP of the $i$-th neuron of layer $k$. Assume that neuron $i$ belongs to the $t$-th LRF-P $\Omega_t$. According to Prop. \ref{pro:rpcp}, all the inputs of the neurons within  $\Omega_t$ should be $\leq 0$. Then, 
%after the $k$-th convolution, we inspect the values of the non-critical neurons within the identified local receptive field ${(\hat{A}_c^{(k)} F^{(k-1)} x)}_j, ~j \in \Omega_i, ~j \neq i$: 
\begin{itemize}
\item if all these inputs are $\leq 0$, the sign of the extracted kernel matrix is correct.
\item if there exists a positive input, the extracted kernel matrix has an inverted sign, and we multiply the matrices $\hat{A}^{(k)}$ and $\hat{B}^{(k)}$ by ``$-1$". % Consequently, the recovered kernel parameters must be multiplied by 
\end{itemize}

%\subsection{Evaluation}

%\subsubsection{Strengths.}
%The proposed method exhibits several key advantages. First, every single RPCP associated with the $k$-th layer enables the direct recovery of the $k$-th layer convolution kernel weights. Second, the identification mechanism is both straightforward and robust: results derived from noise points (non-RPCPs) fail to align with the sparse row vector structure containing the Convolutional Kernel Pattern, causing the pattern matching algorithm to inherently reject them. Furthermore, the subsequent procedures for bias calculation and sign recovery are computationally efficient and simple to implement.

\section{Parameter Extraction with Pooling Switching Point}
\label{sect:PSP Method}

The attack based on the RPCP given in Sect. \ref{sect:RPCP_Method}  relies highly on the successful acquisition of RPCPs. When extending the attack to deeper layers, the proportion of valid RPCPs among all discovered critical points diminishes greatly and the efficiency  of the attack decreases significantly (detailed experiments confirm this phenomenon in Sect. \ref{sect:attack_in_practice}). Therefore, we propose a more efficient parameter extraction attack based on the Pooling Switching Point (PSP). To leverage the PSP, a novel {\em internal differential attack} on CNNs is applied, instead of Carlini {\em et al.}'s  differential attack \cite{DBLP:conf/crypto/CarliniJM20}.  

%recovery phase . %This scarcity leads to a lower hit rate, thereby reducing the overall search efficiency for deep network layers.

\subsection{Internal Differential Extraction Attack on CNN}
\label{subsect:internal differential}
The Internal differential attack \cite{DBLP:conf/crypto/Peyrin10} was proposed  by Peyrin at CRYPTO 2010 against {\sf Gr\o stl}, which was subsequently applied to the cryptanalysis of {\sf Keccak} by Dinur, Dunkelman, and Shamir \cite{DBLP:conf/fse/DinurDS13}. This section proposes the first internal differential attack on CNNs, named the {\em internal differential extraction} attack.

According to Prop. \ref{pro:psp} of the Pooling Switching Point (PSP): Given a PSP $X$ in layer $k$, suppose the $i$-th and $j$-th neurons' inputs within the same $t$-th LRF-P $\Omega_t$ ($0\leq t< d^{(k+1)}$) have the same largest value, which satisfies: 
% \begin{align}
%  Y^{(k)}[i] =  A_{i}^{(k)} (F^{k-1}X+B^{k-1}) + b^{(k)} =  Y^{(k)}[j] =A_{j}^{(k)} (F^{k-1}X+B^{k-1}) + b^{(k)} > 0.
%     \label{eq:PSP}
% \end{align}

\begin{equation} \label{eq:PSP}
    \resizebox{\linewidth}{!}{$\displaystyle
        Y^{(k)}[i] = A_{i}^{(k)} (F^{k-1}X+B^{k-1}) + b^{(k)} = Y^{(k)}[j] = A_{j}^{(k)} (F^{k-1}X+B^{k-1}) + b^{(k)} > 0.
    $}
\end{equation}
%Since these two neurons achieve the maximum value within the same local receptive field, 
Then, $X$ is on the boundary between two linear neighborhoods, across which the selection of the $t$-th LRF-P switches between neuron $i$ and neuron $j$.
It is important to note that $A_{i}^{(k)}$ is a sparse vector according to Eq.~\eqref{eqn:A_row}, which only contains non-zero entries at the indices where the convolution kernel is applied. This sparsity allows us to explicitly extract the positional indices of the convolution kernel's receptive field (LRF-C). 
Denote the vector form of kernel matrix $C^{(k)}$ as $\vec{C}^{(k)}=(C^{(k)}_0, C^{(k)}_1,\cdots,C^{(k)}_{h_{c}^{(k)}-1})^{\mathsf{T}}$. Thus the length of the vector $\vec{C}^{(k)}$ is $l_{c}^{(k)} = h^{(k)}_{c} w^{(k)}_{c}$. Assume that the stride $s_c^{(k)}=1$, and let $\mathcal{K}^{(k)}_i \subset \{0,1,\cdots, d^{(k)}-1\}$ denote the set of indices belonging to the LRF-C of the $i$-th neuron ($0\leq i\leq d^{(k)}_f-1$), then
\begin{equation}  \label{eq:indices K}
    \mathcal{K}_i^{(k)} \mathrel{:=} \left\{w_{in}^{(k)} \cdot \left\lfloor \frac{i}{w^{(k)}_o} \right\rfloor + i \bmod w^{(k)}_o 
   + w_{in}^{(k)} \cdot \left\lfloor \frac{t}{w^{(k)}_{c}} \right\rfloor + t \bmod w^{(k)}_{c},\ 0 \le t < l_{c}\right\}.
\end{equation}
The set $\mathcal{K}^{(k)}_i$ picks out the nonzero entries of $A^{(k)}_i$ according to Eq.~\eqref{eqn:A_row}. 
Then,
\begin{align}
    A_{i}^{(k)} X^{(k)} = \vec{C}^{(k)} \cdot X^{(k)}[\mathcal{K}^{(k)}_i]. 
\end{align}
Similar equations can be deduced for neuron $j$. Then, subtract the two equations, according to Eq.~\eqref{eq:PSP},  we get  
%The property can be rewritten as:
\begin{equation} \label{eqn:internal-differ}
    \vec{C}^{(k)} \cdot (X^{(k)}[\mathcal{K}^{(k)}_i] - X^{(k)}[\mathcal{K}^{(k)}_j]) = 0.
\end{equation}
This is quite similar to the internal-differential attack \cite{DBLP:conf/crypto/Peyrin10}, where the difference between two distinct parts of the unique input $X^{(k)}$ is studied through the cryptographic/CNN operations. The input difference is $\nabla X^{(k)} = X^{(k)}[\mathcal{K}^{(k)}_i] - X^{(k)}[\mathcal{K}^{(k)}_j]$, and the output difference after the convolutional layer is 0, which will be identified as a PSP. Given a PSP, Eq.~\eqref{eqn:internal-differ} is used to construct the system of equations to compute $\vec{C}^{(k)}$. With the deduced parameters of  $k-1$ layers, $X^{(k)}=F^{k-1}X+B^{k-1}$ is deduced. The only unknowns are the positions of the pooling selection switch, {\em i.e.}, $i$ and $j$.

%, the difference bet  ween the $i$-th and the $j$-th part of $X^{(k)}$ constitute the input differential $\Delta X^{(k)} = X^{(k)}_{\mathcal{K}_i} - X^{(k)}_{\mathcal{K}_j}$, and we can calculate $X^{(k)}$ from recovering the previous $k-1$ layers. So when we obtain the positional information $i$ and $j$, we can extract $Ker^{(k)}$ which is the convolutional kernel signature.

\subsection{Extracting  $i$ and $j$ of the Pooling Selection Switch}
\label{sect:psp_ij}
Similar to Carlini {\em et al.}'s differential attack  \cite{DBLP:conf/crypto/CarliniJM20}, we introduce a small perturbation $\vec{\delta}$ such that the points ${X} + \vec{\delta}$ and ${X} - \vec{\delta}$ remain within the linear neighborhoods of the PSP $X$. 
Exactly one of the two points will make the pooling select the $i$-th neuron's output. Assume that this point is ${X} + \vec{\delta}$, and denote the corresponding pooling matrix as $P_{+}^{(k)}$. The other point will make the output of the $j$-th neuron selected, and $P_{-}^{(k)}$ is defined similarly. 
 Then, we have
\begin{align}\label{eq:difference_psp_1}
    \mathcal{F}_{\theta}(X+\vec{\delta})-\mathcal{F}_{\theta}(X) = G^{k+1} P_{+}^{(k)} I^{(k)} A^{(k)} F^{k-1} \vec{\delta},\\
    \label{eq:difference_psp_2}\mathcal{F}_{\theta}(X)-\mathcal{F}_{\theta}(X-\vec{\delta}) = G^{k+1} P_{-}^{(k)} I^{(k)} A^{(k)} F^{k-1} \vec{\delta}.
\end{align} 
Note that because neurons $i$ and $j$ share the same maximum value at $X$, selecting either neuron yields the correct pooling output. Consequently, $P_{+}^{(k)}$ and $P_{-}^{(k)}$ are functionally equivalent at $X$. Hence, for $\mathcal{F}_{\theta}(X)$,  Eq.~\eqref{eq:difference_psp_1} uses $P_{+}^{(k)}$ to maintain the same matrix as $\mathcal{F}_{\theta}(X+\vec{\delta})$, and similarly for Eq.~\eqref{eq:difference_psp_2}.
Assuming the $t$-th ($0\leq t\leq d^{(k+1)}-1$) LRF-P includes neurons $i$ and $j$ ($i<j$) whose input values are $Y^{(k)}[i]$ and $Y^{(k)}[j]$, the matrices $P_{+}^{(k)}$ and $P_{-}^{(k)}$ will be different only in their $t$-th row. %, whose corresponding LRF-P includes . 
%vector corresponding to the pooling window overlapping neuron $i$ and $j$. 
Specifically, in $P_{+}^{(k)}$, this row is $(0\cdots 0, 1_i, 0 \cdots0, 0_j,0\cdots0)$, while in $P_{-}^{(k)}$, the same row becomes $(0 \cdots0, 0_i, 0\cdots 0,$ $ 1_j, 0 \cdots 0)$. Still, the second-order difference function, denoted as $\mathcal{H}(x; \delta)=\mathcal{F}_{\theta}(X+\vec{\delta}) +\mathcal{F}_{\theta}(X-\vec{\delta}) - 2\mathcal{F}_{\theta}(X)$:
\begin{align}\label{eqn:2nd_psp}
    \mathcal{H}(X; \vec{\delta}) = G^{k+1} (P_{+}^{(k)} - P_{-}^{(k)}) I^{(k)} A^{(k)} F^{k-1} \vec{\delta}. 
\end{align}
Similar to the RPCP method in Sect. \ref{sect:RPCP_Method}, when the matrix $G^{k+1}$ is reduced to a single row vector $g = (g_0, g_1, \cdots, g_{d^{(k)}-1})$, then %assuming the $t$-th pooling window contains the neuron $i$ and $j$ after convolutional layer. 
\begin{align}
    G^{k+1} (P_{+}^{(k)} - P_{-}^{(k)}) I^{(k)} A^{(k)}
    =g_t (A_{i}^{(k)} - A_{j}^{(k)}),
\end{align}
where $t$-th row of  $P_{+}^{(k)} - P_{-}^{(k)}$ is $(0 \cdots0, 1_i, 0\cdots0, -1_j,0\cdots0)$, and the other rows are zero vectors. 
 By sampling $n$ distinct directions $\vec{\delta}^{(\ell)}$ ($0\leq \ell \leq n-1$), we can construct a system of equations and recover $g_t (A_{i}^{(k)} - A_{j}^{(k)})$, 
\begin{align}\label{eqn:psp_Aij}
   g_t (A_{i}^{(k)} - A_{j}^{(k)})\cdot F^{k-1}\vec{\delta}^{(\ell)} = \mathcal{H}(X;\vec{\delta}^{(\ell)}).
\end{align}

According to Eq.~\eqref{eqn:A_row}, $A_{i}^{(k)}$ and $A_{j}^{(k)}$ correspond to the $i$-th and $j$-th row vectors of the  matrix $A^{(k)}$, where their non-zero values are offset from each other. Assuming $i<j$, we deduce $i$ and $j$ from the positions of the first and last nonzero entries of the recovered vector $g_t (A_{i}^{(k)} - A_{j}^{(k)})$, respectively,  according to Eq.~\eqref{eqn:A_row}.  % and deduce $j$ from the position of the last nonzero entry of the recovered vector $g_t (A_{i}^{(k)} - A_{j}^{(k)})$, 
%of the first non-zero value, and $j$ from the position $pos_j$ of the last non-zero value by $j = pos_j - l_{pattern} +1$. 
Then $\mathcal{K}_i \text{ and } \mathcal{K}_j$ are deduced from Eq.~\eqref{eq:indices K}.

\subsubsection{Handling vector containing {\sf NaN}.} Sometimes the calculated row vector $\Delta_{ij} \hat{A}^{(k)}$ $\approx g_t (A_{i}^{(k)} - A_{j}^{(k)})$ from Eq.~\eqref{eqn:psp_Aij} is incomplete for reasons similar to those in Sect. \ref{sect:rpcp_nan}. It is necessary to check whether the deduced  $\hat{i}$ and $\hat{j}$ match the practical $g_t (A_{i}^{(k)} - A_{j}^{(k)})$, following the matching principles:
\begin{enumerate}
    \item Union Non-zero Consistency: At indices where either $A_{i}^{(k)}$ or $A_{j}^{(k)}$ is non-zero, the corresponding element in the difference vector $\Delta_{ij} \hat{A}^{(k)}$ should be either non-zero or {\sf NaN}.
    
    \item Common Zero Consistency: At indices where both $A_{i}^{(k)}$ and $A_{j}^{(k)}$ are zero, the corresponding element in $\Delta_{ij} \hat{A}^{(k)}$ should be either $0$ or {\sf NaN}.
\end{enumerate}
Following the above two principles and Eq.~\eqref{eqn:A_row}, we determine $\hat{i}$ by finding the first $w^{(k)}_{c}$ consecutive non-zero or {\sf NaN} entries that contain at least one non-zero value in $\Delta_{ij} \hat{A}^{(k)}$ (denote this partial vector as $\vec{\beta}$), then 
\begin{itemize}
    \item If the first element $\vec{\beta}[0]$ is not {\sf NaN}, return the position index of $\vec{\beta}[0]$ as $\hat{i}$. 
    \item If $\vec{\beta}[0]$ is {\sf NaN}, let $\hat{i}$ choose two possible values, {\em i.e.}, 
    \begin{itemize}
        \item $\hat{i}$ is the index of $\vec{\beta}[0]$ in $\Delta_{ij}$ by assuming the true value $\vec{\beta}[0]$ is not 0.
        \item Increment the index of $\vec{\beta}[0]$ in $\Delta_{ij}$ by 1, and use it as the value of $\hat{i}$, by assuming the true value $\vec{\beta}[0]$ is 0. %Note that with a marginal probability, the $\vec{\beta}[1]$ is of {\sf NaN} and its true value is also 0, and a wrong $\hat{i}$ returned. 
    \end{itemize}
\end{itemize}
Similarly, we determine  $\hat{j}$ by the last $w^{(k)}_{c}$ consecutive non-zero or {\sf NaN} values in $\Delta_{ij} \hat{A}^{(k)}$. 
Only the correct $(\hat{i}, \hat{j})$ will strictly conform to the matching principles; otherwise we discard the PSP and choose a new one. After recovering the pair of $(\hat{i},\hat{j})$, we construct Eq.~\eqref{eqn:internal-differ}.
%After recover the possible pairs of $(\hat{i},\hat{j})$, we solve the Eq. \eqref{eqn:internal-differ} to recover the $\vec{C}^{(k)}$. Note that, with marginal probability, the correct $(\hat{i},\hat{j})$ is not recovered, then, we discard this PSP and choose a new one to build Eq. \eqref{eqn:internal-differ}. 
We need slightly more than $h_{c}^{(k)}w_{c}^{(k)}$ PSPs to build a full-rank system of Eq.~\eqref{eqn:internal-differ} to recover $\vec{C}^{(k)}$ as $\hat{\vec{C}}^{(k)}$.

%\paragraph{Is there {\sf NaN} entries in the recovered $\hat{\vec{C}}^{(k)}$?} According to Eq. \eqref{eqn:internal-differ}, if the We leverage PSPs from the entire input space $\mathbb{R}^{d^{(1)}}$ to build the system of equations, ensuring that it is immune to the specific activation patterns $I^{(k)}$ associated with any individual point. XXXXXXXX This prevents the generation of {\sf NaN} values and ensures a complete recovery of the kernel vector $\vec{C}^{(k)}$.

\subsection{Sign Recovery}
\label{subsect:Sign Recovery from PSP}
Reconstruct the convolutional matrix $\hat{A}^{(k)}$ from the recovered kernel weights $\hat{\vec{C}}^{(k)}$. And once we extract $\hat{i}$ and $\hat{j}$, the involved LRF-P $\Omega_t$ ($0\leq t < d^{(k+1)}$) corresponding to the PSP could be deduced. According to Prop. \ref{pro:psp}, the outputs of the $i$-th and $j$-th neurons should be the largest value in $\Omega_t$. Then, with $\hat{A}^{(k)}$, 
\begin{itemize}
\item if $\hat{A}^{(k)}_i(F^{k-1}X+B^{k-1}) = \hat{A}^{(k)}_j(F^{k-1}X+B^{k-1}) > \hat{A}^{(k)}_{\ell}(F^{k-1}X+B^{k-1}), \ \ell\in \Omega_t$, the sign of the extracted kernel matrix is correct.
\item if $\hat{A}^{(k)}_i(F^{k-1}X+B^{k-1}) = \hat{A}^{(k)}_j(F^{k-1}X+B^{k-1}) < \hat{A}^{(k)}_{\ell}(F^{k-1}X+B^{k-1}), \ \ell\in \Omega_t$, the extracted kernel matrix has an inverted sign, and we multiply the vector $\hat{A}^{(k)}_i$ by ``$-1$".
\end{itemize}

%\paragraph{Is there {\sf NaN} entries in the recovered .} We leverage PSPs from the entire input space $\mathbb{R}^{d^{(1)}}$ to build the system of equations, ensuring that it is immune to the specific activation patterns $I^{(k)}$ associated with any individual point. XXXXXXXX This prevents the generation of {\sf NaN} values and ensures a complete recovery of the kernel vector $\vec{C}^{(k)}$.
%If the pattern matches perfectly, the extracted positional information of $i$ and $j$ is correct. 
%However, mismatches may occur if the boundary elements (the first nonzero or the last nonzero entries of $\Delta_{ij} \hat{A}^{(k)}$) are obscured by {\sf NaN} values, and one can not determine the original value is 0 or nonzero. 

%To address this problem, we apply an adjustment strategy: If the first nonzero element is {\sf NaN} in position $\hat{i}$, then, assign $i=\hat{i}$ or $i=\hat{i}+1$

%, we shift the start index $pos_i$ backward (or the end index $pos_j$ forward) by one position and re-evaluate. If the extended boundary values are both {\sf NaN} or indistinguishable from zero (near-zero), the candidate point is discarded as unreliable.

\subsubsection{On the bias recovery.} Note that the bias term $b^{(k)}$ is eliminated in Eq.~\eqref{eq:PSP} on both sides, making it impossible to derive the bias information from PSP points alone. Consequently, we use the method described in Eq.~\eqref{eqn:rpcp_bias} in Sect.  \ref{sect:RPCP_Method} to recover the bias with a single RPCP.

\section{Attacks in Practice: Unified Model Combining PSP and RPCP}
\label{sect:attack_in_practice}

\subsection{Experiment on RPCP and PSP Methods}\label{sect:experiment_RPCP_PSP} %Suppose we have precisely extracted the CNN up to layer $k$, i.e., the recovered $\hat{f}^{(1 .. k-1)}$ is functionally equivalent to the $f^{(1 .. k-1)}$ of the victim model. 
We first experimentally compare the RPCP and PSP methods on a $(2+1)$-deep CNN with a single input and output channel, trained on the MNIST dataset,
\begin{itemize}
    \item In the first round, the input matrix is convolved with a kernel of size $(1,1,5,5)$, stride $s^{(1)}_c=1$ and padding $pad^{(1)}=0$, followed by ReLU functions, and then downsampled by a pooling kernel of size $2 \times 2$, with stride $s^{(1)}_{\rho}=2$. 
    \item In the second round, the operations are identical to those in the first round.
\end{itemize}

We record the number of model queries, attack runtime, and error upper bound, as summarized in Table~\ref{tab:layer_comparison} and discussed as follows.  %More detailed experimental settings and procedures are provided in subsection \ref{subsect:Experiment Verification}.
\begin{table}[!h]\scriptsize
\centering
\vspace*{-12pt}
\caption{Comparison of RPCP and PSP Methods}
\label{tab:layer_comparison}
\renewcommand{\arraystretch}{1.5} % 调整行高，与样例一致
\setlength{\tabcolsep}{3pt}       % 调整列间距
{
    \begin{threeparttable}
        \begin{tabular}{l c c l l l c c}
        \toprule
        \makecell[l]{\textbf{Layer}/\\ \textbf{Models}} 
        & \makecell[c]{\textbf{Architecture} \\ 
        $d^{(k)}\!-\!d_f^{(k)}\!-\!d^{(k+1)}$}
        & \textbf{Method} 
        & \makecell[l]{\textbf{Time}\\ seconds}
        & \textbf{Queries}
        & \boldmath$(\varepsilon, 0)$ 
        & \boldmath$\max \lvert \theta - \hat{\theta} \rvert$ 
        & \textbf{Number}
        \\
        \midrule
        
        % --- 第1组: (2+1) Layer 1 ---
        \multirow{2}{*}{\makecell[l]{1st/(2+1)}} 
        & \multirow{2}{*}{1024--784--196} 
        & RPCP & $2^{5.39}$ & $2^{15.10}$ & $2^{-12.78}$ & $2^{-16.38}$ & 1400 \\
        & & PSP  & $2^{4.16}$ & $2^{12.62}$ & $2^{-32.09}$ & $2^{-37.10}$ & 6487 \\
           \midrule
        
        % 增加一点间距或使用 \addlinespace 区分不同层，这里直接靠 arraystretch 撑开
        
        % --- 第2组: (2+1) Layer 2 ---
        \multirow{2}{*}{\makecell[l]{2nd/(2+1)}} 
        & \multirow{2}{*}{196--100--25} 
        & RPCP & $2^{6.07}$ & $2^{15.95}$ & $2^{-16.04}$ & $2^{-22.07}$ & 15 \\
        & & PSP  & $2^{4.60}$ & $2^{14.49}$ & $2^{-31.74}$ & $2^{-35.21}$ & 1548 \\
        
        \bottomrule
        \end{tabular}

        \begin{tablenotes}
            \footnotesize
            \item Number: Counting from  10,000 collected critical points.
        \end{tablenotes}
    \end{threeparttable}
}
\vspace*{-12pt}
\end{table}

\begin{itemize}
    \item {\bf Distribution density of different types of critical points.}
Among the identified critical points, the number of PSPs significantly exceeds  that of RPCPs within the same layer. Experimental results for the second layer show that out of 10,000 collected critical points, only 15 correspond to RPCPs, whereas PSP identifies 1548 in Table~\ref{tab:layer_comparison}. Consequently, the PSP method achieves model extraction with shorter runtime and fewer model queries.

%\subsection{Experimental Verification of Different Methods}
%\label{subsect:Experiment Verification}
%We evaluate the RPCP and PSP methods by attacking $(2\!-\!1)$-Deep and $(3-1)$-Deep CNNs trained on the MNIST dataset, in order to compare their practical performance. Table \ref{tab:layer_comparison} summarizes the experimental results, including the number of model queries, runtime, and the resulting $(\varepsilon,0)$-functional equivalence.
%The results support our evaluation on the two different method.

% Since both methods employ an identical strategy for bias recovery, we focus our evaluation solely on the precision of the extracted weights.
%具体说，第k层的$\varepsilon^{(k)}=\delta_ie_i$, $\delta_i$ denotes the largest singular value of $\Delta_i$, $\Delta_i=\hat{\vec{C}}^{(k)}-\vec{C}^{(k)}$, and . 

%\item {\bf Number of RPCPs  in Sect. \ref{sect:RPCP_Method} or PSPs in Sect. \ref{sect:PSP Method} needed to Recover the signature.}

\item {\bf High precision extraction.}
Recall the linear systems of equations used to solve for the convolution kernel in RPCP Eq.~\eqref{eqn:diff_sign} for $A^{(k)}_i$ and PSP Eq.~\eqref{eqn:internal-differ} for $\vec{C}^{(k)}$. For RPCPs, we must solve for $|A^{(k)}_i|$=$d^{(k)}$ variables, whereas for PSPs, the system reduces to only $\lvert \vec{C}^{(k)} \rvert =l_{c}$ variables, and $d^{(k)} \gg l_{c}$ according to Eq.~\eqref{eqn:A_row}. Therefore, the system for RPCP possesses higher degrees of freedom. Note that although the non-zero values in $A^{(k)}_i$ in Eq.~\eqref{eqn:A_row} are the elements of $\vec{C}^{(k)}$, the locations of the zero entries in $A^{(k)}_i$ are unknown before solving the linear system in Eq.~\eqref{eqn:diff_sign}.

According to statistical learning theory, the expected prediction error of a least-squares estimator increases linearly with the number of input parameters \cite[Section 2.5]{DBLP:books/lib/HastieTF09}. Consequently, estimating the $d^{(k)}$ variables tends to introduce a larger variance compared to solving the low-dimensional PSP system. Meanwhile, the method of least squares tends to distribute values across all coefficients \cite{DonohoCS06}, which often `overfits' the noise by assigning spurious non-zero values to the zero-valued entries (note that many entries of $A^{(k)}_i$ are zeros), further exacerbating the extraction error. Therefore, the solution to the PSP system in Eq.~\eqref{eqn:internal-differ} achieves higher precision.

In our practical experiment shown in Table \ref{tab:layer_comparison}, the PSP method significantly enhances the precision of parameter extraction. It achieves a much lower upper bound on the maximum output error (the term $(\varepsilon,0)$) and a much lower maximum parameter error (the term $\max \lvert \theta - \hat{\theta} \rvert$) than the RPCP method with fewer queries and less time.

%and minimizes the error between the model parameters $\theta$ and the extracted parameters $\hat{\theta}$.

%While the precision of RPCPs could theoretically be improved through refinement algorithms, this brings additional excessive model queries, which compromises the stealthiness of the attack. In contrast, PSPs inherently yield solutions with precision comparable to refined RPCPs. Therefore, to prioritize stealth and efficiency, we exclusively employ PSPs for high-precision parameter extraction.
    
\item {\bf Bias recovery.}
A critical limitation remains: 
Since PSPs do not reveal any information regarding the bias, we need to collect at least one RPCP to recover one bias, following the procedure in Eq.~\eqref{eqn:rpcp_bias} in Sect. \ref{sect:RPCP_Method}.

\end{itemize}

\subsection{Unified Model Combining PSP and RPCP}\label{sect:unified_model}

Suppose we have precisely extracted the CNN up to layer $k$, {\em i.e.}, the recovered $\hat{f}^{(1 .. k-1)}$ is functionally equivalent to the $f^{(1 .. k-1)}$ of the victim model. 
As detailed in Sects. \ref{sect:RPCP_Method} and \ref{sect:PSP Method}, the procedures for computing the row vector $\hat{A}^{(k)}_i$ from RPCP and extracting position $i$ and $j$ of the pooling selection switch from PSP are similar (both are based on the computation of the second-order differential function). Meanwhile, we rely on the solution itself to distinguish between RPCPs and PSPs. %We can employ this unified approach without incurring additional costs. 
The full attack is to leverage  the PSP method in Sect. \ref{sect:PSP Method} for high-precision weights solving, and the RPCP method  in Sect. \ref{sect:RPCP_Method} for bias and sign recovery. The detailed procedure is described below.

%Additionally, in deeper layers, NaNs more severely undermine the stability of the PSP Position Matrix Method. 

\subsubsection{Step 1. Identify Critical Points and Filter Feasible Candidates.}
Following Carlini {\em et al.}'s method \cite{DBLP:conf/crypto/CarliniJM20}, we identify a collection of critical points by performing a binary search sweep along continuous linear paths, where the output exhibits a non-linear transition. Then we filter out those belonging to previous layers. 
For each candidate point, we reconstruct the transformation matrix $F^{k-1}$. Let $N^{(k-1)}_{positive}$ denote the number of positive values in the output of the $(k-1)$-th layer. We verify whether the point satisfies the condition: $N^{(k-1)}_{positive} \le \text{rank}(F^{(k-1)})$. According to \cite[Section 3.1]{DBLP:journals/iacr/LiuSELBP26}, points that fail to meet the constraint will result in underdetermined linear systems in Eq.~\eqref{eqn:rpcp_linear_system} or Eq.~\eqref{eqn:psp_Aij}, and therefore will be discarded.

\subsubsection{Step 2. Calculate the Signature.}
For a critical point $X$ (RPCP or PSP or FCP), we sample tiny perturbations $\vec{\delta}$ in different directions, compute the second-order differential function $\mathcal{H}(X;\vec{\delta})=\mathcal{F}_{\theta}(X+\vec{\delta})+\mathcal{F}_{\theta}(X-\vec{\delta})-2\mathcal{F}_{\theta}(X)$, which is used in both the RPCP method (e.g., Eq.~\eqref{eqn:2nd_rpcp}) and the PSP method (e.g., Eq.~\eqref{eqn:2nd_psp}) to establish linear systems, such as Eq.~\eqref{eqn:rpcp_linear_system} for RPCP or Eq.~\eqref{eqn:psp_Aij} for PSP. Let this linear system be $(F^{k-1} \vec{\delta}^{(i)})^{\mathsf{T}} \cdot \vec{\bf sol} = \mathcal{H}(X;\vec{\delta}^{(i)})$, and we solve it by the least squares method.

\subsubsection{Step 3. Distinguish Critical Points and Extract Position Information.}
If the critical point in Step 2 is an RPCP, the solution derived will be the row vector $A^{(k)}_i$. Its non-zero entries are strictly confined to the LRF-C of the $i$-th output neuron (with a length of $l_{c}$).
If the critical point is a PSP, the solution is the difference vector $A^{(k)}_i - A^{(k)}_j$, when two neurons share the same maximum value within an LRF-P. Here, the non-zero entries correspond to the union of the two respective LRF-Cs.

We extract the vector $\vec{sol}$ with all non-zero (excluding {\sf NaN}) values from the calculated row vector $\vec{\bf sol}$. Let $l_{\vec{sol}}$ represents the length of $\vec{sol}$. Since {\sf NaN}s appear in the solutions, the length of $l_{\vec{sol}}$ does not strictly correspond to the above scenarios. We use the relations between  $l_{\vec{sol}}$ and $l_{c}$ to distinguish if the critical point is an RPCP, a PSP, or an FCP: 
\begin{itemize}
    \item $l_{\vec{sol}} \le l_{\text{c}}$: It may correspond to an RPCP or a PSP: we first attempt to match the convolution kernel pattern vector $\vec{V}_i$ defined in Eq.~\eqref{eqn:rpcp_pattern_V} following the method given in Sect. \ref{sect:rpcp_nan}. If no valid result is returned, we treat it as a PSP to extract positional information.  
    \item $l_{\text{c}} < l_{\vec{sol}} \le 2l_{\text{c}}$: Since in $A^{(k)}_i - A^{(k)}_j$, non-zero values of the $i$-th and $j$-th row vectors are offset from each other, the number of non-zero values will not exceed twice $l_{\text{c}}$. The critical point can be assumed as a PSP and we retrieve the positional information.
    \item $l_{\vec{sol}} > 2l_{\text{c}}$: The critical point is an FCP.
\end{itemize}
If a PSP is misclassified as an RPCP, it will be automatically filtered out during the matching phase in Step 5. This is because the RPCP method attempts to directly recover the convolution kernel weights, yielding an invalid solution that fails the matching criteria.
\subsubsection{Step 4. Calculate Precise Solution.}
We employ the PSPs to compute a high-precision solution. After gathering adequate PSPs and their corresponding position information (more than the number of kernel parameter), we establish the linear system of equations in Eq.~\eqref{eqn:internal-differ} %$\mathbf{c}^{(k)} \cdot (X_{\mathcal{K}_i} - X_{\mathcal{K}_j}) = 0$, 
and solve it using the singular value decomposition (SVD) method to extract the convolution kernel weights. 

\subsubsection{Step 5. Recover the Biases and Signs.}
We employ RPCPs to recover the biases and signs. Suppose that a solution derived in Step 2 is recognized as RPCP's in Step 3. Then, it contains a partial signature of the row vector $A^{(k)}_i$ due to the existence of {\sf NaN}  discussed in  Sect. \ref{sect:rpcp_nan}. 
With the full signature recovered in Step 4, we match it with the partial signature (ignoring the positions of {\sf NaN} values) derived from the RPCP 
%according to Eq.~\eqref{eqn:A_row} 
to recognize the row index $i$ of $A^{(k)}_i$. 
Then, we recover the bias according to Eq.~\eqref{eqn:rpcp_bias} and the sign by Sect. \ref{sect:rpcp_sign}. 

%, and thereby recognize the target RPCPs from noisy points. 

When RPCPs are not found in Step 1, we propose a {\em targeted heuristic search strategy for RPCP} with the signature recovered by the PSP method in Step 4.  

\paragraph{Targeted heuristic search strategy for RPCPs.} 
Our objective is to identify an input $X^{(1)}$ that makes a specific neuron $i$ at layer $k$ reach its critical state {\em i.e.}, $\hat{A}_i^{(k)}X^{(k)}+\hat{b}_i^{(k)}=0$.  
Given that the input preprocessing usually normalizes the space of $X^{(1)}$ (e.g., via standardization \cite{he2016deepresidual,Simonyan2015very} or range scaling \cite{DBLP:journals/corr/RadfordMC15}), 
the bias term $\hat{b}^{(k)}$ is not required to compensate for large input offsets \cite{DBLP:conf/icml/IoffeS15},  {\em i.e.}, 
the absolute value of the bias $\hat{b}^{(k)}$ is typically small. The strategy is as follows: 
%This approach leverages the recovered parameters from the preceding $i-1$ layers and the convolutional kernel weights of the $i$-th layer to synthesize inputs.
\begin{enumerate}
    \item Recover the sign by the PSP method in Sect. \ref{subsect:Sign Recovery from PSP}.  Reconstruct the kernel matrix $\hat{A}^{(k)}$ with the correct signs.
    \item With the assumption that the absolute value of the bias $\hat{b}^{(k)}$ is quite small, we employ a gradient-based optimizer to search the input space. The optimization process is governed by three primary constraints: 
        \begin{itemize}
            \item There exists a neuron $i$ such that the term $\hat{A}_i^{(k)}X^{(k)}$ approaches 0.
            \item Within the same LRF-P $\Omega_t$, the values $\hat{A}^{(k)}_jX^{(k)}, \ j\in \Omega_t, \ j \neq i$ must be substantially smaller than 0.
            \item As many neurons as possible in other LRF-Ps have positive values.
        \end{itemize}
    \item Upon reaching a candidate point $X^{(1)}$, execute a randomized binary search within the local neighborhood of $X^{(1)}$ and query the CNN. 
\end{enumerate}
This increases the probability of finding the RPCPs compared to randomly testing the input as in Step 1.

%In the muti-channel setting, we need to match RPCPs to the kernel weights $A^{(k),i}$ ($0\leq i \leq \mathsf{c}_{out}^{(k)}-1$) corresponding to different output channels.

\section{End-to-End Practical Experiments}
\label{Experiments}
% \section{End-to-End Practical Experiments}
Taking the full CNN $\mathcal{F}_{\theta}$ as a black box, with the inputs and the raw outputs, 
we implemented the algebraic attacks on a variety of CNNs. Most experiments were conducted on a server equipped with an NVIDIA RTX 5080 GPU (16 GB memory), although the attack can also be executed on CPUs. 
%The model queries comes from two phases: searching for critical points and quering $\mathcal{F}_{\theta}(x+\vec{\delta})$, $\mathcal{F}_{\theta}(x-\vec{\delta})$ and $\mathcal{F}_{\theta}(x)$ for each identified critical point. 

\subsubsection{LeNet-5 with modern architectures.}
Our target models include modern implementations of LeNet-5 featuring multiple input and output channels. Unlike the original architecture, modern variants typically incorporate ReLU activation and max pooling. We introduce this LeNet-5 style architecture and the algebraic attack on {\em Multiple Channels} in detail in \textsf{Supplementary Material} \ref{sect:LeNet5}.

\subsubsection{Extracting the parameters of the Convolutional Block of CNNs.}
\label{subsect:Experiments on CNNs}
Given a $(m+n)$-deep CNN's inputs and outputs, we first perform the experiment to recover the parameters of the $m$-round {\em Convolutional Block}, which directly proves the efficiency and correctness of our attack. We conduct the experiments on various CNNs with different architectures (including the modern implementation of LeNet-5), with details given in \textsf{Supplementary Material} \ref{supp:detailed model structure}, whose parameters were trained on random data, MNIST or CIFAR 10, respectively. We summarize the results of the experiments in the top 5 experiments in Table \ref{tab:Attack on CNNs} in Sect. \ref{Introduction}.

\subsubsection{End-to-end Attack on Full and Reduced LeNet-5.}
\label{subsect:End-to-end Attack on LeNet-5}
Combining Carlini {\em et al.}'s differential attack \cite{DBLP:conf/crypto/CarliniJM20} and Canales-Mart\'inez {\em et al.}'s Neuron Wiggle sign-recovery method \cite{DBLP:conf/eurocrypt/CanalesMartinezCHRSS24}, we perform a full parameter extraction on both the {\em Convolutional Block} and {\em Fully-connected Block}, as well as the final output layer. We conduct the experiment on a reduced version of the modern $(2+2)$-deep LeNet-5. It is a $(2+1)$-deep CNN with the same $2$-round {\em Convolutional Block} as the full LeNet-5, and a 1-round {\em Fully-connected Block} with the number of neurons reduced from 84 to 20, and the output layer. 

We also attack the full modern LeNet-5, extending the black-box extraction to the Convolutional Block and FCNN Round 1. 
%If the input distribution of random binary search yields samples that only fall into the activated (inactivated) region of a neuron, the neuron is labeled as a persistent (dead) neuron, according to \cite{ito2025hard}. 
Due to the existence of the persistent and dead neurons \cite{ito2025hard}, we recover 93.3\% of the weights in FCNN Round 1. However, a functionally equivalent attack on the FCNN Block could be implemented with the CrossLayer method in \cite{ito2025hard}. 

The details of the reduced and full LeNet-5 are given in \textsf{Supplementary Material} \ref{supp:detailed model structure}. The experiment results and the comparison with Carlini {\em et al.}’s experiment \cite{DBLP:conf/crypto/CarliniJM20} and Foerster {\em et al.}’s
experiment \cite{foerster2024beyond} are given in the last four rows of Table \ref{tab:Attack on CNNs} in Sect. \ref{Introduction}. A more detailed results table, including runtime, is provided in Table \ref{tab:Attack on CNNs full} in \textsf{Supplementary Material} \ref{supp:experiment results}.

\subsubsection{Noise-sensitivity Analysis.}
The PSP method used in the Convolutional Block and the least squares method (introduced by Carlini {\em et al.} \cite{DBLP:conf/crypto/CarliniJM20}) used in the FCNN Block ensure high precision extraction and noise robustness. Take the (2+1)-reduced version of LeNet-5 as an example, as shown in Table \ref{tab:layer_error_comparison} of \textsf{Supplementary Material} \ref{sec:appendix_cnn_noise}. Under the end-to-end black-box setting, the upper bound on the maximum output error increases only marginally across layers: $2^{-27.86}$ (1st layer), $2^{-24.91}$ (2nd layer), $2^{-23.13}$ (3rd layer), and $2^{-22.21}$ (last layer). This accumulated precision loss is only $(27.86-22.21)=5.65$ bits  across four layers, confirming our method's numerical stability. We report the per-layer relative errors and analyze the error propagation in \textsf{Supplementary Material} \ref{sec:appendix_cnn_noise}.

\subsubsection{Persistent and Dead Neurons.} 
According to Ito {\em et al.} \cite{ito2025hard}, if the input distribution during the random binary search exclusively falls into the activated (or inactivated) region of a neuron, it is classified as a persistent (or dead) neuron. 
%This phenomenon severely hinders the discovery of critical points, as the required positive-negative shift rarely occurs.
In CNNs, due to the weight-sharing mechanism, neurons in layer $k$ %to the same output channel 
are structurally symmetric, {\em i.e.}, if a specific neuron in layer $k$ is persistent (or dead), the other neurons will likely exhibit the same behavior, since all the inputs of the neurons of layer $k$ are generated by the same kernel matrix $C^{(k)}$ and bias. Therefore, it is  difficult to find RPCPs in this case. 
For dead neurons, simply discard them since their outputs remain 0. While for persistent neurons, our PSP method does not require the neuron's output to cross zero and can be used to recover the kernel weights (including signs), as introduced in Sect. \ref{sect:PSP Method}. However, RPCPs remain indispensable for bias recovery. 
In Sect. \ref{sect:unified_model}, we introduced a targeted heuristic search strategy to improve the search for RPCPs using the kernel weights recovered via PSPs. If this strategy also fails, we follow  the cross-layer extraction by Ito {\em et al.} \cite{ito2025hard} to recover the bias from
deeper layers.

\section{Discussions and Conclusions}
\label{Conclusions}

With the development of CNNs,  many other structural hyperparameters have emerged, such as zero-padding \cite{krizhevsky2012imagenet,Simonyan2015very,szegedy2015going,he2016deepresidual}, various convolutional strides $s_c>1$ \cite{krizhevsky2012imagenet,szegedy2015going,he2016deepresidual}, various pooling strides $s_{\rho}$, and the dropout technique \cite{krizhevsky2012imagenet,Simonyan2015very,szegedy2015going}. We discuss their effects on our attacks %from those hyperparameters 
in \textsf{Supp.} \ref{supp:other_parameters}. 
Furthermore, we explore a more restricted scenario where the kernel size is unknown to the attacker, and give a potential attack in detail in \textsf{Supp.} \ref{sec:unknown kernel size}.

In this paper, we tackle the challenging problem of cryptanalytic parameter extraction from Convolutional Neural Networks (CNNs) with max pooling for the first time. Unlike existing works focusing only on ReLU-based fully connected neural networks (FCNNs), we extend cryptanalytic extraction attacks to CNNs by overcoming max pooling-induced non-linearity challenges. We propose an algebraic representation of CNNs, identify two novel types of critical points (RPCPs and PSPs), and design corresponding extraction techniques. A unified model integrating RPCPs and PSPs is developed to leverage their strengths, with a heuristic strategy to address RPCP rarity. End-to-end black-box experiments on multiple CNNs validate our method’s effectiveness in recovering full CNN parameters. 
%Our work fills the gap in max pooling-based CNN parameter extraction. 
Future work will extend the method to hard-label scenarios and complex CNN architectures.

\subsubsection{Acknowledgement.} 
We thank the anonymous reviewers from CRYPTO 2026
for their valuable comments. This work is supported by the National Key R\&D
Program of China (2024YFA1013000), the National Natural Science Foundation of China (U25B2014, 62272257, 62302250), the Fundamental and Interdis
ciplinary Disciplines Breakthrough Plan of the Ministry of Education of China
(JYB2025XDXM114), and the Zhongguancun Laboratory.

\bibliographystyle{plain}
\bibliography{bib}

@article{wei2026rnn,
  title={Cryptanalytic Extraction of Recurrent Neural Network Models},
  author={Wei, Longxiang and Lei, Hao and Qi, Xiaokang and Sun, Xiaohan and Gao, Lei and Hu, Kai and Wang, Wei and Wang, Meiqin},
  journal={Cryptology ePrint Archive},
  year={2026}
}

@article{asselineau2026nonlinearactive,
  title={Cryptanalytic Extraction of Deep Neural Networks with Non-Linear Activations},
  author={Asselineau, Roderick and Derbez, Patrick and Fouque, Pierre-Alain and Minaud, Brice},
  journal={Cryptology ePrint Archive},
  year={2026}
}

@article{qi2026various,
  title={Cryptanalytic Extraction of Neural Networks with Various Activation Functions},
  author={Qi, Xiaokang and Lei, Hao and Wei, Longxiang and Sun, Xiaohan and Wang, Meiqin},
  journal={Cryptology ePrint Archive},
  year={2026}
}

@article{liu2026cnn,
  title={Model Extraction of Convolutional Neural Networks with Max-Pooling},
  author={Liu, Haolin and Siproudhis, Adrien and Boura, Christina and Peyrin, Thomas},
  journal={Cryptology ePrint Archive},
  year={2026}
}

@misc{cnn_average_pooling,
      author = {Xiaohan Sun and Hao Lei and Longxiang Wei and Xiaokang Qi and Kai Hu and Meiqin Wang and Wei Wang},
      title = {Cryptanalytic Extraction of Convolutional Neural Networks},
      howpublished = {Cryptology {ePrint} Archive, Paper 2026/139},
      year = {2026},
      url = {https://eprint.iacr.org/2026/139}
}

@inproceedings{krizhevsky2012imagenet,
  title={Imagenet classification with deep convolutional neural networks},
  author={Krizhevsky, Alex and Sutskever, Ilya and Hinton, Geoffrey E},
  booktitle={NIPS 2012},
  pages = {1106--1114},
  year={2012}
}

@inproceedings{boureau2010theoretical,
  title={A theoretical analysis of feature pooling in visual recognition},
  author={Boureau, Y-Lan and Ponce, Jean and LeCun, Yann},
  booktitle={ICML-10},
  pages={111--118},
  year={2010}
}

@article{matoba2023benefits,
  title={Benefits of Max Pooling in Neural Networks: Theoretical and Experimental Evidence},
  author={Matoba, Kyle and Dimitriadis, Nikolaos and Fleuret, Fran{\c{c}}ois},
  journal={Transactions on Machine Learning Research},
  year={2023}
}

@article{fukushima1983neocognitron,
  title={Neocognitron: A neural network model for a mechanism of visual pattern recognition},
  author={Fukushima, Kunihiko and Miyake, Sei and Ito, Takayuki},
  journal={IEEE transactions on systems, man, and cybernetics},
  number={5},
  pages={826--834},
  year={1983},
  publisher={IEEE}
}

@article{ito2025hard,
  title={Is the Hard-Label Cryptanalytic Model Extraction Really Polynomial?},
  author={Ito, Akira and Miura, Takayuki and Todo, Yosuke},
  journal={arXiv preprint arXiv:2510.06692},
  year={2025}
}

@inproceedings{DBLP:conf/latincrypt/CanalesMartinezS25,
  author       = {Isaac A. Canales{-}Mart{\'{\i}}nez and
                  David Santos},
  title        = {Extracting Some Layers of Deep Neural Networks in the Hard-Label Setting},
  booktitle    = {{LATINCRYPT} 2025},
  volume       = {16129},
  pages        = {399--421},
  publisher    = {Springer},
  year         = {2025},
  url          = {https://doi.org/10.1007/978-3-032-06754-8\_15},
  doi          = {10.1007/978-3-032-06754-8\_15},
  bibsource    = {dblp computer science bibliography, https://dblp.org}
}

@inproceedings{DBLP:conf/eurocrypt/CarliniCHRS25,
  author       = {Nicholas Carlini and
                  Jorge Ch{\'{a}}vez{-}Saab and
                  Anna Hambitzer and
                  Francisco Rodr{\'{\i}}guez{-}Henr{\'{\i}}quez and
                  Adi Shamir},
  title        = {Polynomial Time Cryptanalytic Extraction of Deep Neural Networks in the Hard-Label Setting},
  booktitle    = {{EUROCRYPT} 2025, Proceedings, Part {I}},
  volume       = {15601},
  pages        = {364--396},
  publisher    = {Springer},
  year         = {2025},
  url          = {https://doi.org/10.1007/978-3-031-91107-1\_13},
  doi          = {10.1007/978-3-031-91107-1\_13},
  bibsource    = {dblp computer science bibliography, https://dblp.org}
}

@inproceedings{DBLP:conf/asiacrypt/ChenDGSWW24,
  author       = {Yi Chen and
                  Xiaoyang Dong and
                  Jian Guo and
                  Yantian Shen and
                  Anyu Wang and
                  Xiaoyun Wang},
  title        = {Hard-Label Cryptanalytic Extraction of Neural Network Models},
  booktitle    = {{ASIACRYPT} 2024,
                   Proceedings, Part {VIII}},
  volume       = {15491},
  pages        = {207--236},
  url          = {https://doi.org/10.1007/978-981-96-0944-4\_7},
  doi          = {10.1007/978-981-96-0944-4\_7},
  bibsource    = {dblp computer science bibliography, https://dblp.org}
}

@inproceedings{he2015delving,
  title={Delving deep into rectifiers: Surpassing human-level performance on imagenet classification},
  author={He, Kaiming and Zhang, Xiangyu and Ren, Shaoqing and Sun, Jian},
  booktitle={{ICCV} 2015},
  pages={1026--1034}
}

@inproceedings{chen2025delving,
  title={Delving into cryptanalytic extraction of PReLU neural networks},
  author={Chen, Yi and Dong, Xiaoyang and Ma, Ruijie and Shen, Yantian and Wang, Anyu and Yu, Hongbo and Wang, Xiaoyun},
  booktitle={{ASIACRYPT} 2025},
  pages={576--607},
  year={2025},
  organization={Springer}
}

@article{DBLP:journals/iacr/LiuSELBP26,
  author       = {Haolin Liu and
                  Adrien Siproudhis and
                  Samuel Experton and
                  Peter Lorenz and
                  Christina Boura and
                  Thomas Peyrin},
  title        = {Navigating the Deep: End-to-End Extraction on Deep Neural Networks},
  journal      = {To appear in EUROCRYPT 2026},
  url          = {https://eprint.iacr.org/2026/296},
  bibsource    = {dblp computer science bibliography, https://dblp.org}
}

@inproceedings{foerster2024beyond,
  title={Beyond slow signs in high-fidelity model extraction},
  author={Foerster, Hanna and Mullins, Robert and Shumailov, Ilia and Hayes, Jamie},
  booktitle={NeurIPS 2024}
}

@inproceedings{tramer2016stealing,
  title={Stealing machine learning models via prediction $\{$APIs$\}$},
  author={Tram{\`e}r, Florian and Zhang, Fan and Juels, Ari and Reiter, Michael K and Ristenpart, Thomas},
  booktitle={USENIX Security 2016},
  pages={601--618},
  year={2016}
}

@inproceedings{rolnick2020reverse,
  title={Reverse-engineering deep relu networks},
  author={Rolnick, David and Kording, Konrad},
  booktitle={International conference on machine learning},
  pages={8178--8187},
  year={2020},
  organization={PMLR}
}

@article{oliynyk2023know,
  title={I know what you trained last summer: A survey on stealing machine learning models and defences},
  author={Oliynyk, Daryna and Mayer, Rudolf and Rauber, Andreas},
  journal={ACM Computing Surveys},
  volume={55},
  number={14s},
  pages={1--41},
  year={2023},
  publisher={ACM New York, NY}
}

@inproceedings{jagielski2020high,
  title={High accuracy and high fidelity extraction of neural networks},
  author={Jagielski, Matthew and Carlini, Nicholas and Berthelot, David and Kurakin, Alex and Papernot, Nicolas},
  booktitle={USENIX Security 2020},
  pages={1345--1362},
  year={2020}
}

@inproceedings{batina2019csi,
  title={$\{$CSI$\}$$\{$NN$\}$: Reverse engineering of neural network architectures through electromagnetic side channel},
  author={Batina, Lejla and Bhasin, Shivam and Jap, Dirmanto and Picek, Stjepan},
  booktitle={USENIX Security 2019},
  pages={515--532},
  year={2019}
}

@inproceedings{lowd2005adversarial,
  title={Adversarial learning},
  author={Lowd, Daniel and Meek, Christopher},
  booktitle={KDD 2005},
  pages={641--647},
  publisher = {{ACM}},
  year={2005}
}

@article{blum1988training,
  title={Training a 3-node neural network is {NP}-complete},
  author={Blum, Avrim and Rivest, Ronald},
  journal={Advances in neural information processing systems},
  volume={1},
  year={1988}
}

@article{fefferman1994reconstructing,
  title={Reconstructing a neural net from its output},
  author={Fefferman, Charles and others},
  journal={Revista Matem{\'a}tica Iberoamericana},
  volume={10},
  number={3},
  pages={507--556},
  year={1994},
  publisher={Madrid: Revista Matematica Iberoamericana, 1985-}
}

@article{lecun2002gradient,
  title={Gradient-based learning applied to document recognition},
  author={LeCun, Yann and Bottou, L{\'e}on and Bengio, Yoshua and Haffner, Patrick},
  journal={Proceedings of the IEEE},
  volume={86},
  number={11},
  pages={2278--2324},
  year={2002},
  publisher={Ieee}
}

@inproceedings{DBLP:conf/fse/DinurDS13,
  author       = {Itai Dinur and
                  Orr Dunkelman and
                  Adi Shamir},
  editor       = {Shiho Moriai},
  title        = {Collision Attacks on Up to 5 Rounds of {SHA-3} Using Generalized Internal
                  Differentials},
  booktitle    = {{FSE} 2013},
  volume       = {8424},
  pages        = {219--240},
  publisher    = {Springer},
  year         = {2013},
  url          = {https://doi.org/10.1007/978-3-662-43933-3\_12},
  doi          = {10.1007/978-3-662-43933-3\_12},
  bibsource    = {dblp computer science bibliography, https://dblp.org}
}

@inproceedings{DBLP:conf/crypto/Peyrin10,
  author       = {Thomas Peyrin},
  editor       = {Tal Rabin},
  title        = {Improved Differential Attacks for {ECHO} and Gr{\o}stl},
  booktitle    = {{CRYPTO} 2010, Proceedings},
  volume       = {6223},
  pages        = {370--392},
  publisher    = {Springer},
  year         = {2010},
  url          = {https://doi.org/10.1007/978-3-642-14623-7\_20},
  doi          = {10.1007/978-3-642-14623-7\_20},
  bibsource    = {dblp computer science bibliography, https://dblp.org}
}

@inproceedings{DBLP:conf/eurocrypt/CanalesMartinezCHRSS24,
  author       = {Isaac Andr{\'{e}}s Canales-Martinez and
                  Jorge Ch{\'{a}}vez{-}Saab and
                  Anna Hambitzer and
                  Francisco Rodr{\'{\i}}guez{-}Henr{\'{\i}}quez and
                  Nitin Satpute and
                  Adi Shamir},
   title        = {Polynomial Time Cryptanalytic Extraction of Neural Network Models},
  booktitle    = {{EUROCRYPT} 2024, Proceedings, Part {III}},
  volume       = {14653},
  pages        = {3--33},
  publisher    = {Springer},
  year         = {2024},
  url          = {https://doi.org/10.1007/978-3-031-58734-4\_1},
  doi          = {10.1007/978-3-031-58734-4\_1},
  bibsource    = {dblp computer science bibliography, https://dblp.org}
}

@inproceedings{DBLP:conf/crypto/CarliniJM20,
  author       = {Nicholas Carlini and
                  Matthew Jagielski and
                  Ilya Mironov},
  title        = {Cryptanalytic Extraction of Neural Network Models},
  booktitle    = {{CRYPTO} 2020, Proceedings, Part {III}},
  volume       = {12172},
  pages        = {189--218},
  publisher    = {Springer},
  year         = {2020},
  url          = {https://doi.org/10.1007/978-3-030-56877-1\_7},
  doi          = {10.1007/978-3-030-56877-1\_7},
  bibsource    = {dblp computer science bibliography, https://dblp.org}
}

@inproceedings{szegedy2015going,
  author    = {Szegedy, Christian and Liu, Wei and Jia, Yangqing and
               Sermanet, Pierre and Reed, Scott and Anguelov, Dragomir and
               Erhan, Dumitru and Vanhoucke, Vincent and Rabinovich, Andrew},
  title     = {Going Deeper with Convolutions},
  booktitle = {{CVPR} 2015},
  pages     = {1--9},
  doi       = {10.1109/CVPR.2015.7298594}
}

@inproceedings{he2016deepresidual,
  author    = {He, Kaiming and Zhang, Xiangyu and Ren, Shaoqing and Sun, Jian},
  title     = {Deep Residual Learning for Image Recognition},
  booktitle = {{CVPR} 2016},
  year      = {2016},
  pages     = {770--778},
  doi       = {10.1109/CVPR.2016.90}
}

@inproceedings{huang2017densenet,
  author    = {Huang, Gao and Liu, Zhuang and van der Maaten, Laurens and Weinberger, Kilian Q.},
  title     = {Densely Connected Convolutional Networks},
  booktitle = {CVPR 2017},
  year      = {2017},
  pages     = {4700--4708},
  doi       = {10.1109/CVPR.2017.243}
}

@InProceedings{Simonyan2015very,
  author       = "Karen Simonyan and Andrew Zisserman",
  title        = "Very Deep Convolutional Networks for Large-Scale Image Recognition",
  booktitle    = "{ICLR} 2015, San Diego, CA, USA"
}

@inproceedings{DBLP:conf/iclr/0022KDSG17,
  author       = {Hao Li and
                  Asim Kadav and
                  Igor Durdanovic and
                  Hanan Samet and
                  Hans Peter Graf},
  title        = {Pruning Filters for Efficient ConvNets},
  booktitle    = {{ICLR} 2017, Conference Track Proceedings},
  publisher    = {OpenReview.net},
  year         = {2017},
  url          = {https://openreview.net/forum?id=rJqFGTslg},
  bibsource    = {dblp computer science bibliography, https://dblp.org}
}

@inproceedings{DBLP:conf/iri/ParkHC17,
  author       = {Sang{-}Soo Park and
                  Jung{-}Hyun Hong and
                  Ki{-}Seok Chung},
  title        = {Modified Convolution Neural Network for Highly Effective Parallel
                  Processing},
  booktitle    = {{IRI} 2017, San Diego, CA, USA, August 4-6, 2017},
  pages        = {325--331},
  publisher    = {{IEEE} Computer Society},
  year         = {2017},
  url          = {https://doi.org/10.1109/IRI.2017.37},
  doi          = {10.1109/IRI.2017.37},
  bibsource    = {dblp computer science bibliography, https://dblp.org}
}

@inproceedings{DBLP:conf/iclr/FrankleC19,
  author       = {Jonathan Frankle and
                  Michael Carbin},
  title        = {The Lottery Ticket Hypothesis: Finding Sparse, Trainable Neural Networks},
  booktitle    = {{ICLR} 2019},
  publisher    = {OpenReview.net},
  year         = {2019},
  url          = {https://openreview.net/forum?id=rJl-b3RcF7},
  bibsource    = {dblp computer science bibliography, https://dblp.org}
}

@inproceedings{DBLP:journals/jmlr/GlorotB10,
  author       = {Xavier Glorot and
                  Yoshua Bengio},
  title        = {Understanding the difficulty of training deep feedforward neural networks},
  booktitle    = {{AISTATS} 2010, Chia Laguna Resort, Sardinia, Italy, May 13-15, 2010},
  series       = {{JMLR} Proceedings},
  volume       = {9},
  pages        = {249--256},
  url          = {http://proceedings.mlr.press/v9/glorot10a.html},
  bibsource    = {dblp computer science bibliography, https://dblp.org}
}

@inproceedings{DBLP:conf/mm/JiaSDKLGGD14,
  author       = {Yangqing Jia and
                  Evan Shelhamer and
                  Jeff Donahue and
                  Sergey Karayev and
                  Jonathan Long and
                  Ross B. Girshick and
                  Sergio Guadarrama and
                  Trevor Darrell},
   title        = {Caffe: Convolutional Architecture for Fast Feature Embedding},
  booktitle    = {ACM Multimedia 2014},
  pages        = {675--678},
  url          = {https://doi.org/10.1145/2647868.2654889},
  doi          = {10.1145/2647868.2654889},
  bibsource    = {dblp computer science bibliography, https://dblp.org}
}

@InProceedings{Tran2015C3D,
  author    = {Du Tran and Lubomir Bourdev and Rob Fergus and Lorenzo Torresani and Manohar Paluri},
  title     = {Learning Spatiotemporal Features with 3D Convolutional Networks},
  booktitle = {{ICCV} 2015},
  pages     = {4489--4497},
  publisher = {IEEE},
  doi       = {10.1109/ICCV.2015.510},
  url       = {https://doi.org/10.1109/ICCV.2015.510}
}

@article{vanDenOord2016WaveNet,
  author  = {Aaron van den Oord and Sander Dieleman and Heiga Zen and Karen Simonyan and Oriol Vinyals and Alex Graves and Nal Kalchbrenner and Andrew Senior and Koray Kavukcuoglu},
  title   = {WaveNet: A Generative Model for Raw Audio},
  journal = {arXiv preprint arXiv:1609.03499},
  year    = {2016},
  url     = {https://arxiv.org/abs/1609.03499}
}

@InProceedings{Zhang2015CharCNN,
  author    = {Xiang Zhang and
                  Junbo Jake Zhao and
                  Yann LeCun},
  title     = {Character-Level Convolutional Networks for Text Classification},
  booktitle = {NIPS 2015},
  volume    = {28},
  pages     = {649--657},
  url       = {https://papers.nips.cc/paper/5782-character-level-convolutional-networks-for-text-classification}
}

@inproceedings{DBLP:journals/corr/RadfordMC15,
  author       = {Alec Radford and
                  Luke Metz and
                  Soumith Chintala},
  title        = {Unsupervised Representation Learning with Deep Convolutional Generative
                  Adversarial Networks},
  booktitle    = {{ICLR} 2016},
  url          = {http://arxiv.org/abs/1511.06434},
  bibsource    = {dblp computer science bibliography, https://dblp.org}
}

@inproceedings{DBLP:conf/icml/IoffeS15,
  author       = {Sergey Ioffe and
                  Christian Szegedy},
  title        = {Batch Normalization: Accelerating Deep Network Training by Reducing Internal Covariate Shift},
  booktitle    = {{ICML} 2015},
  series       = {{JMLR} Workshop and Conference Proceedings},
  volume       = {37},
  pages        = {448--456},
  publisher    = {JMLR.org},
  year         = {2015},
  url          = {http://proceedings.mlr.press/v37/ioffe15.html},
  bibsource    = {dblp computer science bibliography, https://dblp.org}
}

@book{DBLP:books/lib/HastieTF09,
  author       = {Trevor Hastie and
                  Robert Tibshirani and
                  Jerome H. Friedman},
  title        = {The Elements of Statistical Learning: Data Mining, Inference, and
                  Prediction, 2nd Edition},
  series       = {Springer Series in Statistics},
  publisher    = {Springer},
  year         = {2009},
  url          = {https://doi.org/10.1007/978-0-387-84858-7},
  doi          = {10.1007/978-0-387-84858-7},
  isbn         = {9780387848570},
  bibsource    = {dblp computer science bibliography, https://dblp.org}
}

@ARTICLE{DonohoCS06,
  author={Donoho, D.L.},
  journal={IEEE Transactions on Information Theory}, 
  title={Compressed sensing}, 
  year={2006},
  volume={52},
  number={4},
  pages={1289-1306},
  doi={10.1109/TIT.2006.871582}}

@article{lai2018cmsis,
  title={Cmsis-nn: Efficient neural network kernels for arm cortex-m cpus},
  author={Lai, Liangzhen and Suda, Naveen and Chandra, Vikas},
  journal={arXiv preprint arXiv:1801.06601},
  year={2018}
}

@inproceedings{liu2017oblivious,
  title={Oblivious neural network predictions via minionn transformations},
  author={Liu, Jian and Juuti, Mika and Lu, Yao and Asokan, Nadarajah},
  booktitle={{CCS} 2017},
  pages={619--631},
publisher    = {{ACM}}
}

@inproceedings{kim2022lightweight,
  title={Lightweight convolutional neural network for real-time arrhythmia classification on low-power wearable electrocardiograph},
  author={Kim, Sangkyu and Chon, Sangil and Kim, Jin-Kook and Kim, Joomin and Gil, Yeongjoon and Jung, Sunghoon},
  booktitle={{EMBC} 2022},
  pages={1915--1918},
  organization={IEEE}
}

@article{banbury2021mlperf,
  title={Mlperf tiny benchmark},
  author={Banbury, Colby and Reddi, Vijay Janapa and Torelli, Peter and Holleman, Jeremy and Jeffries, Nat and Kiraly, Csaba and Montino, Pietro and Kanter, David and Ahmed, Sebastian and Pau, Danilo and others},
  journal={arXiv preprint arXiv:2106.07597},
  year={2021}
}

@InProceedings{pmlr-v48-gilad-bachrach16,
  title = 	 {CryptoNets: Applying Neural Networks to Encrypted Data with High Throughput and Accuracy},
  author = 	 {Gilad-Bachrach, Ran and Dowlin, Nathan and Laine, Kim and Lauter, Kristin and Naehrig, Michael and Wernsing, John},
  booktitle = 	 {{ICML} 2016},
  pages = 	 {201--210},
  year = 	 {2016},
  editor = 	 {Balcan, Maria Florina and Weinberger, Kilian Q.},
  volume = 	 {48},
  publisher =    {PMLR},
  url = 	 {https://proceedings.mlr.press/v48/gilad-bachrach16.html}
}

\newpage
\appendix
\renewcommand{\theHsection}{Appendix.\arabic{section}}
\section*{\centering \textsf{Supplementary Material}}

\section{A Toy Example of the Algebraic View of CNN}
\label{sec:appendix_toy_example}

To facilitate a clearer understanding of the algebraic notations introduced in Section \ref{sect:def_notations}, this appendix provides a concrete, step-by-step toy example. We explicitly demonstrate the complete computational pipeline --- including Convolution, ReLU activation, and Max Pooling --- for a single-input and single-output channel scenario ({\em i.e.}, $\mathsf{c}_{in}^{(k)}=\mathsf{c}_{out}^{(k)}=1$). 

\begin{sidewaysfigure}
    \centering
\includegraphics[width=\textwidth]{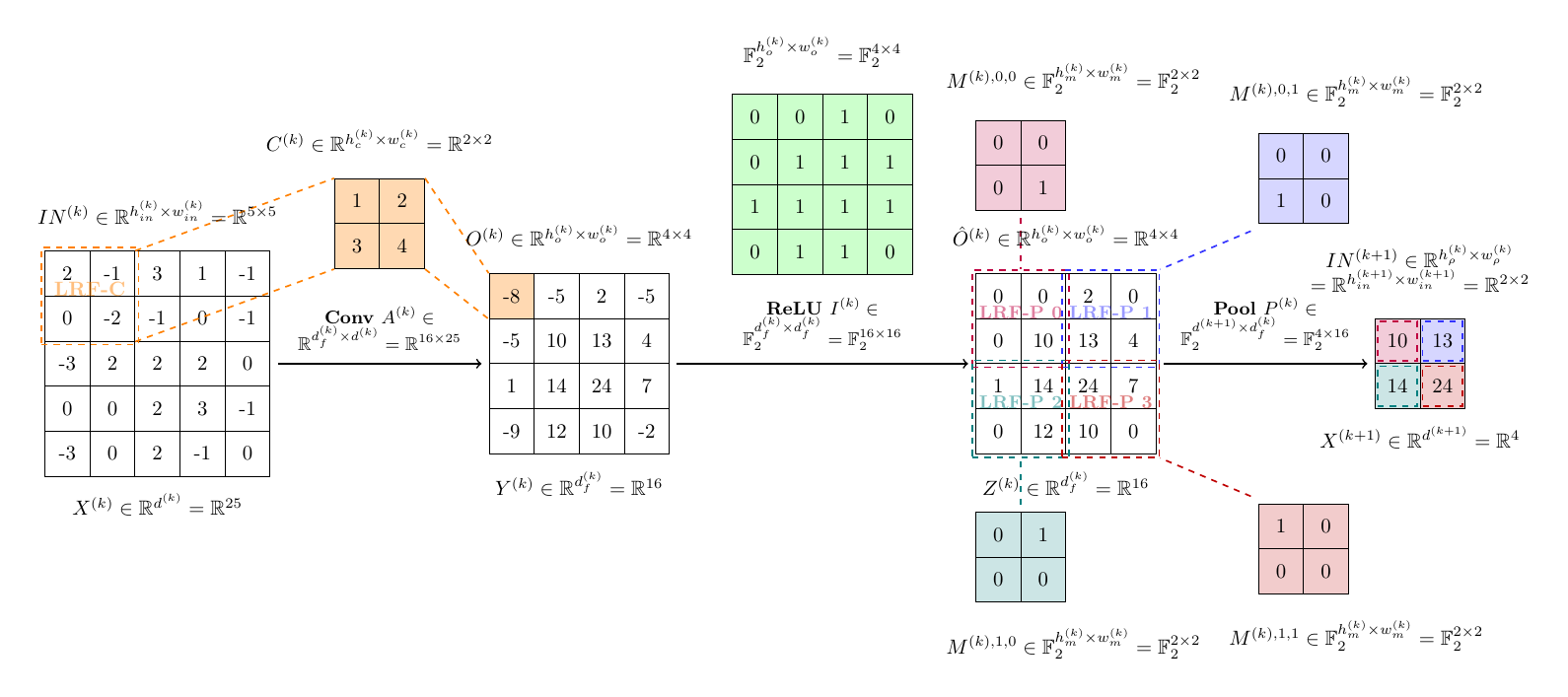} 
    \caption{Overall Visual Illustration of One Convolutional Round.}
    \label{fig:toy_example_diagram}
\end{sidewaysfigure}

Figure \ref{fig:toy_example_diagram} visualizes the spatial mapping mechanism of this process. In the following subsections, we formulate this exact process algebraically by completely expanding the equivalent transformation matrices without any truncation.

We also provide illustrative examples of RPCPs and PSPs in {\sf Supp}. \ref{subsect:Examples of RPCPs and PSPs} to help understand the two novel types of critical points in CNNs introduced in Sect. \ref{sect:critical_points}, along with a warm-up extraction example for the first layer in {\sf Supp}. \ref{subsect:warm-up} to demonstrate the RPCP and PSP methods detailed in Sects. \ref{sect:RPCP_Method} and  \ref{sect:PSP Method}.

\subsection{Setup and Vector Flattening}
Consider an input matrix $IN^{(k)} \in \mathbb{R}^{h_{in}^{(k)} \times w_{in}^{(k)}} = \mathbb{R}^{5 \times 5}$. We apply a convolutional layer $f_c^{(k)}$ with a kernel $C^{(k)} \in \mathbb{R}^{h_c^{(k)} \times w_c^{(k)}} = \mathbb{R}^{2 \times 2}$, a stride of $s_c^{(k)} = 1$, yielding an intermediate output matrix $O^{(k)}$. According to the definitions in Section \ref{sect:def_notations}, the spatial dimensions of $O^{(k)}$ are calculated as follows:
\begin{align}
    h^{(k)}_{o} &= \left\lfloor \frac{h^{(k)}_{in} - h^{(k)}_{c}}{s^{(k)}_c} \right\rfloor + 1 = \left\lfloor \frac{5 - 2}{1} \right\rfloor + 1 = 4, \\
    w^{(k)}_{o} &= \left\lfloor \frac{w^{(k)}_{in} - w^{(k)}_{c}}{s^{(k)}_c} \right\rfloor + 1 = \left\lfloor \frac{5 - 2}{1} \right\rfloor + 1 = 4.
\end{align}
Consequently, the resulting intermediate output matrix is $O^{(k)} \in \mathbb{R}^{4 \times 4}$. Subsequently, a ReLU activation layer $\sigma_c^{(k)}$ is applied to $O^{(k)}$, generating an activated output matrix $\hat{O}^{(k)} \in \mathbb{R}^{4 \times 4}$. 

Following the activation, a max pooling layer $\rho^{(k)}$ with a window size of $h_m^{(k)} \times w_m^{(k)} = 2 \times 2$ and a stride of $s_\rho^{(k)} = 2$ processes $\hat{O}^{(k)}$. The spatial dimensions of the final output matrix $IN^{(k+1)}$ for the subsequent layer are determined by:
\begin{align}
    h^{(k)}_{\rho} &= \left\lfloor \frac{h^{(k)}_{o} - h^{(k)}_{m}}{s^{(k)}_{\rho}} \right\rfloor + 1 = \left\lfloor \frac{4 - 2}{2} \right\rfloor + 1 = 2, \\
    w^{(k)}_{\rho} &= \left\lfloor \frac{w^{(k)}_{o} - w^{(k)}_{m}}{s^{(k)}_{\rho}} \right\rfloor + 1 = \left\lfloor \frac{4 - 2}{2} \right\rfloor + 1 = 2.
\end{align}
This yields the matrix $IN^{(k+1)} \in \mathbb{R}^{2 \times 2}$.

To represent these operations algebraically, we flatten the spatial matrices into vectors using row-major ordering. For the input matrix $IN^{(k)}$, its $i$-th row ($0 \leq i \leq h_{in}^{(k)}-1$) is denoted as $IN^{(k)}_i=(a^{(k)}_{i,0}, a^{(k)}_{i,1}, \dots, a^{(k)}_{i,w_{in}^{(k)}-1})$. Based on the specific numerical values from Figure \ref{fig:toy_example_diagram}, we instantiate the first and the last row vectors as follows:
\begin{equation}
\begin{aligned}
    IN^{(k)}_0 &= (2, -1, 3, 1, -1), \\
    &\ \ \vdots \\
    IN^{(k)}_4 &= (-3, 0, 2, -1, 0).
\end{aligned}
\end{equation}
The flattened input vector $X^{(k)} \in \mathbb{R}^{d^{(k)}} = \mathbb{R}^{h_{in}^{(k)} \cdot w_{in}^{(k)}}= \mathbb{R}^{25}  $ is constructed by concatenating these row vectors as $X^{(k)} = (IN^{(k)}_0, IN^{(k)}_1, \dots, IN^{(k)}_{4})^{\mathsf{T}}$, which explicitly evaluates to:
\begin{equation}
    \small
    X^{(k)} = [2, -1, 3, 1, -1, 0, -2, -1, 0, -1, -3, 2, 2, 2, 0, 0, 0, 2, 3, -1, -3, 0, 2, -1, 0]^{\mathsf{T}}.
\end{equation}

\subsection{The Convolutional Matrix ($A^{(k)}$)}
\begin{figure}
	\centering    \includegraphics[width=0.8\linewidth]{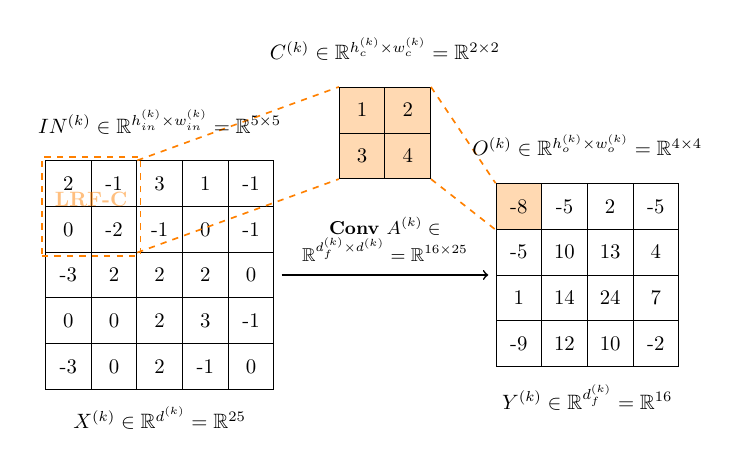}
    \caption{Local view of the convolution operation $f^{(k)}_c$}
	\label{fig:toy_conv}
\end{figure}
Let the $2 \times 2$ convolution kernel matrix be $C^{(k)} = \left[\begin{smallmatrix} 1 & 2 \\ 3 & 4 \end{smallmatrix}\right]$. For simplicity in this example, we assume the bias is $b^{(k)}=0$, yielding an all-zero bias vector $B^{(k)} = \mathbf{0}$. Therefore, the convolution operation translates into a linear matrix multiplication $Y^{(k)} = A^{(k)} X^{(k)}$, where $Y^{(k)} \in \mathbb{R}^{d_{f}^{(k)}} = \mathbb{R}^{h^{(k)}_{o} \cdot w^{(k)}_{o} } = \mathbb{R}^{16}$ is the flattened output vector and $A^{(k)} \in \mathbb{R}^{d_{f}^{(k)} \times d^{(k)} }  = \mathbb{R}^{16 \times 25}$ is the convolutional matrix.

As illustrated by the orange dashed lines in Figure \ref{fig:toy_conv}, the first element of the feature map $O^{(k)}_{0,0}$ (which corresponds to $Y^{(k)}_0$) is computed by the sum of the element-wise multiplications between $C^{(k)}$ and the top-left LRF-C. Formally, this is calculated as:
\begin{equation}
\begin{aligned}
    O^{(k)}_{0,0} &= \sum_{p=0}^{h_{c}^{(k)}-1} \sum_{q=0}^{w_{c}^{(k)}-1} IN^{(k)}_{0+p,0+q} C^{(k)}_{p,q} + b^{(k)} \\
    &= IN^{(k)}_{0,0} C^{(k)}_{0,0} + IN^{(k)}_{0,1} C^{(k)}_{0,1} + IN^{(k)}_{1,0} C^{(k)}_{1,0} + IN^{(k)}_{1,1} C^{(k)}_{1,1} + b^{(k)} \\
    &= (2)(1) + (-1)(2) + (0)(3) + (-2)(4) + 0 = -8.
\end{aligned}
\end{equation}

According to Eq. \eqref{eqn:A_row}, the first row of the convolutional matrix ($i=0$) is systematically constructed by substituting the kernel rows into the padded structure. Given  $w^{(k)}_{in}=5$ and $w^{(k)}_o=4$, the substitution and expansion explicitly demonstrate how the kernel parameters are positioned within $A^{(k)}_0$:
\begin{equation}\label{eq:appendix_A0_example}
\begin{aligned}
    A^{(k)}_0 &= (\underbrace{0, \cdots, 0}_{w^{(k)}_{in}\cdot \lfloor 0/w^{(k)}_{o} \rfloor}, \underbrace{
        \underbrace{0, \cdots, 0}_{0 \bmod w^{(k)}_o}, C^{(k)}_{0}, \underbrace{0, \dots, 0}_{padding}}_{w_{in}^{(k)}}, 
        \underbrace{
        \underbrace{0, \cdots, 0}_{0 \bmod w^{(k)}_o}, C^{(k)}_{1}, \underbrace{0, \dots, 0}_{padding}}_{w_{in}^{(k)}}, 
        \underbrace{0, \cdots, 0}_{padding}) \\
    &= (\underbrace{}_{0}, \underbrace{\underbrace{}_{0}, \mathbf{1, 2}, \underbrace{0, 0, 0}_{3}}_{5}, \underbrace{\underbrace{}_{0}, \mathbf{3, 4}, \underbrace{0, 0, 0}_{3}}_{5}, \underbrace{0, \cdots, 0}_{15}) \\
    &= (1, 2, 0, 0, 0, 3, 4, 0, 0, 0, 0, 0, 0, 0, 0, 0, 0, 0, 0, 0, 0, 0, 0, 0, 0).
\end{aligned}
\end{equation}

By representing the exhaustive sliding window operations over the 2D input matrix, the entire equivalent matrix $A^{(k)}$ is expanded below. We highlight the non-zero elements corresponding to the convolution kernel in each row in bold:

\begin{equation}
\setlength{\arraycolsep}{4pt} 
\resizebox{0.8\textwidth}{!}{$
A^{(k)} = \begin{bmatrix}
\mathbf{1} & \mathbf{2} & 0 & 0 & 0 & \mathbf{3} & \mathbf{4} & 0 & 0 & 0 & 0 & 0 & 0 & 0 & 0 & 0 & 0 & 0 & 0 & 0 & 0 & 0 & 0 & 0 & 0 \\
0 & \mathbf{1} & \mathbf{2} & 0 & 0 & 0 & \mathbf{3} & \mathbf{4} & 0 & 0 & 0 & 0 & 0 & 0 & 0 & 0 & 0 & 0 & 0 & 0 & 0 & 0 & 0 & 0 & 0 \\
0 & 0 & \mathbf{1} & \mathbf{2} & 0 & 0 & 0 & \mathbf{3} & \mathbf{4} & 0 & 0 & 0 & 0 & 0 & 0 & 0 & 0 & 0 & 0 & 0 & 0 & 0 & 0 & 0 & 0 \\
0 & 0 & 0 & \mathbf{1} & \mathbf{2} & 0 & 0 & 0 & \mathbf{3} & \mathbf{4} & 0 & 0 & 0 & 0 & 0 & 0 & 0 & 0 & 0 & 0 & 0 & 0 & 0 & 0 & 0 \\
0 & 0 & 0 & 0 & 0 & \mathbf{1} & \mathbf{2} & 0 & 0 & 0 & \mathbf{3} & \mathbf{4} & 0 & 0 & 0 & 0 & 0 & 0 & 0 & 0 & 0 & 0 & 0 & 0 & 0 \\
0 & 0 & 0 & 0 & 0 & 0 & \mathbf{1} & \mathbf{2} & 0 & 0 & 0 & \mathbf{3} & \mathbf{4} & 0 & 0 & 0 & 0 & 0 & 0 & 0 & 0 & 0 & 0 & 0 & 0 \\
0 & 0 & 0 & 0 & 0 & 0 & 0 & \mathbf{1} & \mathbf{2} & 0 & 0 & 0 & \mathbf{3} & \mathbf{4} & 0 & 0 & 0 & 0 & 0 & 0 & 0 & 0 & 0 & 0 & 0 \\
0 & 0 & 0 & 0 & 0 & 0 & 0 & 0 & \mathbf{1} & \mathbf{2} & 0 & 0 & 0 & \mathbf{3} & \mathbf{4} & 0 & 0 & 0 & 0 & 0 & 0 & 0 & 0 & 0 & 0 \\
0 & 0 & 0 & 0 & 0 & 0 & 0 & 0 & 0 & 0 & \mathbf{1} & \mathbf{2} & 0 & 0 & 0 & \mathbf{3} & \mathbf{4} & 0 & 0 & 0 & 0 & 0 & 0 & 0 & 0 \\
0 & 0 & 0 & 0 & 0 & 0 & 0 & 0 & 0 & 0 & 0 & \mathbf{1} & \mathbf{2} & 0 & 0 & 0 & \mathbf{3} & \mathbf{4} & 0 & 0 & 0 & 0 & 0 & 0 & 0 \\
0 & 0 & 0 & 0 & 0 & 0 & 0 & 0 & 0 & 0 & 0 & 0 & \mathbf{1} & \mathbf{2} & 0 & 0 & 0 & \mathbf{3} & \mathbf{4} & 0 & 0 & 0 & 0 & 0 & 0 \\
0 & 0 & 0 & 0 & 0 & 0 & 0 & 0 & 0 & 0 & 0 & 0 & 0 & \mathbf{1} & \mathbf{2} & 0 & 0 & 0 & \mathbf{3} & \mathbf{4} & 0 & 0 & 0 & 0 & 0 \\
0 & 0 & 0 & 0 & 0 & 0 & 0 & 0 & 0 & 0 & 0 & 0 & 0 & 0 & 0 & \mathbf{1} & \mathbf{2} & 0 & 0 & 0 & \mathbf{3} & \mathbf{4} & 0 & 0 & 0 \\
0 & 0 & 0 & 0 & 0 & 0 & 0 & 0 & 0 & 0 & 0 & 0 & 0 & 0 & 0 & 0 & \mathbf{1} & \mathbf{2} & 0 & 0 & 0 & \mathbf{3} & \mathbf{4} & 0 & 0 \\
0 & 0 & 0 & 0 & 0 & 0 & 0 & 0 & 0 & 0 & 0 & 0 & 0 & 0 & 0 & 0 & 0 & \mathbf{1} & \mathbf{2} & 0 & 0 & 0 & \mathbf{3} & \mathbf{4} & 0 \\
0 & 0 & 0 & 0 & 0 & 0 & 0 & 0 & 0 & 0 & 0 & 0 & 0 & 0 & 0 & 0 & 0 & 0 & \mathbf{1} & \mathbf{2} & 0 & 0 & 0 & \mathbf{3} & \mathbf{4}
\end{bmatrix}
$}
\label{eq:appendix_A}
\end{equation}

Performing the multiplication yields the pre-activation vector $Y^{(k)}$:
\begin{equation}
    \small
    Y^{(k)} = A^{(k)} X^{(k)} = [-8, -5, 2, -5, -5, 10, 13, 4, 1, 14, 24, 7, -9, 12, 10, -2]^{\mathsf{T}}.
\end{equation}

\subsection{The ReLU Activation Matrix ($I^{(k)}$)}

\begin{figure}
	\centering    \includegraphics[width=0.85\linewidth]{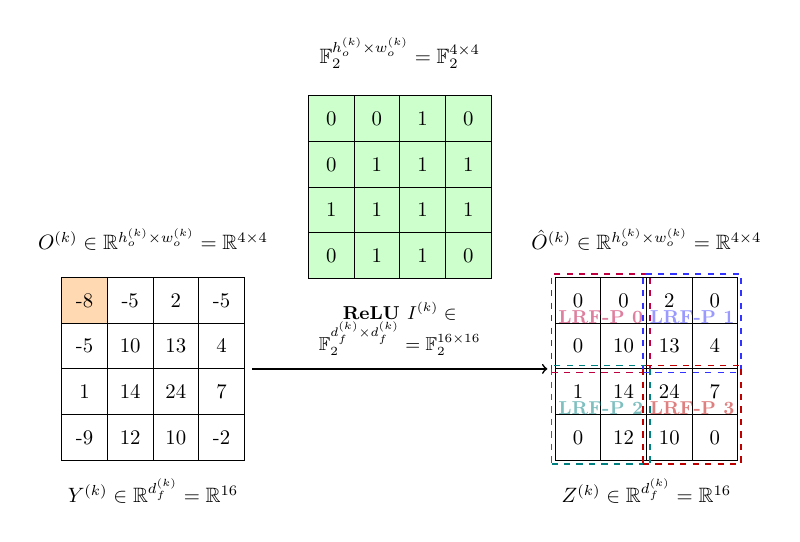}
    \caption{Local view of the ReLU activation $\sigma_c^{(k)}$}
	\label{fig:toy_relu}
\end{figure}

Fig. \ref{fig:toy_relu} shows the operation of ReLU activation.
%The ReLU activation acts as a data-dependent masking operation. 
Algebraically, it is represented as $Z^{(k)} = I^{(k)} Y^{(k)}$, where $I^{(k)} \in \mathbb{F}_2^{d_{f}^{(k)}\times d_{f}^{(k)}}  = \mathbb{F}_2^{16 \times 16}$ is a diagonal matrix. According to Eq. \eqref{eqn:sigma_I}, the $i$-th diagonal entry of $I^{(k)}$ is set to 1 if the corresponding element $Y^{(k)}_i$ is strictly positive, and 0 otherwise. 
Based on the numerical values of $Y^{(k)}$ calculated previously, the fully expanded activation matrix is constructed as follows. We highlight the entire main diagonal in bold:

\begin{equation}
\setlength{\arraycolsep}{4pt}
\resizebox{0.6\textwidth}{!}{$
I^{(k)} = \begin{bmatrix}
\mathbf{0} & 0 & 0 & 0 & 0 & 0 & 0 & 0 & 0 & 0 & 0 & 0 & 0 & 0 & 0 & 0 \\
0 & \mathbf{0} & 0 & 0 & 0 & 0 & 0 & 0 & 0 & 0 & 0 & 0 & 0 & 0 & 0 & 0 \\
0 & 0 & \mathbf{1} & 0 & 0 & 0 & 0 & 0 & 0 & 0 & 0 & 0 & 0 & 0 & 0 & 0 \\
0 & 0 & 0 & \mathbf{0} & 0 & 0 & 0 & 0 & 0 & 0 & 0 & 0 & 0 & 0 & 0 & 0 \\
0 & 0 & 0 & 0 & \mathbf{0} & 0 & 0 & 0 & 0 & 0 & 0 & 0 & 0 & 0 & 0 & 0 \\
0 & 0 & 0 & 0 & 0 & \mathbf{1} & 0 & 0 & 0 & 0 & 0 & 0 & 0 & 0 & 0 & 0 \\
0 & 0 & 0 & 0 & 0 & 0 & \mathbf{1} & 0 & 0 & 0 & 0 & 0 & 0 & 0 & 0 & 0 \\
0 & 0 & 0 & 0 & 0 & 0 & 0 & \mathbf{1} & 0 & 0 & 0 & 0 & 0 & 0 & 0 & 0 \\
0 & 0 & 0 & 0 & 0 & 0 & 0 & 0 & \mathbf{1} & 0 & 0 & 0 & 0 & 0 & 0 & 0 \\
0 & 0 & 0 & 0 & 0 & 0 & 0 & 0 & 0 & \mathbf{1} & 0 & 0 & 0 & 0 & 0 & 0 \\
0 & 0 & 0 & 0 & 0 & 0 & 0 & 0 & 0 & 0 & \mathbf{1} & 0 & 0 & 0 & 0 & 0 \\
0 & 0 & 0 & 0 & 0 & 0 & 0 & 0 & 0 & 0 & 0 & \mathbf{1} & 0 & 0 & 0 & 0 \\
0 & 0 & 0 & 0 & 0 & 0 & 0 & 0 & 0 & 0 & 0 & 0 & \mathbf{0} & 0 & 0 & 0 \\
0 & 0 & 0 & 0 & 0 & 0 & 0 & 0 & 0 & 0 & 0 & 0 & 0 & \mathbf{1} & 0 & 0 \\
0 & 0 & 0 & 0 & 0 & 0 & 0 & 0 & 0 & 0 & 0 & 0 & 0 & 0 & \mathbf{1} & 0 \\
0 & 0 & 0 & 0 & 0 & 0 & 0 & 0 & 0 & 0 & 0 & 0 & 0 & 0 & 0 & \mathbf{0}
\end{bmatrix}
$}
\label{eq:appendix_I}
\end{equation}

Applying this masking matrix to $Y^{(k)}$ zeros out its non-positive components, yielding the activated vector $Z^{(k)}$:
\begin{equation}
    \small
    Z^{(k)} = I^{(k)} Y^{(k)} = [0, 0, 2, 0, 0, 10, 13, 4, 1, 14, 24, 7, 0, 12, 10, 0]^{\mathsf{T}}.
\end{equation}

\subsection{The Max Pooling Matrix ($P^{(k)}$)}
In the pooling layer $\rho^{(k)}$, the $2 \times 2$ max pooling operation with stride 2 partitions the $4 \times 4$ intermediate feature map $\hat{O}^{(k)}$ into 4 disjoint LRF-Ps. For each LRF-P, the pooling function selects the maximum value. 
\begin{figure}
	\centering    \includegraphics[width=0.8\linewidth]{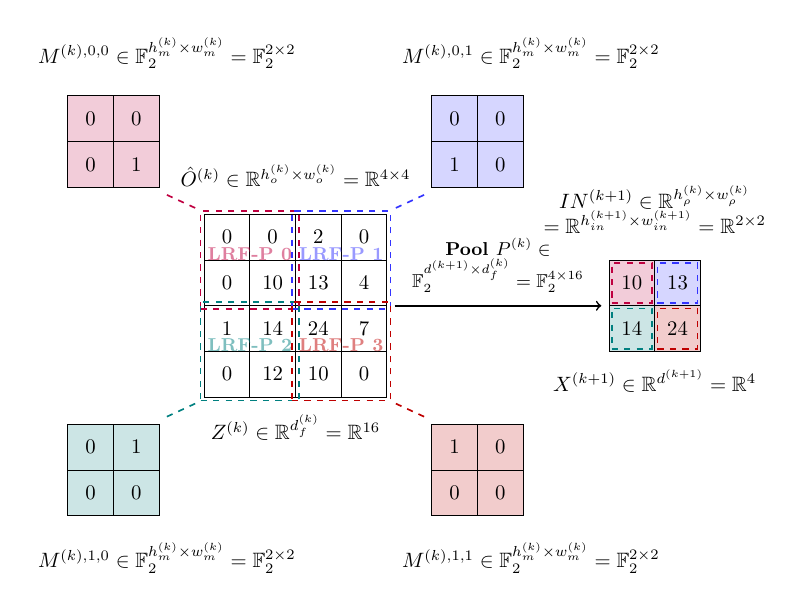}
    \caption{Local view of the Max Pooling $\rho^{(k)}$}
	\label{fig:toy_pooling}
\end{figure}
Fig. \ref{fig:toy_pooling} illustrates the operation of max pooling. According to Eq. \eqref{eqn:max_pool_1}, the first element of the pooled output is the maximum value within the $(i\cdot w^{(k)}_{\rho}+j)=(0\cdot w^{(k)}_{\rho}+0)=0$-th LRF-P denoted as $\Omega_{0,0}$:
\begin{equation}
    IN^{(k+1)}_{0,0} = \max\{\hat{O}^{(k)}_{\Omega_{0,0}}\} = \max(\hat{O}^{(k)}_{0,0}, \hat{O}^{(k)}_{0,1}, \hat{O}^{(k)}_{1,0}, \hat{O}^{(k)}_{1,1}) = \max(0, 0, 0, 10) = 10.
\end{equation}

Since the maximum value ``$10$'' in $\Omega_{0,0}$ is located at local index $(h^*=1, w^*=1)$, the Boolean matrix $M^{(k),0,0}$ assigns $1$ exclusively to this position, making its row vectors $M^{(k),0,0}_0 = (0, 0)$ and $M^{(k),0,0}_1 = (0, 1)$. According to Eq.  \eqref{eqn:P_matrix}, the first row ($r=0$) of the pooling matrix  %, corresponding to spatial indices $i=0, j=0$)
is constructed by mapping these Boolean rows of $M^{(k),0,0}$ into the padded structure. Given $w^{(k)}_o=4$ and $s^{(k)}_\rho=2$, the substitution explicitly demonstrates how the selected bits are positioned within $P^{(k)}_0$:
\begin{equation}
\begin{aligned}
    P^{(k)}_0 &= (\underbrace{0, \cdots, 0}_{w^{(k)}_{o}\cdot 0 \cdot s^{(k)}_{\rho}}, \underbrace{
        \underbrace{0, \cdots, 0}_{0 \cdot s^{(k)}_{\rho}}, M^{(k),0,0}_{0}, \underbrace{0, \dots, 0}_{padding}}_{w_{o}^{(k)}}, 
        \underbrace{
        \underbrace{0, \cdots, 0}_{0 \cdot s^{(k)}_{\rho}}, M^{(k),0,0}_{1}, \underbrace{0, \dots, 0}_{padding}}_{w_{o}^{(k)}}, 
        \underbrace{0, \cdots, 0}_{padding}) \\
    &= (\underbrace{}_{0}, \underbrace{ \underbrace{}_{0}, \mathbf{0, 0}, \underbrace{0, 0}_{2}}_{4}, \underbrace{ \underbrace{}_{0}, \mathbf{0, 1}, \underbrace{0, 0}_{2}}_{4}, \underbrace{0, \cdots, 0}_{8}) \\
    &= (0, 0, 0, 0, 0, 1, 0, 0, 0, 0, 0, 0, 0, 0, 0, 0).
\end{aligned}
\end{equation}

By extending this construction to all four LRF-Ps, the entire equivalent matrix $P^{(k)}$ extracts exactly one element per row. We highlight these selected positions in bold:

\begin{equation}
\resizebox{0.55\textwidth}{!}{$
P^{(k)} = \begin{bmatrix}
0 & 0 & 0 & 0 & 0 & \mathbf{1} & 0 & 0 & 0 & 0 & 0 & 0 & 0 & 0 & 0 & 0 \\
0 & 0 & 0 & 0 & 0 & 0 & \mathbf{1} & 0 & 0 & 0 & 0 & 0 & 0 & 0 & 0 & 0 \\
0 & 0 & 0 & 0 & 0 & 0 & 0 & 0 & 0 & \mathbf{1} & 0 & 0 & 0 & 0 & 0 & 0 \\
0 & 0 & 0 & 0 & 0 & 0 & 0 & 0 & 0 & 0 & \mathbf{1} & 0 & 0 & 0 & 0 & 0
\end{bmatrix}
$}
\label{eq:appendix_P}
\end{equation}

The final output vector for the entire layer is therefore:
\begin{equation}
    X^{(k+1)} = P^{(k)} Z^{(k)} = [10, 13, 14, 24]^{\mathsf{T}}.
\end{equation}

\subsection{Examples of RPCPs and PSPs}
\label{subsect:Examples of RPCPs and PSPs}
\subsubsection{RPCPs.} 
\begin{figure}[htbp]
    \centering
    \includegraphics[width=1\linewidth]{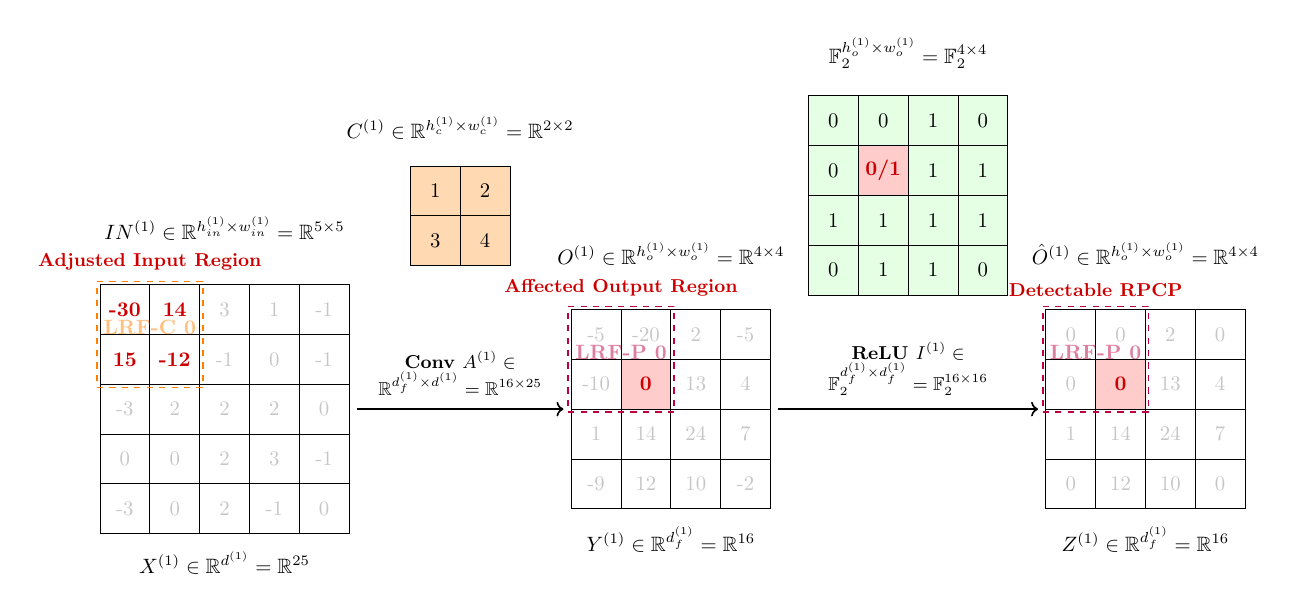}
    \caption{An example of an RPCP. For the targeted input $X^{(1)}$, $Y^{(1)}[5]=0$ and $Y^{(1)}[0],~ Y^{(1)}[1],~Y^{(1)}[4]<0$. Consequently, the corresponding diagonal entry in the activation matrix $I^{(1)}$ can be either $1$ or $0$.}
    \label{fig:toy_rpcp}
\end{figure}

Let $k=1$ in the Figure \ref{fig:toy_example_diagram} and modify inputs of the LRF-C 0 of $IN^{(1)}$ to derive 
Figure \ref{fig:toy_rpcp}, which leads to an example of an RPCP. Given the targeted input $X^{(1)}$,  $O^{(1)}_{1, 1}$ evaluates to exactly $0$ ($Y^{(1)}[5]=0$ in vector form). Meanwhile, the values of the other competing neurons within the LRF-P 0 are strictly negative ($Y^{(1)}_{0}=-5$, $Y^{(1)}_{1}=-20$, and $Y^{(1)}_{4}=-10$). Consequently, the ReLU functions suppress their outputs to $0$.

Consider a small perturbation $\vec{\delta}$ such that the inputs $X+\vec{\delta}$ and $X-\vec{\delta}$ fall into two adjacent linear neighborhoods. Assume that $Y^{(1)}[5]>0$ at $X+\vec{\delta}$ (the active side) and $Y^{(1)}[5]<0$ at $X-\vec{\delta}$ (the inactive side). Then, the max pooling will select neuron $5$ on the active side ($X+\vec{\delta}$). On the inactive side ($X-\vec{\delta}$), the max pooling can still select neuron $5$, since all neurons in the LRF-P 0 output $0$. Consequently, the corresponding diagonal entry (the $5$-th entry) in the activation matrix $I^{(1)}$ can be either $1$ or $0$.

\subsubsection{PSPs.}
\begin{figure}[htbp]
    \centering
    \includegraphics[width=1\linewidth]{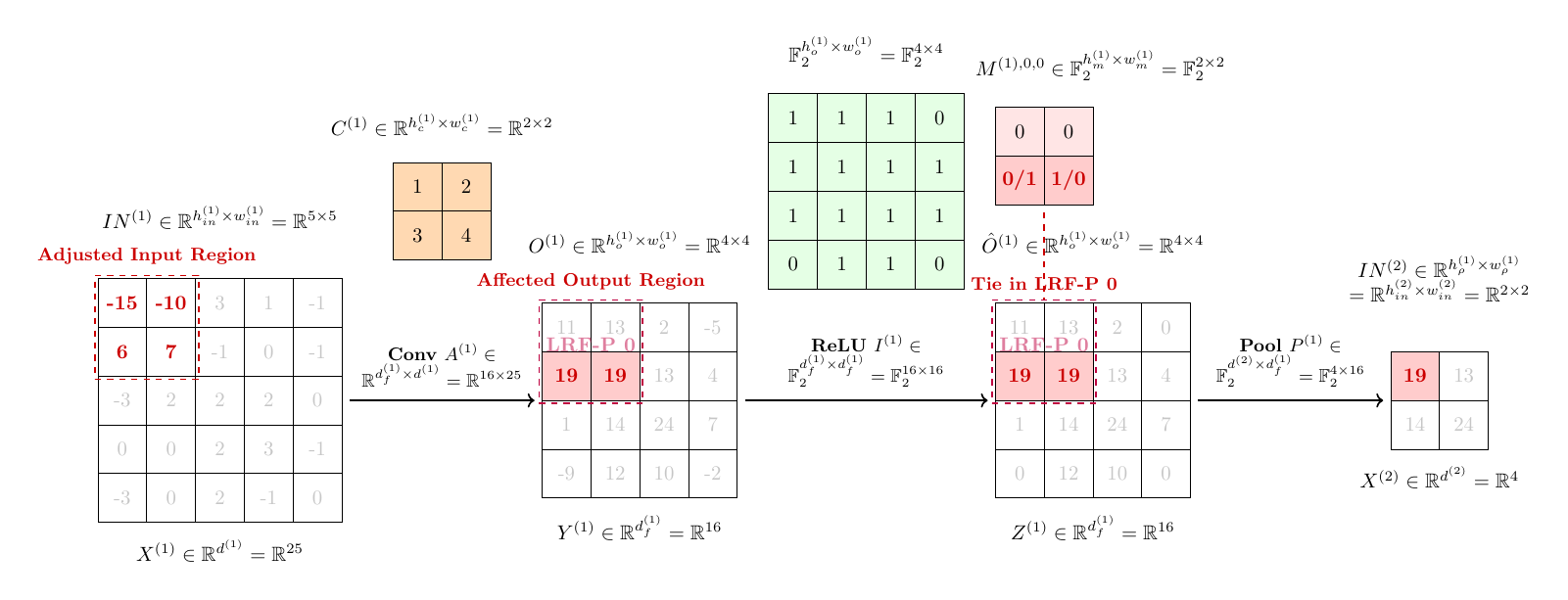}
    \caption{An example of a PSP. The input $X^{(1)}$ induces a tie ($Y^{(1)}[4]=Y^{(1)}[5]=19$) within the LRF-P 0. This makes the pooling selection matrix $M^{(1),0,0}$ to select between either index $i=4$ or $j=5$.}
    \label{fig:toy_psp}
\end{figure}

Figure \ref{fig:toy_psp} demonstrates a PSP instance, where a ``tie'' occurs in the LRF-P 0. Specifically, the pre-activation values $Y^{(1)}[4]$ and $Y^{(1)}[5]$ (corresponding to $O^{(1)}_{1,0}$ and $O^{(1)}_{1,1}$) both evaluate to the exact same maximum value of $19$. 

This equality forces the local pooling selection matrix $M^{(1),0,0}$ to switch between selecting the indices $i=4$ and $j=5$ when subjected to tiny input perturbations, such as $X^{(1)} + \vec{\delta}$ and $X^{(1)} - \vec{\delta}$. At exact PSP $X^{(1)}$, the max pooling can select either neuron $4$ or neuron $5$ since the output is always $19$. Consequently, in $M^{(1),0,0}$, we can have either $M^{(1),1,0}=1$ or $M^{(1),1,1}=1$, with all other entries evaluating to $0$. 

\subsection{Warm-up: Extracting the First Layer}
\label{subsect:warm-up}

%The previous subsections mechanically expand the algebraic matrices in a static forward pass. However, the core of our parameter extraction attack relies on exploiting the dynamic, non-linear shifts within these matrices when inputs cross specific decision boundaries. 
%To explicitly demonstrate how the linear systems are constructed to recover the weights, we set the target layer to $k=1$ as a warm-up scenario. In our attack, the query perturbations $\pm\vec{\delta}$ are always added directly to the network's initial input space. 
%For the first layer, the transition matrix of preceding layers $F^{k-1}$ reduces to an identity matrix ($F^0 = I$). This significantly simplifies the forward propagation of the perturbation, as $F^0\vec{\delta} = \vec{\delta}$, allowing us to directly analyze its impact on the first convolutional matrix $A^{(1)}$. Below, we present two specific input instances that trigger a ReLU-Pooling Critical Point (RPCP) and a Pooling Switching Point (PSP), detailing how their explicitly expanded algebraic structures are exploited.

\subsubsection{The RPCP method.}
As illustrated in Figure \ref{fig:toy_rpcp}, the neuron 5 is in the critical state. By querying the model with perturbed inputs $X^{(1)} + \vec{\delta}$ and $X^{(1)} - \vec{\delta}$, the neuron flips between positive value and negative value, with the corresponding activation matrix having a $1$ at the $5$-th diagonal entry ($I_{+}^{(1)}$) on the active side and $0$ ($I_{-}^{(1)}$) on the inactive side. As formulated in Eq.~\eqref{eqn:2nd_rpcp}, by calculating the second-order differential function $\mathcal{H}(X;\vec{\delta})$ of $\mathcal{H}(X;\vec{\delta}) = \mathcal{F}_{\theta}(X+\vec{\delta}) + \mathcal{F}_{\theta}(X-\vec{\delta}) - 2\mathcal{F}_{\theta}(X)$, we have:
\begin{equation}
\label{eq:RPCP equation of layer 1}
    \mathcal{H}(X^{(1)};\vec{\delta}) = G^{2} P^{(1)} (I_{+}^{(1)}-I_{-}^{(1)}) A^{(1)} \vec{\delta}
\end{equation}
Note that 
\begin{equation}
\setlength{\arraycolsep}{4pt}
\resizebox{0.6\textwidth}{!}{$
I_{+}^{(1)}-I_{-}^{(1)} = \begin{bmatrix}
\mathbf{0} & 0 & 0 & 0 & 0 & 0 & 0 & 0 & 0 & 0 & 0 & 0 & 0 & 0 & 0 & 0 \\
0 & \mathbf{0} & 0 & 0 & 0 & 0 & 0 & 0 & 0 & 0 & 0 & 0 & 0 & 0 & 0 & 0 \\
0 & 0 & \mathbf{0} & 0 & 0 & 0 & 0 & 0 & 0 & 0 & 0 & 0 & 0 & 0 & 0 & 0 \\
0 & 0 & 0 & \mathbf{0} & 0 & 0 & 0 & 0 & 0 & 0 & 0 & 0 & 0 & 0 & 0 & 0 \\
0 & 0 & 0 & 0 & \mathbf{0} & 0 & 0 & 0 & 0 & 0 & 0 & 0 & 0 & 0 & 0 & 0 \\
0 & 0 & 0 & 0 & 0 & \mathbf{1} & 0 & 0 & 0 & 0 & 0 & 0 & 0 & 0 & 0 & 0 \\
0 & 0 & 0 & 0 & 0 & 0 & \mathbf{0} & 0 & 0 & 0 & 0 & 0 & 0 & 0 & 0 & 0 \\
0 & 0 & 0 & 0 & 0 & 0 & 0 & \mathbf{0} & 0 & 0 & 0 & 0 & 0 & 0 & 0 & 0 \\
0 & 0 & 0 & 0 & 0 & 0 & 0 & 0 & \mathbf{0} & 0 & 0 & 0 & 0 & 0 & 0 & 0 \\
0 & 0 & 0 & 0 & 0 & 0 & 0 & 0 & 0 & \mathbf{0} & 0 & 0 & 0 & 0 & 0 & 0 \\
0 & 0 & 0 & 0 & 0 & 0 & 0 & 0 & 0 & 0 & \mathbf{0} & 0 & 0 & 0 & 0 & 0 \\
0 & 0 & 0 & 0 & 0 & 0 & 0 & 0 & 0 & 0 & 0 & \mathbf{0} & 0 & 0 & 0 & 0 \\
0 & 0 & 0 & 0 & 0 & 0 & 0 & 0 & 0 & 0 & 0 & 0 & \mathbf{0} & 0 & 0 & 0 \\
0 & 0 & 0 & 0 & 0 & 0 & 0 & 0 & 0 & 0 & 0 & 0 & 0 & \mathbf{0} & 0 & 0 \\
0 & 0 & 0 & 0 & 0 & 0 & 0 & 0 & 0 & 0 & 0 & 0 & 0 & 0 & \mathbf{0} & 0 \\
0 & 0 & 0 & 0 & 0 & 0 & 0 & 0 & 0 & 0 & 0 & 0 & 0 & 0 & 0 & \mathbf{0}
\end{bmatrix},
$}
\end{equation}
and
\begin{equation}
\setlength{\arraycolsep}{4pt}
\resizebox{0.6\textwidth}{!}{$
P^{(1)} = \begin{bmatrix}
0 & 0 & 0 & 0 & 0 & \mathbf{1} & 0 & 0 & 0 & 0 & 0 & 0 & 0 & 0 & 0 & 0 \\
0 & 0 & 0 & 0 & 0 & 0 & \mathbf{1} & 0 & 0 & 0 & 0 & 0 & 0 & 0 & 0 & 0 \\
0 & 0 & 0 & 0 & 0 & 0 & 0 & 0 & 0 & \mathbf{1} & 0 & 0 & 0 & 0 & 0 & 0 \\
0 & 0 & 0 & 0 & 0 & 0 & 0 & 0 & 0 & 0 & \mathbf{1} & 0 & 0 & 0 & 0 & 0
\end{bmatrix}.
$}
\label{eq:appendix_P}
\end{equation}

The sparse structure of $(I_{+}^{(1)}-I_{-}^{(1)})$ mathematically acts as a column selector. When multiplied by the pooling matrix $P^{(1)}$, it nullifies all columns except the $5$-th one, yielding:
\begin{equation}
\setlength{\arraycolsep}{4pt}
\resizebox{0.6\textwidth}{!}{$
    P^{(1)}(I_{+}^{(1)}-I_{-}^{(1)}) = 
    \begin{bmatrix}
    0 & 0 & 0 & 0 & 0 & \mathbf{1} & 0 & 0 & 0 & 0 & 0 & 0 & 0 & 0 & 0 & 0 \\
    0 & 0 & 0 & 0 & 0 & \mathbf{0} & 0 & 0 & 0 & 0 & 0 & 0 & 0 & 0 & 0 & 0 \\
    0 & 0 & 0 & 0 & 0 & \mathbf{0} & 0 & 0 & 0 & 0 & 0 & 0 & 0 & 0 & 0 & 0 \\
    0 & 0 & 0 & 0 & 0 & \mathbf{0} & 0 & 0 & 0 & 0 & 0 & 0 & 0 & 0 & 0 & 0
    \end{bmatrix}.
$}
\end{equation}

Subsequently, multiplying this resulting matrix by the weight matrix $A^{(1)}$ transforms it into a row selector. Specifically, it perfectly isolates the $5$-th row of $A^{(1)}$ into the top row of the resulting matrix, while leaving all other rows as zeros:
\begin{equation}
\setlength{\arraycolsep}{4pt}
\resizebox{0.6\textwidth}{!}{$
    P^{(1)} (I_{+}^{(1)} - I_{-}^{(1)}) A^{(1)} = 
    \begin{bmatrix}
    A_{5,0}^{(1)} & A_{5,1}^{(1)} & \cdots & A_{5,24}^{(1)} \\
    0 & 0 & \cdots & 0 \\
    0 & 0 & \cdots & 0 \\
    0 & 0 & \cdots & 0
    \end{bmatrix}.
    $}
\end{equation}

Finally, let $G^{2} = [g_0, g_1, g_2, g_3]$ represent the aggregated linear transformation vector from the subsequent layers. Multiplying $G^{2}$ by the isolated matrix projects the $5$-th row of $A^{(1)}$ into a single vector scaled by $g_0$:
\begin{align}
\resizebox{1.0\hsize}{!}{$%
    G^{2} P^{(1)} (I_{+}^{(1)} - I_{-}^{(1)}) A^{(1)} 
    = [g_0, g_1, g_2, g_3] 
    \begin{bmatrix}
    A_{5,0}^{(1)} & A_{5,1}^{(1)} & \cdots & A_{5,24}^{(1)} \\
    0 & 0 & \cdots & 0 \\
    0 & 0 & \cdots & 0 \\
    0 & 0 & \cdots & 0
    \end{bmatrix} \nonumber 
    = (g_0 A_{5,0}^{(1)}, g_0 A_{5,1}^{(1)}, \cdots, g_0 A_{5,24}^{(1)}).
    $}%
\end{align}
{\em i.e.}, 
\begin{align}
    G^{2} P^{(1)} (I_{+}^{(1)} - I_{-}^{(1)}) A^{(1)} 
    = g_0 A_5^{(1)}
\end{align}

By sampling a set of linearly independent perturbations $S = (\vec{\delta}^{(0)}, \vec{\delta}^{(1)}, \dots, \vec{\delta}^{(n-1)})^{\sf T}$, we construct a linear system according to Eq.~\eqref{eq:RPCP equation of layer 1}. Solving this system directly yields a sparse vector proportional to $A_{5}^{(1)}$. Based on the equivalent matrix expansion demonstrated in Eq.~\eqref{eq:appendix_A}, and substituting the true convolutional kernel parameters $C = \left[\begin{smallmatrix} 1 & 2 \\ 3 & 4 \end{smallmatrix}\right]$, the explicit structure of $A_{5}^{(1)}$ is revealed as a 25-dimensional vector:
\begin{equation}
    A_5^{(1)} = (0,0,0,0,0,~ 0, \mathbf{1}, \mathbf{2}, 0,0,~ 0, \mathbf{3}, \mathbf{4}, 0,0,~ 0, \dots, 0)
\end{equation}
This explicit algebraic structure immediately exposes the true parameter ratios ($1:2:3:4$) of the convolution kernel $C^{(1)}$ according to Eqs. \eqref{eq:appendix_A0_example} and \eqref{eq:appendix_A}. %at the specific local receptive field indices $\mathcal{K}^{(1)}_5=\{6, 7, 11, 12\}$.

\subsubsection{The PSP method.}
As illustrated in Fig. \ref{fig:toy_psp}, a PSP $X^{(1)}$ occurs when two competing neurons within the LRF-P 0 reach a same maximum value.

By querying the model with perturbed inputs $X^{(1)} + \vec{\delta}$ and $X^{(1)} - \vec{\delta}$, the max pooling selection switches between neurons $4$ and $5$. 
Consequently, the first row vector of the pooling matrix $P^{(1)}$ has a $1$ at the $4$-th entry ($P_{+}^{(1)}$) for one perturbation direction ({\em e.g., $X^{(1)} + \vec{\delta}$}) and a $1$ at the $5$-th entry ($P_{-}^{(1)}$) for the opposite direction ($X^{(1)} - \vec{\delta}$). As formulated in Eq.~\eqref{eqn:2nd_psp}, by calculating the second-order differential function $\mathcal{H}(X;\vec{\delta})$ defined as $\mathcal{H}(X;\vec{\delta}) = \mathcal{F}_{\theta}(X+\vec{\delta}) + \mathcal{F}_{\theta}(X-\vec{\delta}) - 2\mathcal{F}_{\theta}(X)$, we have:
\begin{equation}
    \mathcal{H}(X^{(1)}; \vec{\delta}) = G^{2} (P_{+}^{(1)} - P_{-}^{(1)}) I^{(1)} A^{(1)} \vec{\delta}. 
\end{equation}

Assuming the max pooling selections in the remaining LRF-Ps are unchanged upon the tiny perturbations, their corresponding rows in $P_{+}^{(1)}$ and $P_{-}^{(1)}$ are identical. Thus, the difference matrix evaluates strictly to zeros except for the first row:
\begin{equation}
\setlength{\arraycolsep}{4pt}
\resizebox{0.6\textwidth}{!}{$
P_{+}^{(1)}-P_{-}^{(1)} = \begin{bmatrix}
0 & 0 & 0 & 0 & \mathbf{1} & \mathbf{-1} & 0 & 0 & 0 & 0 & 0 & 0 & 0 & 0 & 0 & 0 \\
0 & 0 & 0 & 0 & 0 & 0 & \mathbf{0} & 0 & 0 & 0 & 0 & 0 & 0 & 0 & 0 & 0 \\
0 & 0 & 0 & 0 & 0 & 0 & 0 & 0 & 0 & \mathbf{0} & 0 & 0 & 0 & 0 & 0 & 0 \\
0 & 0 & 0 & 0 & 0 & 0 & 0 & 0 & 0 & 0 & \mathbf{0} & 0 & 0 & 0 & 0 & 0
\end{bmatrix}.
$}
\label{eq:appendix_P_diff}
\end{equation}

Because neurons $4$ and $5$ are both strictly positive (achieving the maximum value of $19$), the diagonal entries for these indices in the activation matrix $I^{(1)}$ are both $1$. Thus, left-multiplying the weight matrix $A^{(1)}$ by $(P_{+}^{(1)} - P_{-}^{(1)}) I^{(1)}$ acts as a precise row selector that extracts the difference between the $4$-th and $5$-th rows of $A^{(1)}$:
\begin{equation}
    (P_{+}^{(1)} - P_{-}^{(1)}) I^{(1)} A^{(1)} = \begin{bmatrix} A_4^{(1)} - A_5^{(1)} \\ \mathbf{0} \\ \mathbf{0} \\ \mathbf{0} \end{bmatrix}.
\end{equation}

Subsequently, left-multiplying by the aggregated linear transformation vector $G^{2} = [g_0, g_1, g_2, g_3]$ isolates this row difference, scaled by the scalar $g_0$:
\begin{align}\label{eq:appendix_psp_a4_a5}
    G^{2} (P_{+}^{(1)} - P_{-}^{(1)}) I^{(1)} A^{(1)}
    =g_0 (A_{4}^{(1)} - A_{5}^{(1)}).
\end{align}

By solving the linear system constructed from multiple perturbed queries, we recover a vector proportional to the difference $A_4^{(1)} - A_5^{(1)}$. Substituting the specific kernel parameters $(1,2,3,4)$ into our expanded matrix (Eq.~\eqref{eq:appendix_A}), we can explicitly visualize the alignment of these two rows:
\begin{align}
    A_4^{(1)} &= (0,0,0,0,0,~ \mathbf{1}, \mathbf{2}, 0, 0,0,~ \mathbf{3}, \mathbf{4}, 0, 0,0,~ 0, \dots, 0), \\
    A_5^{(1)} &= (0,0,0,0,0,~ 0, \mathbf{1}, \mathbf{2}, 0,0,~ 0, \mathbf{3}, \mathbf{4}, 0,0,~ 0, \dots, 0).
\end{align}
Consequently, their difference perfectly captures the shifted structural pattern of the convolution window:
\begin{align}
    A_{4}^{(1)} - A_{5}^{(1)} &= (0,0,0,0,0,~ \mathbf{1},~ \mathbf{2-1},~ \mathbf{-2},~ 0,0,~ \mathbf{3},~ \mathbf{4-3},~ \mathbf{-4},~ 0,0,~ 0, \dots, 0) \nonumber \\
    &= (0,0,0,0,0,~ \mathbf{1},~ \mathbf{1},~ \mathbf{-2},~ 0,0,~ \mathbf{3},~ \mathbf{1},~ \mathbf{-4},~ 0,0,~ 0, \dots, 0).
\end{align}

The non-zero elements in this difference vector in Eq. \eqref{eq:appendix_psp_a4_a5} uniquely identify the spatial indices $i=4$ and $j=5$. Once identified, we apply the internal differential property of the PSP defined in Eq.~\eqref{eqn:internal-differ}. This relationship elegantly maps  the kernel weights to the input difference:
\begin{equation} 
    C_{0,0}(X^{(1)}_5 - X^{(1)}_6) + C_{0,1}(X^{(1)}_6 - X^{(1)}_7) + C_{1,0}(X^{(1)}_{10} - X^{(1)}_{11}) + C_{1,1}(X^{(1)}_{11} - X^{(1)}_{12}) = 0.
\end{equation}

To demonstrate its exact correctness, we substitute the specific spatial values from our adjusted input vector $X^{(1)}$ (where $X^{(1)}_5=6, X^{(1)}_6=7, X^{(1)}_7=-1, X^{(1)}_{10}=-3, X^{(1)}_{11}=2, X^{(1)}_{12}=2$):
\begin{align}
    C_{0,0}(6 - 7) + C_{0,1}(7 - (-1)) + C_{1,0}(-3 - 2) + C_{1,1}(2 - 2) &= 0 \nonumber, \\
    -1 \cdot C_{0,0} + 8 \cdot C_{0,1} - 5 \cdot C_{1,0} + 0 \cdot C_{1,1} &= 0.
\end{align}

As expected, the true kernel parameters $(1, 2, 3, 4)$ perfectly satisfy this derived equation: $-1(1) + 8(2) - 5(3) + 0(4) = 0$. By collecting slightly more than $h_c^{(1)} \times w_c^{(1)} = 4$ such independent PSP instances, the adversary establishes a full-rank linear system. This directly and explicitly yields the exact kernel weights $C^{(1)}$.

\section{Attack on LeNet-5 Style Architectures}
\label{sect:LeNet5}

% LeNet-5 is widely regarded as the seminal convolutional neural network that established the foundation for modern deep learning. 

%In 1998, LeCun officially proposed CNN, which produced the concept of convolutional neural  networks. \cite{lecun2002gradient}

This section focuses on the practical convolutional neural network adopting the LeNet-5 Style Architecture \cite{lecun2002gradient}, which was first proposed by LeCun in 1998. We demonstrate the complete process of applying our proposed algebraic attack to extract the model parameters.

\begin{figure}
	\centering
        \includegraphics[width=0.9\linewidth]{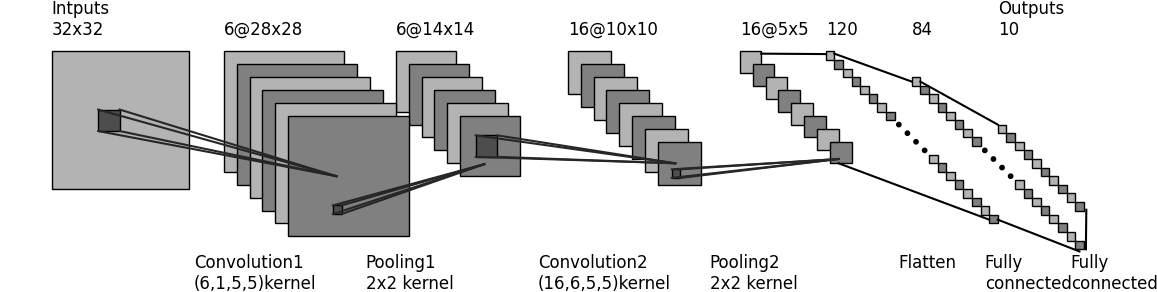}
	\caption{LeNet-5 Architecture}
	\label{fig:LeNet-5}
\end{figure}
\subsection{LeNet-5 Style Architecture}
\label{subsect: modern LeNet-5}
As shown in Figure \ref{fig:LeNet-5}, the classic LeNet-5 \cite{lecun2002gradient} consists of two Convolutional Rounds and two FCNN Rounds, employing a sparse channel connection in its second convolutional layer, Tanh activation and Average Pooling. However, the sparse structure becomes a computational bottleneck for parallel GPU processing \cite{DBLP:conf/iclr/0022KDSG17,DBLP:conf/iri/ParkHC17} and restricts complex feature extraction \cite{DBLP:conf/iclr/FrankleC19}. Activation functions (Sigmoid/Tanh) pose a higher risk of vanishing and exploding gradients \cite{DBLP:journals/jmlr/GlorotB10} and slow convergence \cite{krizhevsky2012imagenet}. 
% Originally optimized to minimize computational cost and memory footprint, the classic LeNet-5 employed a partial (sparse) connection scheme in its second convolutional layer. However, given the massive parallel computing capabilities of modern GPUs, this sparse structure has become a bottleneck, restricting the model's ability to capture complex features through cross-channel combinations. Furthermore, the legacy activation functions (Sigmoid/Tanh) pose a higher risk of vanishing gradients and slow convergence; specifically, in saturation regions where inputs are large, gradients approach zero, hindering weight updates in the later stages of training. 
Consequently, modern implementations of LeNet-5 and other CNNs with similar architecture typically incorporate the following modifications \cite{krizhevsky2012imagenet,DBLP:conf/mm/JiaSDKLGGD14,Simonyan2015very,szegedy2015going}:
\begin{itemize}
\item \textbf{Convolutional Connectivity:} The original asymmetric sparse connections are replaced with full channel connections. In modern frameworks, the second layer kernel is standardized to a size of $(16, 6, 5, 5)$, with standard convolution stride $s_c^{(k)}=1$ and pooling stride $s_{\rho}^{(k)}=2$, ($k=1,2$).  
\item \textbf{Activation Function:} The Tanh function is replaced by ReLU.
\item \textbf{Pooling Strategy:} Average pooling is substituted with max pooling to capture the most prominent features.
%\item \textbf{Raw Output Accessible.}
\end{itemize}

% This modernized LeNet-5 (incorporating Full Conv, ReLU, MaxPool) achieves accuracies exceeding 99.2\% with faster training convergence, enhanced stability, and reduced susceptibility to local optima.
%\input{figures/table_Table_Version.tex}
% As shown in Table \ref{tab:model_structure}, 
In this section, our practical attack targets this modern architecture with raw output accessible. By definition, the victim model is a $(2+2)$-Deep CNN.
% $f=f^{(3)} \circ \sigma^{(2)} \circ f^{(2)} \circ \sigma^{(1)} \circ f^{(1)} \,
% \circ p^{(2)} \circ \sigma_c^{(2)} \circ f_c^{(2)} \circ p^{(1)} \circ \sigma_c^{(1)} \circ f_c^{(1)}$.

\subsection{Multiple Channels}
\begin{figure}
	\centering
        \includegraphics[width=0.8\linewidth]{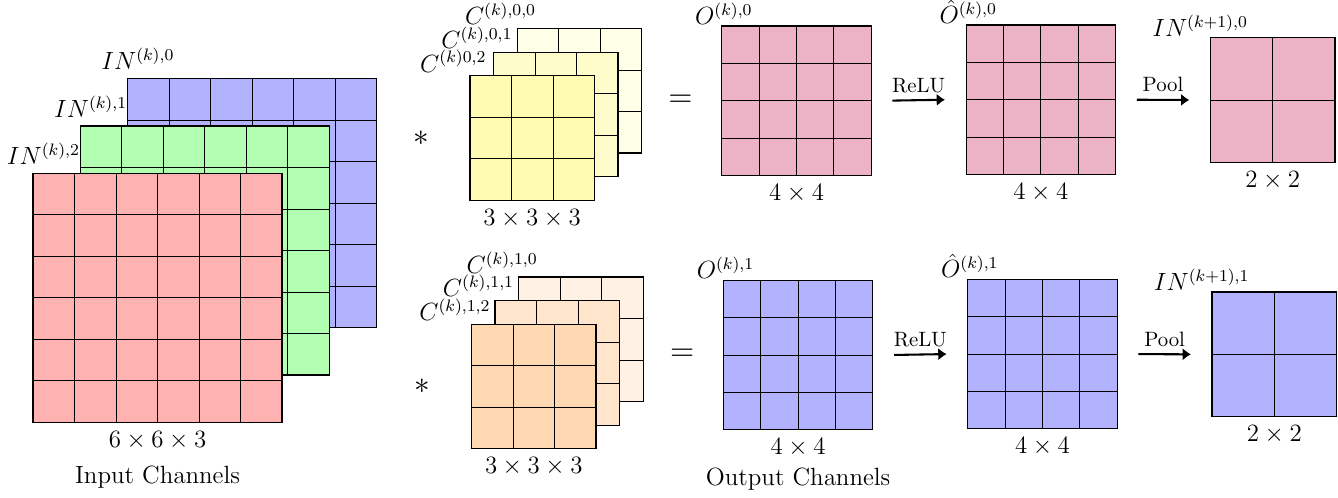}
	\caption{Multiple Input and Output Channels:  $\mathsf{c}_{in}^{(k)}=3, ~\mathsf{c}_{out}^{(k)}=2$.}
	\label{fig:multi_channel_Conv}
\end{figure}

% \zirui{The problem here is that if we include the number of input and output channels in the shape of the convolution kernel (input channels, output channels, height, width), then the entire convolutional layer has only one convolution kernel, which makes it difficult to describe the convolution kernel corresponding to each channel.}
LeNet-5 adopts multiple input and output channels in the Convolutional Block. %A complete convolutional layer is composed of several feature maps (with different kernel matrices), so that multiple features can be extracted at each location. 
Suppose in layer $k$, the convolution layer has  \(\mathsf{c}_{in}^{(k)}\) input channels and \(\mathsf{c}_{out}^{(k)}\) output channels, {\em i.e.}, there are $\mathsf{c}_{in}^{(k)}$ input matrices, denoted as $IN^{(k),0},\cdots, IN^{(k),\mathsf{c}_{in}^{(k)}-1}$, and  $\mathsf{c}_{out}^{(k)}$ output matrices, denoted as $O^{(k),0},\cdots, O^{(k),\mathsf{c}_{out}^{(k)}-1}$. 
An example of convolutional layer with $\mathsf{c}_{in}^{(k)}=3$ input channels, $\mathsf{c}_{out}^{(k)}=2$ output channels  is shown in Figure \ref{fig:multi_channel_Conv}. %Simply, we can consider that the number of output channels is equal to that of convolutional kernels. 
Formally, denote the kernel matrices as \(C^{(k),m,n}=[C^{(k),m,n}_{i,j}]_{h_{c}^{(k)}\times w_{c}^{(k)}}\), where \(0\le m \le \mathsf{c}_{out}^{(k)}-1\) and \(0 \le n \le \mathsf{c}_{in}^{(k)}-1\). 
The entries \(O^{(k),m}_{i,j}\) (\(0\le m \le \mathsf{c}_{out}^{(k)}-1,~ 0\le i \le h^{(k)}_{o}-1,~0\le j \le w^{(k)}_{o}-1\)) of the output matrices generated by the convolution operation are defined as:
\begin{equation}\label{eqn:conv_multi_channels}\footnotesize
O^{(k),m}_{i,j}=\sum_{n=0}^{\mathsf{c}_{in}^{(k)}-1}\sum_{p=0}^{h_{c}^{(k)}-1}\sum_{q=0}^{w^{(k)}_{c}-1}IN_{i+p,j+q}^{(k),n}C^{(k),m,n}_{p,q} + b^{(k),m}, 
\end{equation}
%where \(IN_{i+p,j+q}^{(k),n}\) denotes the input element of the n-th input channel and 
where \(b^{(k),m}\) are the biases for the $m$-th output channel, and the stride $s_c^{(k)}=1$.

\subsubsection{First Layer of LeNet-5.} 
\begin{figure}
	\centering
        \includegraphics[width=0.8\linewidth]{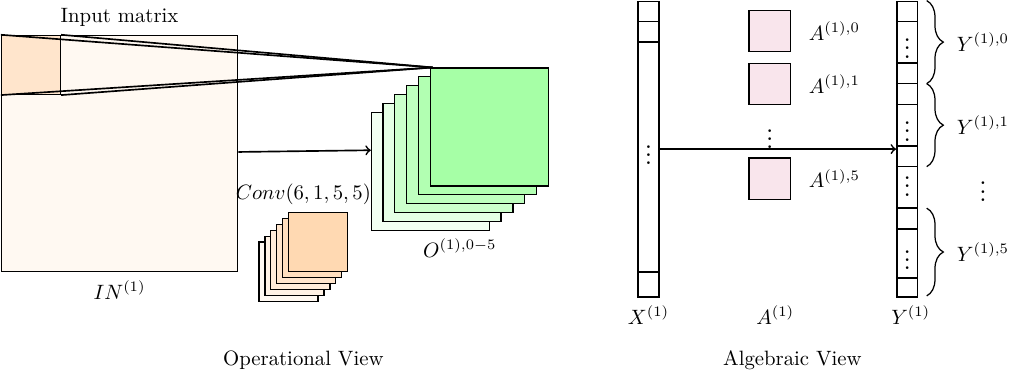}
	\caption{Operational View and Algebraic View,  $(\mathsf{c}_{out}^{(k)},\mathsf{c}_{in}^{(k)}, h_c^{(k)}, w_c^{(k)})=(6,1,5,5).$}
	\label{fig:OV_AV_1}
\end{figure}

As shown in Figure \ref{fig:OV_AV_1}, the input and output channels of the first layer LeNet-5 are $\mathsf{c}_{in}^{(k)}=1, ~\mathsf{c}_{out}^{(k)}=6$. The sizes of the output matrices \(O^{(1),m}\) (\(0\le m \le 5\)) are $h^{(k)}_{o}\times w^{(k)}_{o}= 5 \times 5$. 
%are organized in six feature maps. Each unit in a feature map has 25 inputs (from a $5 \times 5$ area in the input) and therefore 25 trainable coefficients plus a trainable bias.
In vector form, the operation of the first convolutional layer can be modeled as applying six distinct convolution kernel matrices to the input vector $X^{(1)}$. The 6 output matrices are then flattened and sequentially concatenated to form a single vector $Y^{(1)}$ as shown on the right side of Figure \ref{fig:OV_AV_1}.
Consequently, the convolutional matrix $A^{(1)}$ for this layer can be constructed by vertically stacking (or concatenating) the six corresponding convolutional sub-matrices, as shown in Figure \ref{fig:OV_AV_1}. More details are given in the following 2nd layer of LeNet-5. 

%\begin{figure}
%	\centering
%        \includegraphics[width=1\linewidth]{figures/OV_AV.pdf}
%	\caption{Operational View and Algebraic View}
%	\label{fig:OV_AV}
%\end{figure}

%While the first layer has one input channel and multiple output channels, the second layer expands to multiple input and output channels. Specifically, the second layer receives 6 input feature maps and produces 16 output feature maps. Each output is computed by element-wise summation of the responses from six distinct single-channel convolutional kernels (each applied to the same spatial region of its corresponding input feature map), followed by adding a learnable bias term. 

\subsubsection{Second Layer of LeNet-5.} 
\begin{figure}
	\centering
        \includegraphics[width=0.8\linewidth]{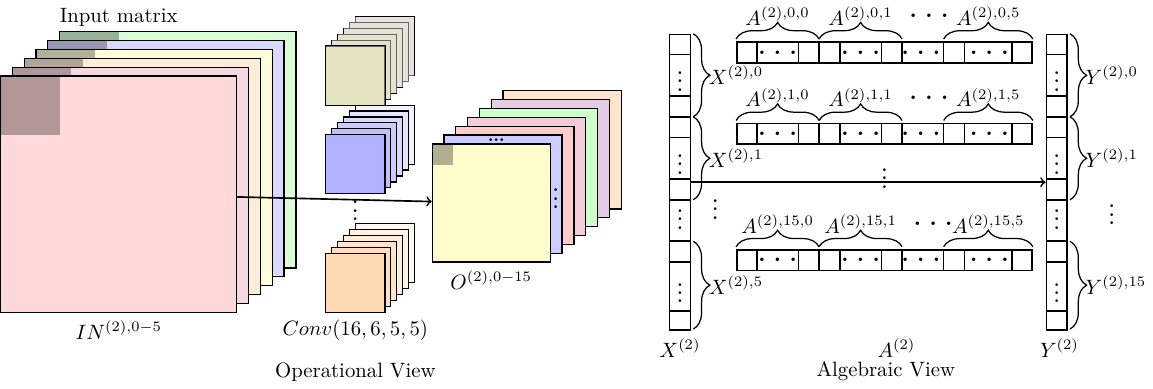}
	\caption{Operational View and Algebraic View of multiple input and output channels convolutional operations. $(\mathsf{c}_{out}^{(k)},\mathsf{c}_{in}^{(k)}, h_c^{(k)}, w_c^{(k)})=(16,6,5,5).$}
	\label{fig:multi_input_multi_output_OV_AV}
\end{figure}

Similarly, from an algebraic perspective, the operation of the second convolutional layer involves 6 input channels and 16 output channels, {\em i.e.}, \(\mathsf{c}_{in}^{(2)}=6, \mathsf{c}_{out}^{(2)}=16\). The 6 input matrices \(IN^{(2),0-5}\) are transformed into a single vector \(X^{(2)}=(X^{(2),0},X^{(2),1},...,X^{(2),n},...,X^{(2),5})^{\sf T}\), where \(X^{(2),n}=(IN^{(2),n}_{0},..,IN^{(2),n}_{h_{in}^{(2)}-1})^{\sf T}\). %The convolutional computation is performed on the corresponding input channels, followed by element-wise summation. 
According to Eq. \eqref{eqn:conv_multi_channels}, \(X^{(2)}\) is convolved with \(A^{(2),m}\) (the kernel corresponding to the $m$-th output channel), and the 6 input channels are accumulated element-wise to generate the target 16 feature matrices \(O^{(2),m}\) ($0\leq m\leq 15$). %Eventually, 16 output feature maps are derived from this operation.
Flattening the feature matrix \(O^{(2),m}\), \(Y^{(2),m}=(O^{(2),m}_0,...,O^{(2),m}_{h_o^{(2)}-1})^{{\sf T}}\). Consequently, the single (\(d_{f}^{(2)}=\mathsf{c}_{out}^{(2)}\cdot h_{o}^{(2)}\cdot w_{o}^{(2)}\))-dimensional output vector \(Y^{(2)}\), which is composed of multiple output channels, is denoted as \(Y^{(2)}=(Y^{(2),0},Y^{(2),1},\cdots,Y^{(2),15})^{{\sf T}}\), as shown in Figure \ref{fig:multi_input_multi_output_OV_AV}.

For the $i$-th output channel, this operation is equivalent to horizontally concatenating 6 individual convolutional matrices into a matrix $A^{(2),i}=(A^{(2),i,0},$ $A^{(2),i,1},\cdots, A^{(2),i,5})$ ($0\leq i\leq 15$), where $A^{(2),i,j}$ ($0\leq j\leq 5$) is derived by Def. \ref{def:conv_matrix}. For the full output of the second convolutional layer, 16 such matrices $A^{(2),i}$ ($0\leq i\leq 15$) are vertically stacked to form the final convolutional matrix $A^{(2)}$.

Based on our analysis, we can derive the algebraic expression of the multi-channel convolutional layer:
Suppose that for the $i$-th output channel and the $j$-th input channel, the convolutional matrix of the corresponding single convolution kernel is $A^{(k),i,j}$ derived by Def. \ref{def:conv_matrix}, then the full convolutional matrix of the entire $k$-th convolutional layer is
\begin{align}\small
    A^{(k)}=
    \begin{pmatrix}
        A^{(k),0,0} & A^{(k),0,1} & \cdots & A^{(k),0,\mathsf{c}_{in}^{(k)}-1} \\
        A^{(k),1,0} & A^{(k),1,1} & \cdots & A^{(k),1,\mathsf{c}_{in}^{(k)}-1} \\
        \vdots & \vdots & \ddots & \vdots \\
        A^{(k),\mathsf{c}_{out}^{(k)}-1,0} & A^{(k),\mathsf{c}_{out}^{(k)}-1,1} & \cdots & A^{(k),\mathsf{c}_{out}^{(k)}-1,\mathsf{c}_{in}^{(k)}-1}
    \end{pmatrix}.
\end{align}
Therefore, the multiple channels can also be transformed into the regular operations {\em i.e.}, the input vector is multiplied with a matrix to derive the output vector. Hence, Def. \ref{def:layer_merging} also holds in this setting,  and our algebraic attacks based on the RPCP or PSP method proposed in Sect. \ref{sect:RPCP_Method} and \ref{sect:PSP Method} also work.

%\section{Evaluating the $(\varepsilon,0)$-Functional Equivalence \cite{DBLP:conf/crypto/CarliniJM20}}
%\label{supp:carlini_error_eval}
%\input{supp/carlini_error_eval}
\section{More Detailed Information in Experiments}
\label{supp:model structure}

In this section, we demonstrate the full details of our experimental results in Sect. \ref{supp:experiment results} and describe the detailed structure information of the models we experimented on in Sect. \ref{supp:detailed model structure}.%and Table \ref{tab:Attack on LeNet-5}.
\subsection{Experimental Results}
\label{supp:experiment results}
We include a more concise version of our experimental results in Section \ref{Introduction} ({\em i.e.}, Table \ref{tab:Attack on CNNs}). For the complete experimental result metrics -- including runtime, model queries, and error evaluation $(\varepsilon,0)$ and $\max \lvert \theta-\hat{\theta} \rvert$ -- we provide these full details in Table \ref{tab:Attack on CNNs full}.
\begin{table}[h!] \tiny
\centering
\caption{Experiments on different CNNs. }
\label{tab:Attack on CNNs full}
% 1. 设置表格整体行高 (Models 之间的距离)
\renewcommand{\arraystretch}{1.5} 
\setlength{\tabcolsep}{4pt}

% 2. 定义一个统一的换行间距命令，方便统一调整内部行高
% 修改这个值 (比如 6pt, 1em, 12pt) 可以同时改变两列的内部行间距
\newcommand{\ggap}{\\[3pt]} 

\begin{threeparttable}
    \begin{tabular}{l c l c c l l l l} % 将 Architecture 和 Kernel 改为左对齐(l)会更整齐
    \toprule
    \makecell[l]{\textbf{Models} \\
    ($m+n$) 
    }
    & \textbf{Trainset}
    & \makecell[c]{\textbf{Architecture}\\ 
    $d^{(k)}$--$d_f^{(k)}$--$d^{(k+1)}$}
    & \textbf{Kernel}$^\dagger$ 
    & \textbf{Param} 
    & \makecell[c]{\textbf{Runtime}\\
    Seconds}
    & \textbf{Queries}
    & $(\varepsilon, 0)$$^\ddag$ 
    & $\max \lvert \theta - \hat{\theta} \rvert$$^\ddag$ 
    \\
    \midrule
    
    (\textcolor{red}{\bf 1}+1)
    & Random 
    & Conv R1: 64--36--9
    & (1,1,3,3)
    & 10 & $2^{-0.26}$ & $2^{9.98}$ & $2^{-43.19}$ & $2^{-47.52}$
    \\
    \midrule % 添加分割线让不同模型区分更明显
    
    (\textcolor{red}{\bf 2}+1)
    & MNIST 
    & \makecell[l]{ % 左对齐
        Conv R1: 1024--784--196 \ggap % 使用自定义间距
        Conv R2: 196--100--25
      }
    & \makecell[c]{ % 对应使用左对齐和相同的间距
        (1,1,5,5) \ggap
        (1,1,5,5)
      }
    & 52 & $2^{5.97}$ & $2^{15.06}$ & $2^{-30.20}$ & $2^{-34.65}$
    \\
    \midrule
    
    (\textcolor{red}{\bf 2}+2)$^{\ast}$ & MNIST 
    & \makecell[l]{
        Conv R1: 1024--4704--1176 \ggap
        Conv R2: 1176--1600--400
      }
    & \makecell[c]{
        (1,6,5,5) \ggap
        (6,16,5,5)
      }
    & 2572 & $2^{17.21}$ & $2^{27.29}$ & $2^{-23.38}$ & $2^{-27.31}$
    \\
    \midrule
    
    (\textcolor{red}{\bf 3}+1)${}^{p_2}$
    & MNIST 
    & \makecell[l]{
        Conv R1: 1024--1024--256 \ggap
        Conv R2: 256--256--64 \ggap
        Conv R3: 64--64--16
      } 
    & \makecell[c]{
        (1,1,5,5) \ggap
        (1,1,5,5) \ggap
        (1,1,5,5)
      }
    & 78 & $2^{8.76}$ & $2^{16.90}$ & $2^{-27.54}$ & $2^{-31.15}$
    \\
    \midrule

    (\textcolor{red}{\bf 2}+2)${}^{p_1}$  
    & CIFAR10 
    & \makecell[l]{
        Conv R1: 3072--3072--768 \ggap
        Conv R2: 768--1024--256
      }
    & \makecell[c]{
        (3,3,3,3) \ggap
        (3,4,3,3)
      }
    & 196 & $2^{10.96}$ & $2^{18.37}$ & $2^{-29.13}$ & $2^{-36.19}$
    \\\midrule[1pt]
    
    (\textcolor{red}{\bf 2+1})$^{\clubsuit}$ & MNIST &
    \makecell[l]{
        Conv R1: 1024--4704--1176 \ggap
        Conv R2: 1176--1600--400 \ggap
        FCNN: 400--20--10 
      } 
      
    & \makecell[c]{ % 对应使用左对齐和相同的间距
        (1,6,5,5) \ggap
        (6,16,5,5) \ggap
        --
      }
    & \makecell[c]{10772 \ggap (2572+8230)} 
    & $2^{17.38}$ & $2^{27.65}$ & $2^{-22.06}$ & $2^{-27.14}$
    \\ \midrule

    (\textcolor{red}{\bf 2+2})$^{\ast}$$^{\blacktriangle}$ & MNIST &
    \makecell[l]{
        \textcolor{red}{$\bf C_1$}: 1024--4704--1176 \ggap
        \textcolor{red}{$\bf C_2$}: 1176--1600--400 \ggap
        FCNN: \textcolor{red}{\bf 400--120}--84--10
      } 
      
    & \makecell[c]{ % 对应使用左对齐和相同的间距
        (1,6,5,5) \ggap
        (6,16,5,5) \ggap
        --
      }
    & \makecell[c]{61706 \ggap (2572+59134)} 
    & $2^{18.46}$
    & $2^{27.31}$ & -- & $2^{-26.03}$
    \\ \midrule

    (0+\textcolor{red}{\bf 3}) \cite{DBLP:conf/crypto/CarliniJM20} & MNIST &
    \makecell[l]{
        FCNN: 40--20--10--10--1 
      } 
      
    & --
    & \makecell[c]{1110} 
    & -- & $2^{17.8}$ & $2^{-23.4}$ & $2^{-27.1}$
    \\ 
    \midrule

    (0+\textcolor{red}{\bf 3})
    \cite{foerster2024beyond} $^{\spadesuit}$& MNIST &
    \makecell[l]{
        FCNN: 784--16--\textcolor{red}{\bf 16--16}--1 
      } 
      
    & --
    & \makecell[c]{272} 
    & $2^{16.0}$ & $2^{25.4}$ & -- & --\\

    \bottomrule

    \end{tabular}
    \begin{tablenotes}[flushleft] 
    \item[$\ast$:] LeNet-5 modern model, specified in in Sect. \ref{subsect: modern LeNet-5}.\\
    
    \item[$p_1$, $p_2$:] $p_i$ indicates $pad^{(k)}=i$ defined as ``same padding'' in  \textsf{Supplementary Material} \ref{supp:other_parameters}; otherwise, $pad^{(k)}=0$.\\
    
    \item[$\dag$:] Convolutional kernel size of $(\mathsf{c}_{in}^{(k)}, \mathsf{c}_{out}^{(k)}, h_{c}^{(k)}, w_{c}^{(k)})$, with stride $s^{(k)}_c=1$.\\
        
   %\multicolumn{9}{l}{Pooling kernels are of size $2 \times 2$ with stride $s^{(k)}_{\rho}=2$.}\\
    \item[$\ddag$:] The accuracy term $(\varepsilon,0)$ is defined in Sect. \ref{sect:assumptions_1}, $\max \lvert \theta - \hat{\theta} \rvert$ directly measures the maximum extraction error of model parameters.\\
    
     \item[$\clubsuit$:] Reduced version of LeNet-5 with only one FCNN Round.\\

     \item[$\blacktriangle$:] Red bold layer in Architecture indicates end-to-end extracting the full Convolutional Block and 93.33\% weights of FCNN Round 1 (not full due to existing dead neurons and persistent neurons according to \cite{ito2025hard}).\\  

     % \multicolumn{9}{l}{$l_3$: $l_i$ indicates extracting the $i$-th layer with prior layers’ extractions assumed to be correct.}\\
     \item[$\spadesuit$]: Red bold layer in Architecture indicates extracting the single layer, with prior layers’ extractions assumed to be correct.
     \end{tablenotes}
\end{threeparttable}
\end{table}

\subsection{Detailed Model Structure Information in Experiments}
\label{supp:detailed model structure}
\paragraph{(1+1)-Deep CNN:}
\begin{enumerate}
    \item Convolutional Round 1: 
    \begin{itemize}
        \item \textbf{Convolutional layer $f_c^{(1)}$}: $8 \times 8$ Input matrix $IN^{(1)}$, convolved with $(1,1,3,3)$ convolution kernel, with stride $s_c^{(1)}=1$, padding $pad^{(1)}=0$, produces $6 \times6$ output matrix $O^{(1)}$, followed by ReLU activation.
        \item \textbf{Max Pooling layer $\rho^{(1)}$}: %$6 \times 6$ Input matrix $\hat{O}^{(1)}$, downsampled by
        $2 \times 2$ max pooling kernel, with stride $s_{\rho}^{(1)}=2$, produces $3 \times 3$ output matrix $IN^{(2)}$.
    \end{itemize}
    \item FCNN Round 1: 
    \begin{itemize}
        \item Flatten the matrix $IN^{(2)}$ and get the 9-dimensional vector $X^{(1)}$.
        \item \textbf{Fully-connected Layer} $f^{(1)}$: $d^{(1)}=9$ input neurons and $d^{(2)}=5$ output neurons, followed by ReLU activation.
    \end{itemize}
    \item Last Layer $f^{(2)}$: $d^{(2)}=5$ input neurons and $d^{(3)}=10$ output neurons.
\end{enumerate}

\paragraph{(2+1)-Deep CNN:}
\begin{enumerate}
    \item Convolutional Round 1: 
    \begin{itemize}
        \item \textbf{Convolutional layer} $f_c^{(1)}$: $32 \times 32$ Input matrix $IN^{(1)}$, convolved with $(1,1,5,5)$ convolution kernel, with stride $s_c^{(1)}=1$, padding $pad^{(1)}=0$, produces $28 \times 28$ output matrix $O^{(1)}$, followed by ReLU activation.
        \item \textbf{Max Pooling layer} $\rho^{(1)}$: %$28 \times 28$ Input matrix $\hat{O}^{(1)}$, downsampled by 
        $2 \times 2$ max pooling kernel, with stride $s_{\rho}^{(1)}=2$, produces $14 \times 14$ output matrix $IN^{(2)}$.
    \end{itemize}
    
    \item Convolutional Round 2:
    \begin{itemize}
        \item \textbf{Convolutional layer} $f_c^{(2)}$: $14 \times 14$ Input matrix $IN^{(2)}$, convolved with $(1,1,5,5)$ convolution kernel, with stride $s_c^{(2)}=1$, padding $pad^{(2)}=0$, produces $10 \times 10$ output matrix $O^{(2)}$, followed by ReLU activation.
        \item \textbf{Max Pooling layer} $\rho^{(2)}$: %$10 \times 10$ Input matrix $\hat{O}^{(2)}$, downsampled by 
        $2 \times 2$ max pooling kernel, with stride $s_{\rho}^{(2)}=2$, produces $5 \times 5$ output matrix $IN^{(3)}$.
    \end{itemize}

    \item FCNN Round 1: 
    \begin{itemize}
        \item Flatten the matrix $IN^{(3)}$ and get the 25-dimensional vector $X^{(1)}$.
        \item \textbf{Fully-connected Layer} $f^{(1)}$: $d^{(1)}=25$ input neurons and $d^{(2)}=18$ output neurons, followed by ReLU activation.
    \end{itemize}
    
    \item Last Layer $f^{(2)}$: $d^{(2)}=18$ input neurons and $d^{(3)}=10$ output neurons.
\end{enumerate}

\paragraph{(2+2)-Deep LeNet-5:}
\begin{enumerate}
    \item Convolutional Round 1: 
    \begin{itemize}
        \item \textbf{Convolutional layer} $f_c^{(1)}$: $32 \times 32$ Input matrix $IN^{(1)}$, convolved with $(6,1,5,5)$ convolution kernel, with stride $s_c^{(1)}=1$, padding $pad^{(1)}=0$, produces 6 output matrices  $O^{(1)}$ of size $28 \times 28$ , followed by ReLU activation.
        \item \textbf{Max Pooling layer} $\rho^{(1)}$: %6 Input matrix $\hat{O}^{(1)}$ of size $28 \times 28$, downsampled by 
        $2 \times 2$ max pooling kernel, with stride $s_{\rho}^{(1)}=2$, produces 6 output matrices $IN^{(2)}$ of size $14 \times 14$ .
    \end{itemize}
    
    \item Convolutional Round 2:
    \begin{itemize}
        \item \textbf{Convolutional layer} $f_c^{(2)}$: 6 Input matrix $IN^{(2)}$ of size $14 \times 14$, convolved with $(6,16,5,5)$ convolution kernel, with stride $s_c^{(2)}=1$, padding $pad^{(2)}=0$, produces 16 output matrix $O^{(2)}$ of size $10 \times 10$, followed by ReLU activation.
        \item \textbf{Max Pooling layer} $\rho^{(2)}$: %$10 \times 10$ Input matrix $\hat{O}^{(2)}$, downsampled by 
        $2 \times 2$ max pooling kernel, with stride $s_{\rho}^{(2)}=2$, produces 16 output matrix $IN^{(3)}$ of size $5 \times 5$.
    \end{itemize}

    \item FCNN Round 1: 
    \begin{itemize}
        \item Flatten the matrix $IN^{(3)}$ and get the 400-dimensional vector $X^{(1)}$.
        \item \textbf{Fully-connected Layer} $f^{(1)}$: $d^{(1)}=400$ input neurons and $d^{(2)}=120$ output neurons, followed by ReLU activation.
    \end{itemize}

    \item FCNN Round 2:
    \begin{itemize}
        \item \textbf{Fully-connected Layer} $f^{(2)}$: $d^{(2)}=120$ input neurons and $d^{(3)}=84$ output neurons, followed by ReLU activation.
    \end{itemize}
    
    \item Last Layer $f^{(3)}$: $d^{(3)}=84$ input neurons and $d^{(3)}=10$ output neurons.
\end{enumerate}

\paragraph{(3+1)-Deep CNN:}
\begin{enumerate}
    \item Convolutional Round 1: 
    \begin{itemize}
        \item \textbf{Convolutional layer} $f_c^{(1)}$: $32 \times 32$ Input matrix $IN^{(1)}$, convolved with $(1,1,5,5)$ convolution kernel, with stride $s_c^{(1)}=1$, padding $pad^{(1)}=2$, produces $32 \times 32$ output matrix $O^{(1)}$, followed by ReLU activation.
        \item \textbf{Max Pooling layer} $\rho^{(1)}$: 
        $2 \times 2$ max pooling kernel, with stride $s_{\rho}^{(1)}=2$, produces $16 \times 16$ output matrix $IN^{(2)}$.
    \end{itemize}
    
    \item Convolutional Round 2:
    \begin{itemize}
        \item \textbf{Convolutional layer} $f_c^{(2)}$: $16 \times 16$ Input matrix $IN^{(2)}$, convolved with $(1,1,5,5)$ convolution kernel, with stride $s_c^{(2)}=1$, padding $pad^{(2)}=2$, produces $16 \times 16$ output matrix $O^{(2)}$, followed by ReLU activation.
        \item \textbf{Max Pooling layer} $\rho^{(2)}$: 
        $2 \times 2$ max pooling kernel, with stride $s_{\rho}^{(2)}=2$, produces $8 \times 8$ output matrix $IN^{(3)}$.
    \end{itemize}

    \item Convolutional Round 3:
    \begin{itemize}
        \item \textbf{Convolutional layer} $f_c^{(3)}$: $8 \times 8$ Input matrix $IN^{(3)}$, convolved with $(1,1,5,5)$ convolution kernel, with stride $s_c^{(3)}=1$, padding $pad^{(3)}=2$, produces $8 \times 8$ output matrix $O^{(3)}$, followed by ReLU activation.
        \item \textbf{Max Pooling layer} $\rho^{(3)}$: 
        $2 \times 2$ max pooling kernel, with stride $s_{\rho}^{(3)}=2$, produces $4 \times 4$ output matrix $IN^{(4)}$.
    \end{itemize}

    \item FCNN Round 1: 
    \begin{itemize}
        \item Flatten the matrix $IN^{(4)}$ and get the 16-dimensional vector $X^{(1)}$.
        \item \textbf{Fully-connected Layer} $f^{(1)}$: $d^{(1)}=16$ input neurons and $d^{(2)}=25$ output neurons, followed by ReLU activation.
    \end{itemize}
    
    \item Last Layer $f^{(2)}$: $d^{(2)}=25$ input neurons and $d^{(3)}=10$ output neurons.
\end{enumerate}

\paragraph{(2+2)-Deep CNN trained on CIFAR 10:}
\begin{enumerate}
    \item Convolutional Round 1: 
    \begin{itemize}
        \item \textbf{Convolutional layer} $f_c^{(1)}$: 3 Input matrix $IN^{(1)}$ of size $32 \times 32$, convolved with $(1,1,3,3)$ convolution kernel, with stride $s_c^{(1)}=1$, padding $pad^{(1)}=1$, produces 3 output matrix $O^{(1)}$ of size $32 \times 32$ , followed by ReLU activation.
        \item \textbf{Max Pooling layer} $\rho^{(1)}$: 
        $2 \times 2$ max pooling kernel, with stride $s_{\rho}^{(1)}=2$, produces 3 output matrix $IN^{(2)}$ of size $16 \times 16$ .
    \end{itemize}
    
    \item Convolutional Round 2:
    \begin{itemize}
        \item \textbf{Convolutional layer} $f_c^{(2)}$: 3 Input matrix $IN^{(2)}$ of size $16 \times 16$, convolved with $(3,4,3,3)$ convolution kernel, with stride $s_c^{(2)}=1$, padding $pad^{(2)}=1$, produces 4 output matrix $O^{(2)}$ of size $16 \times 16$, followed by ReLU activation.
        \item \textbf{Max Pooling layer} $\rho^{(2)}$: 
        $2 \times 2$ max pooling kernel, with stride $s_{\rho}^{(2)}=2$, produces 4 output matrix $IN^{(3)}$ of size $8 \times 8$.
    \end{itemize}

    \item FCNN Round 1: 
    \begin{itemize}
        \item Flatten the matrix $IN^{(3)}$ and get the 256-dimensional vector $X^{(1)}$.
        \item \textbf{Fully-connected Layer} $f^{(1)}$: $d^{(1)}=256$ input neurons and $d^{(2)}=256$ output neurons, followed by ReLU activation.
    \end{itemize}

    \item FCNN Round 2:
    \begin{itemize}
        \item \textbf{Fully-connected Layer} $f^{(2)}$: $d^{(2)}=256$ input neurons and $d^{(3)}=120$ output neurons, followed by ReLU activation.
    \end{itemize}
    
    \item Last Layer $f^{(3)}$: $d^{(3)}=120$ input neurons and $d^{(3)}=10$ output neurons.
\end{enumerate}

\paragraph{(2+1)-Deep LeNet-5:}
\label{paragraph:(2+1)-Deep LeNet-5}
\begin{enumerate}
    \item Convolutional Round 1: 
    \begin{itemize}
        \item \textbf{Convolutional layer} $f_c^{(1)}$: $32 \times 32$ Input matrix $IN^{(1)}$, convolved with $(6,1,5,5)$ convolution kernel, with stride $s_c^{(1)}=1$, padding $pad^{(1)}=0$, produces 6 output matrices $O^{(1)}$ of size $28 \times 28$ , followed by ReLU activation.
        \item \textbf{Max Pooling layer} $\rho^{(1)}$: %6 Input matrix $\hat{O}^{(1)}$ of size $28 \times 28$, downsampled by 
        $2 \times 2$ max pooling kernel, with stride $s_{\rho}^{(1)}=2$, produces 6 output matrices $IN^{(2)}$ of size $14 \times 14$ .
    \end{itemize}
    
    \item Convolutional Round 2:
    \begin{itemize}
        \item \textbf{Convolutional layer} $f_c^{(2)}$: 6 Input matrices $IN^{(2)}$ of size $14 \times 14$, convolved with $(6,16,5,5)$ convolution kernel, with stride $s_c^{(2)}=1$, padding $pad^{(2)}=0$, produces 16 output matrices $O^{(2)}$ of size $10 \times 10$, followed by ReLU activation.
        \item \textbf{Max Pooling layer} $\rho^{(2)}$: %$10 \times 10$ Input matrix $\hat{O}^{(2)}$, downsampled by 
        $2 \times 2$ max pooling kernel, with stride $s_{\rho}^{(2)}=2$, produces 16 output matrix $IN^{(3)}$ of size $5 \times 5$.
    \end{itemize}

    \item FCNN Round 1: 
    \begin{itemize}
        \item Flatten the matrix $IN^{(3)}$ and get the 400-dimensional vector $X^{(1)}$.
        \item \textbf{Fully-connected Layer} $f^{(1)}$: $d^{(1)}=400$ input neurons and $d^{(2)}=20$ output neurons, followed by ReLU activation.
    \end{itemize}
    
    \item Last Layer $f^{(2)}$: $d^{(2)}=20$ input neurons and $d^{(3)}=10$ output neurons.
\end{enumerate}

\paragraph{Modern (2+2)-Deep LeNet-5:}
\label{paragraph:(2+2)-Deep LeNet-5}
\begin{enumerate}
    \item Convolutional Round 1: 
    \begin{itemize}
        \item \textbf{Convolutional layer} $f_c^{(1)}$: $32 \times 32$ Input matrix $IN^{(1)}$, convolved with $(6,1,5,5)$ convolution kernel, with stride $s_c^{(1)}=1$, padding $pad^{(1)}=0$, produces 6 output matrices $O^{(1)}$ of size $28 \times 28$ , followed by ReLU activation.
        \item \textbf{Max Pooling layer} $\rho^{(1)}$: %6 Input matrix $\hat{O}^{(1)}$ of size $28 \times 28$, downsampled by 
        $2 \times 2$ max pooling kernel, with stride $s_{\rho}^{(1)}=2$, produces 6 output matrices $IN^{(2)}$ of size $14 \times 14$ .
    \end{itemize}
    
    \item Convolutional Round 2:
    \begin{itemize}
        \item \textbf{Convolutional layer} $f_c^{(2)}$: 6 Input matrices $IN^{(2)}$ of size $14 \times 14$, convolved with $(6,16,5,5)$ convolution kernel, with stride $s_c^{(2)}=1$, padding $pad^{(2)}=0$, produces 16 output matrices $O^{(2)}$ of size $10 \times 10$, followed by ReLU activation.
        \item \textbf{Max Pooling layer} $\rho^{(2)}$: %$10 \times 10$ Input matrix $\hat{O}^{(2)}$, downsampled by 
        $2 \times 2$ max pooling kernel, with stride $s_{\rho}^{(2)}=2$, produces 16 output matrix $IN^{(3)}$ of size $5 \times 5$.
    \end{itemize}

    \item FCNN Round 1: 
    \begin{itemize}
        \item Flatten the matrix $IN^{(3)}$ and get the 400-dimensional vector $X^{(1)}$.
        \item \textbf{Fully-connected Layer} $f^{(1)}$: $d^{(1)}=400$ input neurons and $d^{(2)}=120$ output neurons, followed by ReLU activation.
    \end{itemize}

    \item FCNN Round 2: 
    \begin{itemize}
        \item \textbf{Fully-connected Layer} $f^{(2)}$: $d^{(2)}=120$ input neurons and $d^{(3)}=84$ output neurons, followed by ReLU activation.
    \end{itemize}
    
    \item Last Layer $f^{(3)}$: $d^{(3)}=84$ input neurons and $d^{(4)}=10$ output neurons.
\end{enumerate}

\section{Noise Robustness Analysis on the Reduced LeNet-5}
\label{sec:appendix_cnn_noise}

In this section, we take the black-box end-to-end attack on a reduced version of LeNet-5 as an example. The target CNN consists of two Convolutional blocks, one Fully-connected Block, and a final fully-connected layer, yielding exactly four layers of parameters to extract. 
While the overall end-to-end performance is summarized in Table \ref{tab:Attack on CNNs} (Section \ref{subsec:our contributions}) and detailed model structures are given in \textsf{Supplementary Material} \ref{supp:detailed model structure}, this section presents a detailed layer-by-layer extraction record. 

% It is important to note that the parameters of the CNN are not mathematically unique for given inputs and outputs. There exist inherent symmetries within the network architecture: (1) {\em Permutation Invariance}: permuting the order of the output channels within a convolutional layer, or hidden neurons within a fully-connected layer, along with their corresponding incoming and outgoing weights, does not change the network's output; (2) {\em Positive Scaling Invariance}: multiplying a convolution kernel's weights and bias by a positive scalar $c>0$ while dividing the corresponding weights in the subsequent layer by $c$ yields an identical network output.

% Consequently, our attack recovers the parameters up to these permutations and positive scalars. To directly evaluate the extraction error, we first apply the parameter alignment procedure introduced by Carlini {\em et al.} in \cite{DBLP:conf/crypto/CarliniJM20} to align the recovered kernel weight $\hat{C}^{(k)}$ and bias $\hat{b}^{(k)}$) with the original parameters $C^{(k)}$ and $b^{(k)}$. We then construct the corresponding convolutional matrices and bias vectors to compare the recovered $\hat{A}^{(k)}$ and $\hat{B}^{(k)}$ against the original $A^{(k)}$ and $B^{(k)}$.

\begin{table}[H] % 使用 !ht 强化定位
\centering
% \captionsetup{font={color=purple}}
\caption{Layer-wise Recovery Errors of the (2+1) reduced LeNet-5 Model}
\label{tab:layer_error_comparison}
\renewcommand{\arraystretch}{1.5}
\setlength{\tabcolsep}{6pt} % 稍微加宽一点，3pt太挤了

\begin{threeparttable}
    \begin{tabular}{c c c c c c} 
    \toprule
    \makecell[c]{\textbf{Layer}} 
    & \makecell[c]{\textbf{Architecture}$^{*}$ \\ $d^{(k)}\!-\!d_f^{(k)}\!-\!d^{(k+1)}$}
    & \makecell[c]{\textbf{Relative} \\ \textbf{Error}} 
    & \makecell[c]{\textbf{Max Rel} \\ \textbf{Error}}
    & \boldmath$\max \lvert \theta - \hat{\theta} \rvert$
    & \boldmath $(\varepsilon, 0)$ \\
    \midrule
    
    Conv R1
    & 1024--4704--1176 
    & $2^{-31.78}$ & $2^{-28.97}$ & $2^{-33.45}$ &  $2^{-27.86}$\\
    \midrule
    
    Conv R2
    & 1176--1600--400 
    & $2^{-27.55}$ & $2^{-23.94}$ & $2^{-30.78}$ &  $2^{-24.91}$\\
    \midrule
    
    FCNN R1
    & 400--20 
    & $2^{-23.21}$ & $2^{-19.61}$ & $2^{-27.26}$ &  $2^{-23.13}$\\
    \midrule
    
    FCNN R2
    & 20--10 
    & $2^{-26.26}$ & $2^{-22.61}$ & $2^{-27.26}$ &  $2^{-22.21}$\\
    \bottomrule
    \end{tabular}
    \begin{tablenotes}
        \footnotesize
        \item $^{*}$ : The architectures of the layers in FCNN Block (FCNN Round 1 and Round 2) are denoted in the form of $d^{(k)}\!-\!d^{(k+1)}$.
    \end{tablenotes}
\end{threeparttable}
\end{table}

Table \ref{tab:layer_error_comparison} reports the layer-wise error propagation evaluated across four metrics: Relative Error, element-wise Maximum Relative Error, element-wise Maximum Absolute Error ($\max \lvert \theta - \hat{\theta} \rvert$ \cite{DBLP:conf/crypto/CarliniJM20}), and the $(\varepsilon, 0)$-functional equivalence \cite{DBLP:conf/crypto/CarliniJM20}.

{\em Relative Error} is defined as the maximum \text{L2}-based relative error between the parameters and the aligned recovered parameters, 
\begin{equation}
\text{Relative Error} = \max\left(
\frac{\left\|\hat{A}^{(k)}-A^{(k)}\right\|_{2}}{\left\|A^{(k)}\right\|_{2}},
\frac{\left\|\hat{b}^{(k)}-b^{(k)}\right\|_{2}}{\left\|b^{(k)}\right\|_{2}}
\right).
\end{equation}

{\em Max Rel Error} denotes the element-wise Maximum Relative Error,
\begin{equation}
\text {Max Rel Error}=\max\Bigg(\max\left| \frac {\hat {A}^{(k)}_{i,j}-A^{(k)}_{i,j}}{A^{(k)}_{i,j}} \right|,\max\left| \frac{\hat{b}^{(k)}_{i}-b^{(k)}_{i}}{b^{(k)}_{i}} \right|\Bigg).
\end{equation}

The $\max \lvert \theta - \hat{\theta} \rvert$ denotes the global maximum absolute element-wise deviation between the original and recovered parameters across the entire recovered $k$-layers network. Formally,
\begin{equation}
\max \lvert \theta - \hat {\theta} \rvert= \max_{1 \le t \le k}\Bigg (\max\big| \hat {A}^{(t)}_{i,j}-A^{(t)}_{i,j}\big|, ~\max\big| \hat{B}^{(t)}_{i}-B^{(t)}_{i}\big|\Bigg).
\end{equation}

As described in Section \ref{sect:assumptions_1}, $(\varepsilon, 0)$-functional equivalence denotes the maximum output error bound derived via forward error propagation. 
After aligning the recovered and original parameters, let $e_{k}$ denote the input error to layer $k$ and $\delta_i$ denote the largest singular value of $\hat{A}^{(k)}-A^{(k)}$, the error propagates according to \cite{DBLP:conf/crypto/CarliniJM20}:
\begin{equation}
e_{k+1} \leq 
\delta_k
\cdot e_{k}+\left\| \hat{B}^{(k)}-B^{(k)}\right\|_{2}.
\end{equation}

All four error metrics demonstrate that our proposed attack achieves extremely high recovery precision with well-controlled error accumulation.

\section{Analysis on Unknown Kernel Size}
\label{sec:unknown kernel size}

This section explores a more restricted scenario where the convolutional kernel size is unknown to the attacker.
We argue that our proposed attack remains effective, albeit requiring additional offline analysis and computational attempts. It is worth noting that this process -- inferring the bounds of the kernel size and enumerating potential configurations -- relies solely on the extracted solution vectors. Therefore, it is conducted entirely offline and does not require any further queries to the victim model.

\subsubsection{Analyzing the Bound of the Kernel Size.}
First, an attacker can simply guess the kernel size based on common practices. Modern CNNs tend to use small, standardized kernel sizes to achieve better feature extraction (e.g., $3 \times 3$, $5 \times 5$). Consequently, the search space is highly restricted. An attacker can simply enumerate the possible kernel sizes and strides based on empirical values.

Furthermore, our attack inherently extracts additional structural information, significantly reducing the reliance on blind guessing. The solution vectors ($A^{(k)}_i$ of the linear system in Eq.~\eqref{eqn:diff_sign} and $A^{(k)}_i - A^{(k)}_j$ of the system in \eqref{eqn:psp_Aij}) expose the structural information of the convolution kernel. From these vectors, we can infer the bounds of the kernel size by identifying the length of contiguous non-zero segments.

Specifically, in practice, the row vector $A^{(k)}_i$ of a convolutional matrix for an RPCP is partially recovered because some neurons in the $(k-1)$-th layer are suppressed by ReLU. Recall from Eq.~\eqref{eqn:A_row_nan} that the recovered $\hat{A}_i^{(k)}$ could be 
\begin{equation*} \tiny
    \hat{A}^{(k)}_i = (\underbrace{\bf 0}_{w^{(k)}_{in}\cdot \bar{i}},\underbrace{
    \underbrace{\bf 0}_{\underline{i}}, \hat{C}^{(k)}_{0,0}, {\sf NaN}, \cdots, \hat{C}^{(k)}_{0,w_{c}^{(k)}-1},    \underbrace{\bf 0}_{pad}}_{w_{in}^{(k)}}, 
    \cdots, 
    \underbrace{
    \underbrace{\bf 0}_{\underline{i}}, 
    {\sf NaN}, {\sf NaN}, \cdots, \hat{C}^{(k)}_{h_{c}^{(k)}-1,w_{c}^{(k)}-1},   \underbrace{\bf 0}_{pad}}_{w_{in}^{(k)}},  
    \underbrace{\bf 0}_{pad}),
\end{equation*}
where $\bar{i}=\left\lfloor \frac{i}{w^{(k)}_{o}} \right\rfloor$, $\underline{i}=i\bmod w^{(k)}_o$. Note that, the length of the contiguous non-zero values in each row vectors $A^{(k)}_i$ reflects the kernel width. However, any entry in the recovered $\hat{A}_i^{(k)}$ may contain a {\sf NaN} value, including those entries whose actual value should be 0. Assuming there are $n$ contiguous non-zero segments in $\hat{A}_i^{(k)}$, with lengths denoted by $l_j$ for $0 \leq j \leq n-1$, the actual kernel width $w^{(k)}_c$ is less than or equal to the length of any such segment, {\em i.e.}, 
\begin{align}
    \label{eq:kernel width bound}
    \omega^{(k)}_c \leq \min{\{ l_j, ~ 0 \leq j \leq n-1 \}}.
\end{align}
The kernel height $h^{(k)}_c$ is bounded by the total number of these non-zero segments, {\em i.e.},
\begin{align}
    \label{eq:kernel height bound}
    h^{(k)}_c \leq n.
\end{align}

For a PSP, the non-zero entries in the calculated row vector $\Delta_{ij} \hat{A}^{(k)} \approx g_t (A_{i}^{(k)} - A_{j}^{(k)})$, solved from the linear system in Eq.~\eqref{eqn:psp_Aij}, represents the union of those in $A^{(k)}_i$ and $A^{(k)}_j$ because their non-zero values are are offset from each other. Consequently, Eqs.~\eqref{eq:kernel width bound} and \eqref{eq:kernel height bound} still hold, {\em i.e.}, the length of each contiguous non-zero segment is greater than or equal to the kernel width ($w^{(k)}_c \leq \min{\{ l_j, ~ 0 \leq j \leq n-1 \}}$), and similarly, the kernel height is bounded by the number of these segments ($h^{(k)}_c \leq n$).

\subsubsection{Enumerating the Kernel Size.}
Recall from the {\em Pattern Matching Method} in Sect. \ref{sect:rpcp_nan} that we determine the correct neuron index $i$ by verifying whether the convolution kernel pattern vector $\vec{V}_i$ and the recovered $\hat{A}^{(k)}_i$ strictly follow the {\em matching principle} (requiring ``Non-zero Consistency'' and ``Zero Consistency''). When the kernel size is unknown, we can enumerate candidate values for the kernel width $w^{(k)}_c$ and height $h^{(k)}_c$ within the previously inferred bounds. For each guessed kernel size, we attempt to find a matching neuron $i$. The guessed sizes $(w^{(k)}_c, h^{(k)}_c)$ that successfully yield a valid neuron $i$ form a candidate set of kernel sizes, denoted as $\mathcal{K}$.

Similarly, for a PSP, recall from Sect. \ref{sect:psp_ij} that we extract the competing neurons' indices $i$ and $j$ by checking whether the difference vector $A^{(k)}_i-A^{(k)}_j$ and the recovered $\Delta_{ij} \hat{A}^{(k)}$ satisfy the {\em matching principle} (requiring ``Union Non-zero Consistency'' and ``Common Zero Consistency''). By enumerating the candidate dimensions $w^{(k)}_c$ and $h^{(k)}_c$, we evaluate potential pairs $(i, j)$. The guessed kernel sizes that successfully yield a valid pair $(i, j)$ form the candidate size set $\mathcal{K}$ for this specific PSP.

With multiple critical points (including both RPCPs and PSPs), we can uniquely determine the true kernel size. By taking the intersection of the candidate size sets ($\bigcap \mathcal{K}$) derived above, we rapidly filter out incorrect guesses, ultimately isolating the exact structural dimensions of the convolution kernel.

\section{Analysis of Other Structural Hyperparameters}
\label{supp:other_parameters}

In this section, we discuss some variants of CNN. Specifically, we focus on modifications to those structural hyperparameters, including padding, stride (in both convolutional and pooling layers), and dropout mechanisms.  Naturally, this raises a question: \textit{Will our attack still succeed with the modifications to the hyperparameters?}

% \textit{Can we still successfully recover the model parameters without prior knowledge of these concrete hyperparameter configurations?}

\subsection{Zero-padding in the Convolutional Layer}
To %maintain the spatial dimensions of the feature map after convolution and 
fully capture the features near the boundaries,  zeros are usually padded around the input matrix $IN^{(k)}$.  For example,  $pad^{(k)}=1$ denotes adding one row/column of zeros to all four sides of $IN^{(k)}$. Consequently, the output matrix $O^{(k)}$ after convolution with $pad^{(k)}$ has a shape of $h_o^{(k)} \times w_o^{(k)}$, where
\begin{align}
        h^{(k)}_{o} = \left\lfloor \frac{h^{(k)}_{in} + 2 \cdot pad^{(k)}- h^{(k)}_{c}}{s^{(k)}_c} \right\rfloor + 1, \ 
        w^{(k)}_{o} = \left\lfloor \frac{w^{(k)}_{in} + 2 \cdot pad^{(k)}- w^{(k)}_{c}}{s^{(k)}_c} \right\rfloor + 1
    \end{align}
Two common types of padding are widely used in practice:
\begin{enumerate}
    \item \textbf{Valid padding.} No zero padding is applied ($pad^{(k)}=0$). The convolution kernel is restricted to visit only positions where it is fully contained within the input boundaries.
    
    \item \textbf{Same padding.} Zeros are padded around the input matrix such that the size of the output feature map is identical to that of the input.
    
    % \item \textbf{Full padding.} Sufficient zeros are added for every position of the input matrix to be visited the same times, resulting in a feature map larger than the input.
\end{enumerate}

\begin{figure}
	\centering
        \includegraphics[width=0.5\linewidth]{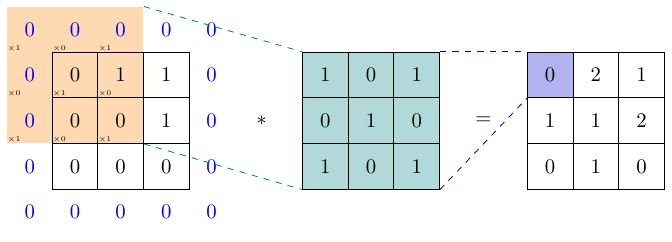}
	\caption{Example of Convolution Operation with zero padding}
	\label{fig:conv_f_padding}
\end{figure}
Figure \ref{fig:conv_f_padding} illustrates a scenario using same padding, with $\mathsf{c}_{in}^{(k)}=\mathsf{c}_{out}^{(k)}=s_c^{(k)}=1$, where the input matrix $IN^{(k)}$ has dimensions $h_{in}^{(k)} \times w_{in}^{(k)} = 3 \times 3$, {\em i.e.}, $d^{(k)}=9$. Given a $h^{(k)}_{c} \times w^{(k)}_{c} = 3 \times 3$ kernel matrix $C^{(k)}$, with the padding $pad^{(k)}=1$, the output matrix $O^{(k)}$ is of dimension $h_o^{(k)}\times w_o^{(k)}=3 \times 3$, {\em i.e.}, $d^{(k)}_{f}=9$. 
In the algebraic view, the padding operation can be modeled as padding zeros to the input vector $X^{(k)}$ and get the zero-padded input vector $X_{pad}^{(k)} \in \mathbb{R}^{d_{pad}^{(k)}}$, $d_{pad}^{(k)}=(h_{in}^{(k)}+2 \cdot pad^{(k)})(w_{in}^{(k)}+2 \cdot pad^{(k)})$, {\em i.e.}, $d_{pad}^{(k)}=25$. 
According the Def. \ref{def:conv_matrix}, the convolutional matrix corresponding to $X_{pad}^{(k)}$ is:
\begin{equation*}
    % 1. 设置列间距极小，减少不必要的空白
    \setlength{\arraycolsep}{1.5pt} 
    % 2. 强制压缩整个公式到版心宽度
    \resizebox{\textwidth}{!}{%
    $ % 注意：resizebox 内部进入了文本模式，需要重新用 $ 包裹数学模式
    \tiny % 保持字体尽可能小
    A_{pad}^{(k)} = \left(
    \begin{array}{*{25}{c}} % 25列
        c^{(k)}_{0,0} &c^{(k)}_{0,1} &c^{(k)}_{0,2} &0 &0
        &c^{(k)}_{1,0} &c^{(k)}_{1,1} &c^{(k)}_{1,2} &0 &0 &c^{(k)}_{2,0} &c^{(k)}_{2,1} &c^{(k)}_{2,2} &0 &0
        &0 &0 &0 &0 &0 
        &0 &0 &0 &0 &0 \\ %1

        0 &c^{(k)}_{0,0} &c^{(k)}_{0,1} &c^{(k)}_{0,2} &0
        &0 &c^{(k)}_{1,0} &c^{(k)}_{1,1} &c^{(k)}_{1,2} &0 
        &0 &c^{(k)}_{2,0} &c^{(k)}_{2,1} &c^{(k)}_{2,2} &0
        &0 &0 &0 &0 &0 
        &0 &0 &0 &0 &0 \\ %2

        0  &0 &c^{(k)}_{0,0} &c^{(k)}_{0,1} &c^{(k)}_{0,2}
        &0 &0 &c^{(k)}_{1,0} &c^{(k)}_{1,1} &c^{(k)}_{1,2} 
        &0 &0 &c^{(k)}_{2,0} &c^{(k)}_{2,1} &c^{(k)}_{2,2}
        &0 &0 &0 &0 &0 
        &0 &0 &0 &0 &0 \\ %3

        0 &0 &0 &0 &0 
        &c^{(k)}_{0,0} &c^{(k)}_{0,1} &c^{(k)}_{0,2} &0 &0
        &c^{(k)}_{1,0} &c^{(k)}_{1,1} &c^{(k)}_{1,2} &0 &0 &c^{(k)}_{2,0} &c^{(k)}_{2,1} &c^{(k)}_{2,2} &0 &0
        &0 &0 &0 &0 &0 \\ %4

        0 &0 &0 &0 &0
        &0 &c^{(k)}_{0,0} &c^{(k)}_{0,1} &c^{(k)}_{0,2} &0
        &0 &c^{(k)}_{1,0} &c^{(k)}_{1,1} &c^{(k)}_{1,2} &0 
        &0 &c^{(k)}_{2,0} &c^{(k)}_{2,1} &c^{(k)}_{2,2} &0 
        &0 &0 &0 &0 &0 \\ %5

        0 &0 &0 &0 &0 
        &0 &0 &c^{(k)}_{0,0} &c^{(k)}_{0,1} &c^{(k)}_{0,2}
        &0 &0 &c^{(k)}_{1,0} &c^{(k)}_{1,1} &c^{(k)}_{1,2} 
        &0 &0 &c^{(k)}_{2,0} &c^{(k)}_{2,1} &c^{(k)}_{2,2}
        &0 &0 &0 &0 &0 \\ %6
        
        0 &0 &0 &0 &0
        &0 &0 &0 &0 &0 
        &c^{(k)}_{0,0} &c^{(k)}_{0,1} &c^{(k)}_{0,2} &0 &0
        &c^{(k)}_{1,0} &c^{(k)}_{1,1} &c^{(k)}_{1,2} &0 &0 &c^{(k)}_{2,0} &c^{(k)}_{2,1} &c^{(k)}_{2,2} &0 &0 
        \\ %7
        
        0 &0 &0 &0 &0 
        &0 &0 &0 &0 &0 
        &0 &c^{(k)}_{0,0} &c^{(k)}_{0,1} &c^{(k)}_{0,2} &0
        &0 &c^{(k)}_{1,0} &c^{(k)}_{1,1} &c^{(k)}_{1,2} &0 
        &0 &c^{(k)}_{2,0} &c^{(k)}_{2,1} &c^{(k)}_{2,2} &0
        \\ %8

        0 &0 &0 &0 &0 
        &0 &0 &0 &0 &0 
        &0 &0 &c^{(k)}_{0,0} &c^{(k)}_{0,1} &c^{(k)}_{0,2}
        &0 &0 &c^{(k)}_{1,0} &c^{(k)}_{1,1} &c^{(k)}_{1,2} 
        &0 &0 &c^{(k)}_{2,0} &c^{(k)}_{2,1} &c^{(k)}_{2,2}
        \\ %9 
    \end{array}
    \right)
    $
    }.
\end{equation*}

However, for the padded input vector $X_{pad}^{(k)}\in \mathbb{R}^{d_{pad}^{(k)}}$, we can only change the $d^{(k)}$ dimensions ($d^{(k)}<d_{pad}^{(k)}$), as the padded entries must remain fixed at zero. This constraint results in a loss of full control over the entire input space $\mathbb{R}^{d_{pad}^{(k)}}$. Instead, we continue to use the raw input vector $X^{(k)} \in \mathbb{R}^{d^{(k)}}$, and the convolutional matrix $A^{(k)}$ is obtained by extracting the columns of $A_{pad}^{(k)}$ that correspond to the elements of the raw input vector $X^{(k)}$ before padding. Let ${Col}^{(k)} \subset \{0,1,\cdots,d_{pad}^{(k)}\}$ denote the set of corresponding column indices, then 
\begin{equation}
\resizebox{1.0\hsize}{!}{$%
{Col}^{(k)}\mathrel{:=}\{(w^{(k)}_{in}+2pad^{(k)}) \cdot i+j, ~ pad^{(k)} \leq i \leq pad^{(k)}+h^{(k)}_{in}-1, ~ pad^{(k)} \leq j \leq pad^{(k)}+w^{(k)}_{in}-1\}.
$}%
\end{equation}

In this example, the column indices ${Col}^{(k)}=\{6, \dots, 8, 11, \dots, 13, 16, \dots, 18\}$,
\begin{equation*}
    %\label{eqn:conv matrix in padding}
    %\setlength{\arraycolsep}{3pt}
    A^{(k)} = \left(
    \begin{array}{*{9}{c}}
        c^{(k)}_{1,1} &c^{(k)}_{1,2} &0  
        &c^{(k)}_{2,1} &c^{(k)}_{2,2} &0 
        &0 &0 &0\\ %1

        c^{(k)}_{1,0} &c^{(k)}_{1,1} &c^{(k)}_{1,2} 
        & c^{(k)}_{2,0} &c^{(k)}_{2,1} &c^{(k)}_{2,2}
        &0 &0 &0 \\ %2

        0 &c^{(k)}_{1,0} &c^{(k)}_{1,1} 
        &0 &c^{(k)}_{2,0} &c^{(k)}_{2,1}
        &0 &0 &0\\ %3

        c^{(k)}_{0,1} &c^{(k)}_{0,2} &0
        &c^{(k)}_{1,1} &c^{(k)}_{1,2} &0 
        &c^{(k)}_{2,1} &c^{(k)}_{2,2} &0 \\ %4

        c^{(k)}_{0,0} &c^{(k)}_{0,1} &c^{(k)}_{0,2} 
        &c^{(k)}_{1,0} &c^{(k)}_{1,1} &c^{(k)}_{1,2} &c^{(k)}_{2,0} &c^{(k)}_{2,1} &c^{(k)}_{2,2}\\ %5

        0 &c^{(k)}_{0,0} &c^{(k)}_{0,1}
        &0 &c^{(k)}_{1,0} &c^{(k)}_{1,1}
        &0 &c^{(k)}_{2,0} &c^{(k)}_{2,1}\\ %6
        
        0 &0 &0
        &c^{(k)}_{0,1} &c^{(k)}_{0,2} &0 
        &c^{(k)}_{1,1} &c^{(k)}_{1,2} &0\\ %7
        
        0 &0 &0
        &c^{(k)}_{0,0} &c^{(k)}_{0,1} &c^{(k)}_{0,2}  
        &c^{(k)}_{1,0} &c^{(k)}_{1,1} &c^{(k)}_{1,2}\\ %8

        0 &0 &0
        &0 &c^{(k)}_{0,0} &c^{(k)}_{0,1}
        &0 &c^{(k)}_{1,0} &c^{(k)}_{1,1}\\ %9

    \end{array}
    \right).
\end{equation*}
%Directly padding zeros to the input vector $X^{(k)}$ will lose full control of the entire input space. Instead, we continue to flatten the raw input matrix to the vector. Accordingly, the convolutional kernel matrix needs to be augmented with additional rows. These rows correspond to positions where the kernel overlaps the padded zeros. Only part of the kernel interacts with the valid input data, so these addtional rows contains truncated subset of the kernel weights, while strictly preserving their relative spatial positions. 

% Figure \ref{} illustrates a $3 \times 3$ kernel applied to the top-left corner of a $5 \times 5$ input matrix with $pad^{(k)}=1$, 
% $$A^{(k)}_0=\left( c^{(k)}_5 \ c^{(k)}_6 \ 0 \ 0 \ 0 \ 0 \ c^{(k)}_8 \ c^{(k)}_9 \ \underbrace{0 \ 0 \cdots \ 0 \ 0}_{17~zeros} \right)$$
% when the kernel acts on the top side of the input matrix,
% $$A^{(k)}_1=\left( c^{(k)}_4 \ c^{(k)}_5 \ c^{(k)}_6 \ 0 \ 0 \ 0 \ c^{(k)}_7 \ c^{(k)}_8 \ c^{(k)}_9 \ \underbrace{0 \cdots \ 0 \ 0}_{16~zeros} \right)$$

In our attack, the RPCP and PSP methods are designed for critical points corresponding to vector $A^{(k)}_i$ and $A^{(k)}_i-A^{(k)}_j$, respectively. Both $A^{(k)}_i$ and $A^{(k)}_j$ need to contain the full information about the kernel matrix (e.g., $A^{(k)}_{4}$ in this case). Other row vectors with truncated subset of the kernel weights (e.g., $A^{(k)}_{0} {\sim} A^{(k)}_{3}$ and $A^{(k)}_{5} {\sim} A^{(k)}_{8}$ in this case) will interfere with the merging of partial weights in Sect. \ref{sect:rpcp_nan} and the computation of $\vec{C}^{(k)}$ in Sect. \ref{subsect:internal differential}.

However, the pattern matching method for RPCPs in Sect. \ref{sect:rpcp_nan} is designed to identify the row vector with full kernel parameters, this will filter out noisy RPCPs. For PSPs, after extracting $i$ and $j$ from $\Delta_{ij} \hat{A}^{(k)}=A^{(k)}_i - A^{(k)}_j$ in Sect. \ref{sect:psp_ij}, we can deduce $\mathcal{K}_i$ and $\mathcal{K}_j$ which denote the set of indices belonging to the LRF-C of the $i$-th and $j$-th neuron, respectively, according to Eq. \ref{eq:indices K}. Denote the first element in $\mathcal{K}_i$ as $\mathcal{K}_i[0]$, and the last element in $\mathcal{K}_j$ as $\mathcal{K}_j[l^{(k)}_c-1]$, where $l^{(k)}_c=h^{(k)}_c w^{(k)}_c$). Appropriate $i$ and $j$ follow the principle: $\frac{\Delta_{ij} \hat{A}^{(k)}[\mathcal{K}_i[0]]}{\Delta_{ij} \hat{A}^{(k)}[\mathcal{K}_j[l^{(k)}_c-1]]}=\frac{\vec{C}^{(k)}[0]}{\vec{C}^{(k)}[l^{(k)}_c-1]}$.
This enables us to cluster the $\frac{\Delta_{ij} \hat{A}^{(k)}[\mathcal{K}_i[0]]}{\Delta_{ij} \hat{A}^{(k)}[\mathcal{K}_j[l^{(k)}_c-1]]}$ from each PSP and discard noisy PSPs (if $\Delta_{ij} \hat{A}^{(k)}[\mathcal{K}_i[0]]={\sf NaN}$ or $\Delta_{ij} \hat{A}^{(k)}[\mathcal{K}_j[l^{(k)}_c-1]]={\sf NaN}$, also discard them).

A small fraction of the differences $A^{(k)}_i - A^{(k)}_j$ from PSPs may yield row vectors structurally identical to the $A^{(k)}_i$ in RPCP method which satisfy the matching principle in Sect. \ref{sect:rpcp_nan} and are therefore identified as RPCPs. We filter these out via Step 5 in Sect. \ref{sect:unified_model}. Therefore we can effectively discard these false-positives cases, excluding them from the set of valid information used for reconstruction.

\subsection{Stride in Convolutional Layer}
If the convolution stride $s_c > 1$, the convolution kernel will skip certain positions when sliding over the input matrix, which can be regarded as downsampling the output of the full convolution ($s_c = 1$). However, this does not undermine our method. While it results in a lower density of available critical points, the algebraic integrity of the extracted row vector (specifically, $A^{(k)}_i$ derived from RPCPs in Sect. \ref{subsect: RPCP_signature} and $A_i^{(k)}-A_j^{(k)}$ derived from PSPs in Sect. \ref{sect:psp_ij}) remains complete and valid.

\subsection{Stride in Pooling Layer}
Different values of the pooling stride $s^{(k)}_\rho$ lead to two distinct scenarios:
\begin{enumerate}
    \item \textbf{Disjoint Pooling} ($s^{(k)}_\rho \geq h^{(k)}_m, w^{(k)}_m$).
    Some elements of the output $\hat{O}^{(k)}$ after convolution and activation may not be covered by the pooling window. 

    \item \textbf{Overlapping Pooling} ($s^{(k)}_\rho < h^{(k)}_m, w^{(k)}_m$).
    Pooling windows intersect, meaning a single neuron may belongs to several local receptive fields. 
\end{enumerate}

Under Disjoint Pooling, when $s^{(k)}_\rho > h^{(k)}_m, w^{(k)}_m$, the pooling kernel skips certain neurons. Note that neuron $i$ in the convolutional layer selects the $i$-th column of the pooling matrix $P^{(k)}$. If neuron $i$ shows the maximum value in the $t$-th LRF-P, then we have $P^{(k)}_{t,i}=1$, otherwise $P^{(k)}_{t,i}=0$. Therefore, these skipped neurons correspond to constant all-zero column vectors. However, this does not interfere with the identification and calculation of RPCPs and PSPs, as they will not be detected.

\begin{figure}[htbp] % htbp为浮动位置参数，保证图片排版合理
	\centering
	\begin{minipage}{0.45\linewidth}
		\centering
		\includegraphics[width=\linewidth]{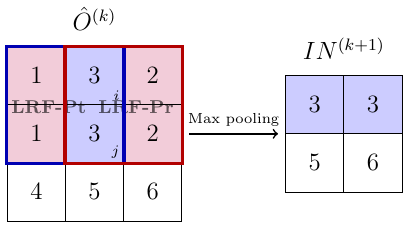}
		\caption{Overlapping Pooling case 1}
		\label{fig:Overlapping_Same}
	\end{minipage}
	\vrule width 1pt 
	\hspace{1pt}     
	\begin{minipage}{0.45\linewidth}
		\centering
		\includegraphics[width=\linewidth]{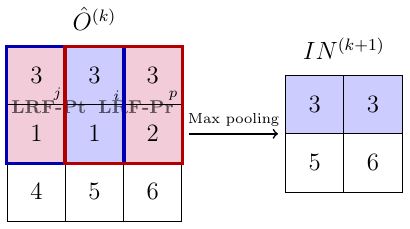}
		\caption{Overlapping Pooling case 2}
		\label{fig:Overlapping_Diff}
	\end{minipage}
\end{figure}

% Unlike FCNN, where two different critical neurons in the same layer will lead to weight combination, the calculation fails.
Under Overlapping Pooling, for RPCPs, a neuron in the critical state ({\em i.e.}, input=0) may show the maximum value across multiple overlapping pooling windows, resulting in multiple ones in its column vector. Any pooling window that selects this critical neuron, is sufficient to compute the row vector $A^{(k)}_i$. The simultaneous selection by multiple windows does not induce calculation failure. In the PSP method, suppose in the $t$-th LRF-P, neuron $i$ and $j$ achieve the same maximum value, as shown in the top-left of Fig. \ref{fig:Overlapping_Same} and Fig. \ref{fig:Overlapping_Diff} (outlined in blue).
\begin{enumerate}
    \item As illustrated in figure \ref{fig:Overlapping_Same}, if both neurons $i$ and $j$ also show the maximum value in another $r$-th LRF-P (outlined in red), the output difference still presents the gradient information $A^{(k)}_i-A^{(k)}_j$.
    \item As illustrated in figure \ref{fig:Overlapping_Diff} , if neuron $i$ ($j$) and another neuron $p$ show the maximum value in $r$-th LRF-P (outlined in red), with high probability we will get a combination of $A^{(k)}_i-A^{(k)}_j$ and $A^{(k)}_i-A^{(k)}_p$ ($A^{(k)}_j-A^{(k)}_p$). However, the extraction in Sect. \ref{sect:psp_ij} returns no valid result and therefore filters out this noise.
\end{enumerate}

\subsection{Dropout}
The dropout technique in CNN was first introduced by Krizhevsky {\em et al.}  \cite{krizhevsky2012imagenet} in 2012, working by setting the output of each hidden neuron to zero with a predefined probability during the training period. The neurons which are ``drop out'' in this way do not contribute to the forward  and backward procedure. However, this technique does not delete neurons from the network topology. The model function $f$ is still static and deterministic. From a practical attack perspective, this mechanism makes no difference to our attack.

\end{document}